\documentclass[prd,reprint,superscriptaddress,longbibliography,eqsecnum,nofootinbib,amsmath,amssymb,aps,floatfix]{revtex4-2}

\usepackage[utf8]{inputenc}
\usepackage[T1]{fontenc}
\usepackage{multirow}
\usepackage[dvipsnames]{xcolor}
\usepackage{graphicx}
\usepackage{bm}
\usepackage[colorlinks=true,allcolors=blue]{hyperref}
\usepackage{mathrsfs}
\usepackage{leftidx}

\newcommand{\mem}{\mathrm{mem}}
\newcommand{\PN}{\mathrm{PN}}
\newcommand{\surr}{\mathrm{surr}}
\newcommand{\insp}{\mathrm{insp}}
\newcommand{\intr}{\mathrm{int}}
\newcommand{\rd}{\mathrm{rd}}
\newcommand{\step}{\mathrm{step}}

\newcommand{\model}{\mathrm{model}}

\newcommand{\UVA}{Department of Physics, University of Virginia, P.O.~Box 400714, Charlottesville, Virginia 22904-7414, USA}

\begin{document}

\title{Waveform models for the gravitational-wave memory effect:\\
II.~Time-domain and frequency-domain models for nonspinning binaries}

\author{Arwa Elhashash}
\email{aze5tn@virginia.edu}
\affiliation{\UVA}%

\author{David A.~Nichols}%
\email{david.nichols@virginia.edu}
\affiliation{\UVA}

\date{\today}

\begin{abstract}

The nonlinear gravitational-wave (GW) memory effect\textemdash a permanent shift in the GW strain that arises from nonlinear GW interactions in the wave zone\textemdash is a prediction of general relativity which has not yet been observed.
The amplitude of the GW memory effect from binary-black-hole (BBH) mergers is small compared to that of primary (oscillatory) GWs and is unlikely to be detected by current ground-based detectors.
Evidence for its presence in the population of all the BBH mergers is more likely, once thousands of detections are made by these detectors.
Having an accurate and computationally efficient waveform model of the memory signal will assist detecting the memory effect with current data-analysis pipelines.
In this paper, we build on our prior work to develop analytical time-domain and frequency-domain models for the dominant nonlinear memory multipole signal ($l=2$, $m=0$) from nonspinning BBH mergers in quasicircular orbits.
The model is calibrated for mass ratios between one and eight.
There are three parts to the time-domain signal model: a post-Newtonian inspiral, a quasinormal-mode-based ringdown, and a phenomenological signal during the late inspiral and merger (which interpolates between the inspiral and ringdown).
The time-domain model also has an analytical Fourier transform, which we compute in this paper.
We assess the accuracy of our model using the mismatch between our waveform model and the memory signal computed from the oscillatory modes of a numerical-relativity surrogate model.
We use the advanced LIGO sensitivity curve from the fourth observing run and find that the mismatch increases with the total mass of the system and is of order $10^{-2}$\textendash$10^{-4}$.
\end{abstract}

\maketitle

\tableofcontents

\section{Introduction} \label{sec:intro}

Gravitational waves (GWs) from the mergers of nearly 100 binary black holes have been announced by the Laser Interferometer Gravitational-Wave Observatory (LIGO) and Virgo collaborations~\cite{LIGOScientific:2018mvr,LIGOScientific:2020ibl,LIGOScientific:2021djp}.
Over 200 more detection candidates have been announced during the fourth observing run of the LIGO-Virgo-KAGRA (LVK) collaboration to date~\cite{web:GraceDB}.
The detections have been used to test the predictions of general relativity in the strong-gravity and high-luminosity regime of the theory~\cite{LIGOScientific:2019fpa,LIGOScientific:2020tif,LIGOScientific:2021sio}.
The works~\cite{LIGOScientific:2019fpa,LIGOScientific:2020tif,LIGOScientific:2021sio} contain a large suite of tests, most of which look for parametrized (or, in some cases, unparametrized) deviations from general relativity, which are often described as being ``agnostic'' about the underlying theory that produces any such deviations.

An alternate approach to testing general relativity is to verify that all components of the GW signal predicted by general relativity do indeed appear in the observed signals.
One such example of such a test was the confirmation of the presence of higher harmonics (i.e., multipole moments) of the dominant quadrupolar GW signal with the exceptional event GW190814~\cite{LIGOScientific:2020zkf}.
As GW measurements improve, more of these predictions of general relativity will become accessible to observational study.
One such prediction of general relativity that has not yet been detected, is a nonlinear gravitational effect that arises from a nonlinear interaction of GWs in the wave zone far from an isolated source.
It is known as the nonlinear or Christodoulou GW memory effect~\cite{Christodoulou:1991cr,Blanchet:1992br}.
This paper will focus on this GW phenomenon and aspects of the effort to detect this effect.

The nonlinear GW memory effect (as well as the linear memory effect~\cite{Zeldovich:1974gvh}) produces a lasting offset in the GW strain following a burst of GWs from an isolated system.
Many decades prior to the first detection of gravitational waves, the memory effect was identified as a possible source that GW interferometers could detect~\cite{Braginsky:1985vlg,Braginsky:1987kwo,Kennefick:1994nw}.
Interest in detecting the memory effect has grown because of the detection of GWs by LIGO and Virgo, and because of theoretical investigations of the memory effects that showed its close connections to infrared properties of gravitational physics. 
Specifically, the realization that the memory effect is closely connected to the Bondi-Metzner-Sachs supertranslation symmetries~\cite{Bondi:1962px,Sachs:1962wk,Sachs:1962zza} (and their conserved charges the supermomentum~\cite{Geroch:1977big}) and Weinberg's soft graviton theorem~\cite{Weinberg:1965nx} (see, e.g.,~\cite{Strominger:2013jfa,Strominger:2014pwa,Flanagan:2015pxa,Strominger:2017zoo}) in an ``infrared triangle'' has provided more compelling theoretical reasons to search for the GW memory effect.

The limited frequency response of ground-based GW interferometers makes the lasting (i.e., zero-frequency) memory signal challenging to detect with interferometers, such as LIGO and Virgo.
There is, however, a time-dependent signal (which contains signal power at higher frequencies), which is associated with the nonlinear GW memory effect and which GW interferometers can measure more readily~\cite{Grant:2022bla}.
Its detection will be challenging because the memory signal still remains small compared to both the dominant quadrupolar waves and even some of the higher harmonics that LIGO and Virgo have measured.
Thus, it is unlikely that the LVK collaboration will detect the memory effect from an individual black hole merger~\cite{Pollney:2010hs}, given what is currently known about the population of merging binary black holes~\cite{LIGOScientific:2018jsj,LIGOScientific:2020kqk} and the detectors' observation plans and timelines to be upgraded~\cite{KAGRA:2013rdx}.\footnote{The LISA mission~\cite{Amaro-Seoane:2017ADS} has better prospects for detecting the memory effect~\cite{Favata:2009ii,Islo:2019qht,Gasparotto:2023fcg,Inchauspe:2024ibs} as do next-generation ground-based GW detectors~\cite{Grant:2022bla,Goncharov:2023woe}.}

It is much more likely, however, that the presence of the nonlinear memory effect can be inferred in the population of all the binary-black-hole (BBH) mergers~\cite{Lasky:2016knh,Boersma:2020gxx,Hubner:2019sly,Grant:2022bla} measured by the LVK collaboration.
A Bayesian search for the nonlinear memory effect already has been implemented and applied to the data from the first three observing runs~\cite{Hubner:2019sly,Hubner:2021amk,Cheung:2024zow}.
No significant evidence for the nonlinear memory effect in the population of BBH mergers has been found, which is consistent with forecasts~\cite{Boersma:2020gxx,Hubner:2019sly,Grant:2022bla}.
The searches for the GW memory effect require a well-defined notion of the nonlinear GW memory signal from a BBH merger, as discussed above.
Computing this signal from a set of multipole moments of the GW strain requires numerically differentiating the modes, integrating different products of the modes, and summing the different modes to obtain the final result.
While the works~\cite{Hubner:2019sly,Hubner:2021amk,Cheung:2024zow} have shown that it is possible to perform such searches with existing GW memory waveform models, it would be advantageous for the search for the GW memory effect to have a waveform model that does not require as many steps to compute.\footnote{Having a stand-alone model for the GW memory signal can also allow the signal to be evaluated more efficiently. Numerically differentiating and integrating the oscillatory ringdown modes to compute the memory signal requires sampling in time many times per orbital period to accurately evaluate the derivatives. The memory model for nonprecessing binaries evolves on a slower timescale (namely, that of radiation reaction), so its waveform will not need to be evaluated as frequently in time as the oscillatory modes to obtain the same accuracy requirement on the signal.}

In the paper~\cite{Elhashash:2024thm} (henceforth, Paper I), the authors made a first step towards building such a model.
Paper I contained a fit of the final memory strain offset, which included results from a high post-Newtonian (PN) order calculation in the extreme mass-ratio limit.
These results will be used in this current work.
The main aim of this paper is to build a time-domain model for the $l=2$, $m=0$ spin-weighted spherical-harmonic mode of the memory signal from nonspinning BBH mergers.
We use a PN model of the memory signal during the inspiral stage and a superposition of products of quasinormal modes (QNMs) during the ringdown.
Unlike the minimal waveform model for nonspinning equal-mass binaries in~\cite{Favata:2009ii}, we need to add a phenomenological ``intermediate'' model that bridges between the PN inspiral model and the QNM ringdown model.
Similar to the minimal waveform model of~\cite{Favata:2009ii}, we find that the time-domain model has an analytic Fourier transform in terms of transcendental and other special functions.
This allows us to obtain an analytical frequency-domain model of the nonlinear GW memory effect.

While this work was in progress, several waveform models for the $l=2$, $m=0$ spin-weighted spherical-harmonic mode of the gravitational waveform were developed.
These include a phenomenological frequency-domain waveform model of the waveform~\cite{Valencia:2024zhi} and three time-domain models: a numerical-relativity (NR) surrogate model using Cauchy-characteristic extraction (CCE)~\cite{Yoo:2023spi}, a phenomenological nonprecessing waveform model~\cite{Rossello-Sastre:2024zlr}, and an effective-one-body model for the plunge and ringdown~\cite{Albanesi:2024fts} (which could be joined with a related calculation during the inspiral~\cite{Grilli:2024lfh}, but was not in~\cite{Albanesi:2024fts}).
The nonprecessing model~\cite{Rossello-Sastre:2024zlr} was generalized to include precession~\cite{Rossello-Sastre:2025dep} and was used to perform parameter estimation~\cite{Rossello-Sastre:2025dep,Rossello-Sastre:2025gtq}.\footnote{It had previously been shown that including the memory signal in parameter estimation can help break the distance-inclination degeneracy~\cite{Xu:2024ybt}.}
All of these waveforms contain the memory signal as part of the $l=2$, $m=0$ mode, but the mode also includes other features in this signal, such as the QNMs of the remnant Kerr black hole formed during the merger.
These QNMs are not related to the nonlinear GW memory effect, so they would need to be removed from these models, if they were to be used in searches for the GW memory effect alone.
Our time-domain and its Fourier transform does not need any such modifications to be used in searches for the GW memory signal.

\subsection{Summary and organization of this paper}

We close the introduction by summarizing the organization and main results in this paper.
First, in Sec.~\ref{sec:review}, we review some results from Paper I, which describe how we compute the memory signal from multipole moments of the gravitational-wave strain and how we compute the late-time memory offset strain.
In Sec.~\ref{sec:time-domain}, we discuss the modeling techniques we use for the inspiral, ringdown and intermediate stages of the memory signal, respectively.
The inspiral model uses results from PN theory, the ringdown model uses multimode QNM fitting, and the intermediate model is a phenomenological approach that enforces continuity and some smoothness between the inspiral and ringdown.
The resulting time-domain model has a relative error of a few percent compared with a calculation of the memory signal from a NR surrogate model.
Section~\ref{sec:freq-domain} covers several topics related to the frequency-domain representation of our time-domain model: namely, the analytical calculation of the Fourier transform of the time-domain model, a method to remove windowing artifacts from the calculation of the fast Fourier transform (FFT) of the time-domain model, and finally the performance of the model, as quantified by the mismatch.
We conclude in Sec.~\ref{sec:conclusions}.
Several more technical results related to the phase of the complex frequency-domain signal, a more general ringdown model, the fitting coefficients in parts of our model, and the calculation of the FFT are given in four Appendices~\ref{app:memory_phase}--\ref{app:hdot-calc}.

\section{Review of results from Paper I} \label{sec:review}

We begin by reviewing some aspects of the calculation of the memory signal from Paper I. 
Specifically, we summarize the expansion of the memory signal in spin-weighted spherical harmonics, and computation of the memory signal from the NR surrogate model. 
We note the small differences from Paper I in some aspects of these calculations where applicable.

\subsection{Multipolar expansion of the memory}

We start by writing the multipolar expansion of the complex strain $h \equiv h_+ - i h_\times$ in terms of spin-weighted spherical harmonics.
We denote the multipole moments by $h_{lm}$, so that
\begin{equation}\label{eq:strain_modes_Ylm}
    h \equiv h_+ - i h_\times = \sum_{l=2}^\infty \sum_{m=-l}^l h_{lm} (_{-2}Y_{lm}) \, .
\end{equation}
The multipoles $h_{lm}$ are functions of the retarded time $u$, and the spin-weighted spherical harmonics are functions of the polar and azimuthal angles $(\theta,\phi)$, respectively.
The spin-weighted spherical harmonics with fixed spin $s$ form a complete basis for functions on the two-sphere of spin weight $s$.
They are orthonormal,
\begin{equation}\label{eq:Ylm_orthogonality}
    \int d^2\Omega \, (_{s}\bar{Y}_{lm})(_{s}{Y}_{l'm'}) = \delta_{ll'}\delta_{mm'} ,
\end{equation}
where the overline represents complex conjugation.

Evaluating the multipole moments for the GW memory signal involves computing the integral of three spin-weighted spherical harmonics (see, e.g.,~\cite{Nichols:2017rqr}, for more details).
We use the notation of~\cite{Nichols:2017rqr} for these integrals:
\begin{align} \label{eq:CldefIntegral}
& C_l(s',l',m';s'',l'',m'') \equiv \nonumber\\
& \int d^2\Omega \, (_{s'+s''}\bar{Y}_{lm'+m''})(_{s'}{Y}_{l'm'})(_{s''}{Y}_{l''m''}) .
\end{align}
These coefficients are nonvanishing when the index $l$ is in the set $\Lambda$,
\begin{align} \label{eq:Lambda-set}
&\Lambda \equiv \nonumber \\
&\{\max(|l'-l''|,|m'+m''|,|s'+s'' |),...,l'+l''-1,l'+l''\} .
\end{align}
The coefficients can also be written in terms of Clebsch-Gordon coefficients:
\begin{align} \label{eq:CldefClebschGordon}
C_l(s',l',m';s'',l'',m'')= (-1)^{l+l'+l''} \sqrt{\frac{(2l'+1)(2l''+1)}{4 \pi (2l+1)}} \nonumber\\
\times \left< l',s';l'',s''|l,s'+s'' \right> \left< l',m';l'',m''|l,m'+m'' \right> .
\end{align}
We use the same conventions for the Clebsch-Gordon coefficients as those implemented in \textsc{Mathematica}.
One useful transformation property of the coefficients $C_l(s',l',m';s'',l'',m'')$ that we will use is
\begin{align}
\label{eq:Cl_transformation}
    C_l(s',l',m';s'',l'',m'') &\nonumber\\
    = (-1)^{l+l'+l''} & C_l(s',l',-m';s'',l'',-m'') \, .
\end{align}
%

%
%
%
%
It was shown in~\cite{Grant:2022bla}, for example, 
that the nonlinear memory signal, when written in terms of the GW strain multipole moments $h_{lm}$, is given by
\begin{align} \label{eq:hlmMemory_hlm}
    h_{lm}^{\mem}(u) = {} & r \sqrt{\frac{(l-2)!}{(l+2)!}} \sum_{l',l'',m',m''} \int_{-\infty}^{u} du' \, \dot{h}_{l'm'}\dot{\bar{h}}_{l''-m''}  \nonumber\\
    & \quad \times  (-1)^{m''} C_l(-2,l',m';2,l'',m'') .
\end{align}
The sum over the indices $l'$, $l''$, $m'$ and $m''$ in Eq.~\eqref{eq:hlmMemory_hlm} must satisfy the constraints that $l'$, $l''\geq 2$ as well as $|m'| \leq l'$ and $|m''| \leq l''$. 
For fixed values of $l$ and $m$ on the left-hand side of Eq.~\eqref{eq:hlmMemory_hlm}, the coefficients $C_l(-2,l',m';2,l'',m'')$ in the sum will only be nonzero when $m = m' + m''$ as well as when $l$, $l'$ and $l''$ satisfy the relationships in Eq.~\eqref{eq:Lambda-set}.

As in Paper I, we will specialize to the $l=2$, $m=0$ mode of the memory signal.
This requires that $m'' = -m'$, and Eq.~\eqref{eq:hlmMemory_hlm} reduces in this case to
%
%
%
\begin{align} \label{eq:h20Memory_hlm}
    h_{20}^{\mem}(u) = {} & \frac{r}{2\sqrt{6}} \sum_{l',l'',m'}  (-1)^{m'} C_l(-2,l',m';2,l'',-m')\nonumber\\
    & \times \int_{-\infty}^{u} du' \, \dot{h}_{l'm'}\dot{\bar{h}}_{l''m'}  .
\end{align}
%

For the nonspinning binaries that we will consider in this paper, 
the complex conjugate of $h_{lm}$ is given by $\bar h_{lm} = (-1)^l h_{l-m}$.
We can use this transformation of $h_{lm}$ for nonspinning binaries along with the transformation of the coefficients $C_l(s',l',m';s'',l'',m'')$ in Eq.~\eqref{eq:Cl_transformation} to rewrite the sum over $m'$ in Eq.~\eqref{eq:h20Memory_hlm} to be over only positive values of $m'$:
\begin{align}
\label{eq:h20Memory_hlm_nonprecessing}
    h_{20}^{\mem}(u) = 
    &\frac{r}{\sqrt{6}} \sum_{l',l''} \sum_{m'=1}^{l'} (-1)^{m'} C_l(-2,l',m';2,l'',-m')\nonumber\\
    \times & \int_{-\infty}^{u} du' \, \Re\Big[ \dot{h}_{l'm'}\dot{\bar{h}}_{l''m'}\Big]\, .
\end{align}
Because the integrand is the real part of $\dot{h}_{l'm'}\dot{\bar{h}}_{l''m'}$, this implies that the memory signal $h_{20}^\mem$ is also real.

\subsection{Surrogate memory signal and hybridization} \label{subsec:surr_hybrid}

\begin{table*}[t]
    \centering
    \caption{The $m$ values (for each $l$) of the oscillatory modes that use used in constructing the time-domain memory model (first row) and that are generated by the NRHybSur3dq8 surrogate model (second row).}
    \begin{tabular}{lcccc}
    \hline
    \hline
        & $l=2$ & $l=3$ & $l=4$ & $l=5$ \\
       \hline
        $m$ values (memory model) & $\{\pm 1, \pm 2\}$ & $\{\pm 2\}$ & --- & --- \\
        $m$ values (NR surrogate)  & $\{\pm 1, \pm 2\}$ & $\{\pm 1, \pm 2 , \pm 3\}$ & $\{\pm 2, \pm 3, \pm 4\}$ & $\{\pm 5\}$ \\
    \hline
    \hline
    \end{tabular}
    \label{tab:lm-modes-sur}
\end{table*}

As in Paper I, we will construct our memory model using the spin-weighted spherical-harmonic moments $h_{lm}$ that are output from the NR hybrid surrogate model NRHybSur3dq8~\cite{Varma:2018mmi}.
To obtain the initial offset for the memory during the long inspiral, we found it more efficient to hybridize the memory computed from the surrogate model with a PN memory waveform than to evaluate the surrogate model for very long periods of time.
We perform a hybridization procedure similar to the one done in Paper I, except here we hybridize using the 3.5PN memory waveform computed in~\cite{Cunningham:2024dog}, rather than the 3PN memory waveform used in Paper I. 
The PN memory waveforms are written in terms of the PN parameter $x$, which is defined to be
\begin{align} \label{eq:xPN}
    x \equiv (M\Omega)^{2/3} \, .
\end{align}
We use the notation $M=m_1+m_2$ for the total mass of the binary, where the primary mass is $m_1$ and secondary mass is $m_2$, and the orbital frequency is $\Omega$.
As in Paper I, we write the PN parameter $x$ in terms of the coordinate time $t$ at Newtonian order, for simplicity:
\begin{align} \label{eq:x-of-t}
    x(t) = \frac 14 \left[ \frac{\eta}{5M} (t_c-t) \right]^{-1/4} .
\end{align}
The parameter $t_c$ is the time of coalescence.
The symmetric mass ratio $\eta$ has several equivalent expressions, which can be given in terms of the total mass $M$ and the individual masses ($m_1$ and $m_2$), the mass ratio ($q=m_1/m_2$), or the reduced mass ($\mu=m_1m_2/M$):
\begin{equation}
    \eta = \frac{q}{(q+1)^2} = \frac{m_1 m_2}{M^2} = \frac{\mu}M \, .
\end{equation}
The 3.5PN memory waveform, when written in terms of the PN parameter $x$, is given by the lengthy expression in ~\cite{Cunningham:2024dog}:
\begin{widetext}
\begin{align}
    \label{eq:PN_memory}
    h_{20}^{\PN}(x) = {} & \frac{4M}{7r} \sqrt{\frac{5\pi}{6}} \eta x \bigg\{1+ x \bigg(-\frac{4075}{4032}+\eta \frac{67}{48}\bigg)
    + x^2 \bigg(-\frac{151877213}{67060224} - \eta \frac{123815}{44352} + \eta^2 \frac{205}{352}\bigg) + \pi x^{5/2} \bigg(-\frac{253}{336} + \eta  \frac{253}{84}\bigg) \nonumber\\
    &  + x^3 \bigg[-\frac{4397711103307}{532580106240} + \eta\bigg(\frac{700464542023}{13948526592} - \frac{205}{96} \pi^2\bigg) + \eta^2 \frac{69527951}{166053888} + \eta^3 \frac{1321981}{5930496} \bigg] \nonumber\\
    & + \pi x^{7/2} \bigg( \frac{38351671}{28740096} - \eta \frac{3486041}{598752} - \eta^2  \frac{652889}{598752}\bigg)
\bigg\} \, .
\end{align}
\end{widetext}

Following the procedure in Paper I, we hybridize over a time interval of time $t\in [t_1,t_2]$.
We again choose the time interval of the hybridization to be between $t_1=-5000M$ and $t_2=-4000 M$ (which was the time interval over which the oscillatory modes of the surrogate were hybridized).
To hybridize, we allow $t_c$ to be a free parameter in the PN memory waveform, and we add a constant $h_0$ to the surrogate memory waveform, which represents the memory signal accumulated from before the initial time of $t/M = -10^4$, at which we first evaluate the surrogate model.
The main difference between this paper and Paper I, is that in addition to computing the memory signal from all the oscillatory waveform modes in the surrogate model as in Paper I (see the second row of Table~\ref{tab:lm-modes-sur}), here we also hybridize the surrogate with the three lowest $l$ and $m$ modes that contribute to the memory signal (see the first row of Table~\ref{tab:lm-modes-sur}).
This latter choice is related to our modeling of the ringdown part of the signal, which we discuss in more detail in Sec.~\ref{subsec:ringdown}.

To perform the hybridization we minimize the following cost function involving the surrogate waveform $h^{(A)} = h^\surr_{20}$ and the PN waveform $h^{(B)} = h^\PN_{20}$ for a specific mass ratio $q$: 
\begin{align}
\label{eq:cost_function}
    C_q[h_{(A)},h_{(B)}] \equiv \frac{\displaystyle\int_{t_1}^{t_2} dt \, |h^{(A)}_{20}(t;q)-h^{(B)}_{20}(t;q)|^2}{\displaystyle\int_{t_1}^{t_2} dt \, |h^{(A)}_{20}(t;q)|^2} \, .
\end{align}
The free parameters that can be tuned in this cost function are $t_c$ and $h_0$.
We use the \texttt{SciPy} \emph{minimize} function to perform the optimization and to obtain estimates of the best-fit parameters $h_0$ and $t_c$ for each mass ratio.

\begin{table*}[t]
    \centering
    \caption{Coefficients for the $\Delta h_{20}(\eta)$ polynomial fits in Eq.~\eqref{eq:Deltahmem_fit}.
    These fits were constructed through a least-squares procedure using $50$ BBH systems with mass ratios equally spaced in $\eta$ for a range of mass ratios with the range of validity of the NRHybSur3dq8: $1\leq q\leq 8$. 
    The data in the first row are for the memory fit computed from the three plus-minus pairs of oscillatory modes in the first row of Table~\ref{tab:lm-modes-sur}.
    The data in the second row are for the memory fit computed from all oscillatory modes available in the NR surrogate model.
    They are similar to the results given in Paper I, except for the fact that a 3.5PN memory signal was used here rather than the 3PN signal used in Paper I.
    Both fits have the linear term fixed to be the value computed in the EMRI limit, which was derived in Paper I.}
    \begin{tabular}{lcccccc}
    \hline
    \hline
       $(l,m)$ mode cases & $c_1$ & $c_2$ & $c_3$ & $c_4$ & $c_5$ & $c_6$ \\
       \hline
       Memory model  & 0.102414 & 0.824195 & -2.3413 & 22.486 & -58.276 & 105.885 \\
       NR surrogate  & 0.102414 & 0.770384 & -1.70081 & 18.1139 & -34.4687 & 52.7548 \\
    \hline
    \hline
    \end{tabular}
    \label{tab:Delta-hmem-fit}
\end{table*}

As in Paper I, we use the values of $t_c$ and $h_0$ to evaluate the final memory offset at 50 equally spaced points in $\eta$.
We similarly assume a sixth-order polynomial function in the symmetric mass ratio $\eta$ as our fitting ansatz.
We do not include an coefficient with no powers of $\eta$, and we fix the coefficient linear in $\eta$ to be given by the value of the memory offset in the extreme mass-ratio limit, which was computed in Paper I.
This leaves five undetermined parameters in our fitting functions.
Specifically, we write the final memory offset as
\begin{align}\label{eq:Deltahmem_fit}
    \Delta h_{20}(\eta) = \frac Mr \sum_{j=1}^6 c_j \eta^j \, ,
\end{align}
where $c_1 = 0.102414$ is fixed.
The values of the coefficients $c_j$ from the least-squares fit, using just $l$ and $m$ modes in the first line of Table~\ref{tab:lm-modes-sur}, are summarized in the first row of Table~\ref{tab:Delta-hmem-fit}.
The coefficients of the fit using all surrogate modes (as in Paper I, except using the 3.5PN memory signal to hybridize) is given in the second row of Table~\ref{tab:Delta-hmem-fit} for comparison.
Using the 3.5PN rather than the 3PN waveform for performing the hybridization did not change the values of the coefficients $c_j$ at the accuracy at which we give the results.
Note that there is a more significant difference in the coefficients from using just the modes in the first line of Table~\ref{tab:lm-modes-sur} rather than the modes in the second line.

\section{Time-domain model} \label{sec:time-domain}

As discussed in Sec.~\ref{sec:intro}, we divide the memory signal into three parts: an inspiral part (during which the memory slowly grows in amplitude on the inspiral time scale), an intermediate part (which includes the late inspiral and merger), and ringdown part (which follows the peak of the amplitude of the $l=2$, $m=2$ mode).
We use different analytical or phenomenological models to model the memory signal in each part of the binary's evolution.
During the inspiral phase, we use the PN approximation.
In the late inspiral phase, however, the PN approximation begins to deviate more from the NR surrogate memory; therefore, we stop using the PN waveform before any significant differences arise.
During the ringdown part, we use a superposition of QNMs with different overtone numbers to model oscillatory $(l,m)$ modes, and their products to model the memory signal. 
We also require that the final accumulated memory should match the value computed from our fit for the final memory strain $\Delta h_{20}$ defined in Eq.\eqref{eq:Deltahmem_fit}.
This implies that our time-domain memory model can be written naturally as a piecewise function as follows:
\begin{equation}
\label{eq:h20_insp_int_rd}
    h_{20}(t) = \begin{cases}
        h^{\insp}_{20}(t) & \mbox{for} \ t<t_\mathrm{int}, \\
        h^{\intr}_{20}(t) & \mbox{for} \ t_\mathrm{int} \leq t \leq t_\mathrm{rd}, \\
        \Delta h_{20} -  h_{20}^{\rd}(t) & \mbox{for} \ t \geq t_\mathrm{rd} .
    \end{cases}
\end{equation}
The times $t_\mathrm{int}$ and $t_\mathrm{rd}$ define the start of the intermediate and ringdown memory signals, respectively.
Notice that we wrote the memory accumulated during the ringdown as the difference between the final memory $\Delta h_{20}$ and the memory accumulated from some time $t$ up to infinity $h_{20}^{\rd}(t)$, so as to enforce that the final time-domain memory signal matches our fit for the final memory for $t$ much greater than $t_\mathrm{rd}$.

\subsection{PN memory signal during inspiral} \label{subsec:PNinspiral}

During the inspiral phase, we use the 3.5PN memory waveform given in Eq.~\eqref{eq:PN_memory} as our model for the time-domain memory signal, as we did for our hybridization procedure in Sec.~\ref{subsec:surr_hybrid}.
To make the inspiral signal faster to evaluate, we would like to avoid having to perform a hybridization procedure with the surrogate at each mass ratio to obtain the optimal $t_c$ for that mass ratio.
Instead, we construct a fitting function for $t_c$ as a function of mass ratio (for $q\in[1,8]$), which minimizes a residual between the hybridized surrogate and the 3.5PN waveform.

As in Paper I, we will assume that $t_c$ can be fit using a quartic polynomial in the symmetric mass ratio, $\eta$,
\begin{align}\label{eq:tc_expansion}
    t_c=\sum_{i=0}^4 t_{c,i} \, \eta^{i} .
\end{align}
The five coefficients $t_{c,i}$ in the polynomial function of $\eta$ are the parameters for which we will fit.
In this paper, however, we will use a somewhat different approach to find the values of the $t_{c,i}$ parameters from what was performed in Paper I.
Here, we determine these coefficients by minimizing the cost function
\begin{align}
\label{eq:cost_function_inspiral}
    C[h_{20}^{\surr},h_{20}^{\insp}] ={}& \sum_{q=1}^{8} C_q[h_{20}^{\surr},h_{20}^{\insp}] , 
\end{align}
where the cost function $C_q[h_{20}^{\surr},h_{20}^{\insp}]$ was defined in Eq.~\eqref{eq:cost_function}.
We obtain the optimized values of the parameters $t_{c,i}$ by using the \texttt{SciPy} \emph{minimize} function again.
Note that unlike the hybridization in Sec.~\ref{subsec:surr_hybrid}, which determined the value of $t_c$ by minimizing a cost function for a single mass ratio over a time window of duration $1000M$, here we modify the procedure in two ways.
First, the hybridization is done over a longer time interval between $t_1=-10^4M$ and $t_2=-2000M$.
Second, the cost function is defined as the sum over eight integer mass ratios $q=1,2,\ldots,8$ of the cost functions used for the hybridization for a single mass ratio in Sec.~\ref{subsec:surr_hybrid}. 

\begin{table*}[t]
    \centering
    \caption{Coefficients for the $t_c$ polynomial fit in Eq.~\eqref{eq:tc_expansion}. More details about how the fits were constructed is given in the text of Sec.~\ref{subsec:PNinspiral}.}
    \begin{tabular}{lccccc}
    \hline
    \hline
        Coefficient & $i=0$ & $i=1$ & $i=2$ & $i=3$ & $i=4$ \\
       \hline
        $t_{c,i}$ & $2.1558 \times 10^{3}$&  $-2.2124 \times 10^{4}$&  $1.1760 \times 10^{5}$&
       $-3.3106 \times 10^{5}$&  $3.80724 \times 10^{5}$\\
    \hline
    \hline
    \end{tabular}
    \label{tab:tc-fit}
\end{table*}

The result of this optimization are the five coefficients $t_{c,i}$, the values for which are given in Table~\ref{tab:tc-fit}.
The inspiral memory waveform model can be computed using the values of the coefficients $t_{c,i}$ in the expansion of $t_c$ in Eq.~\eqref{eq:tc_expansion} and substituting by $t_c$ into the 3.5PN memory waveform in Eq.~\eqref{eq:PN_memory} through the expression for $x(t)$ in Eq.~\eqref{eq:x-of-t}.
This allows the inspiral memory signal to be evaluated rapidly for $q\in[1,8]$ without additional hybridization, but to still have the appropriate offset at the initial time at which it is evaluated.

\begin{figure*}
    \centering
    \includegraphics[width=0.48\textwidth]{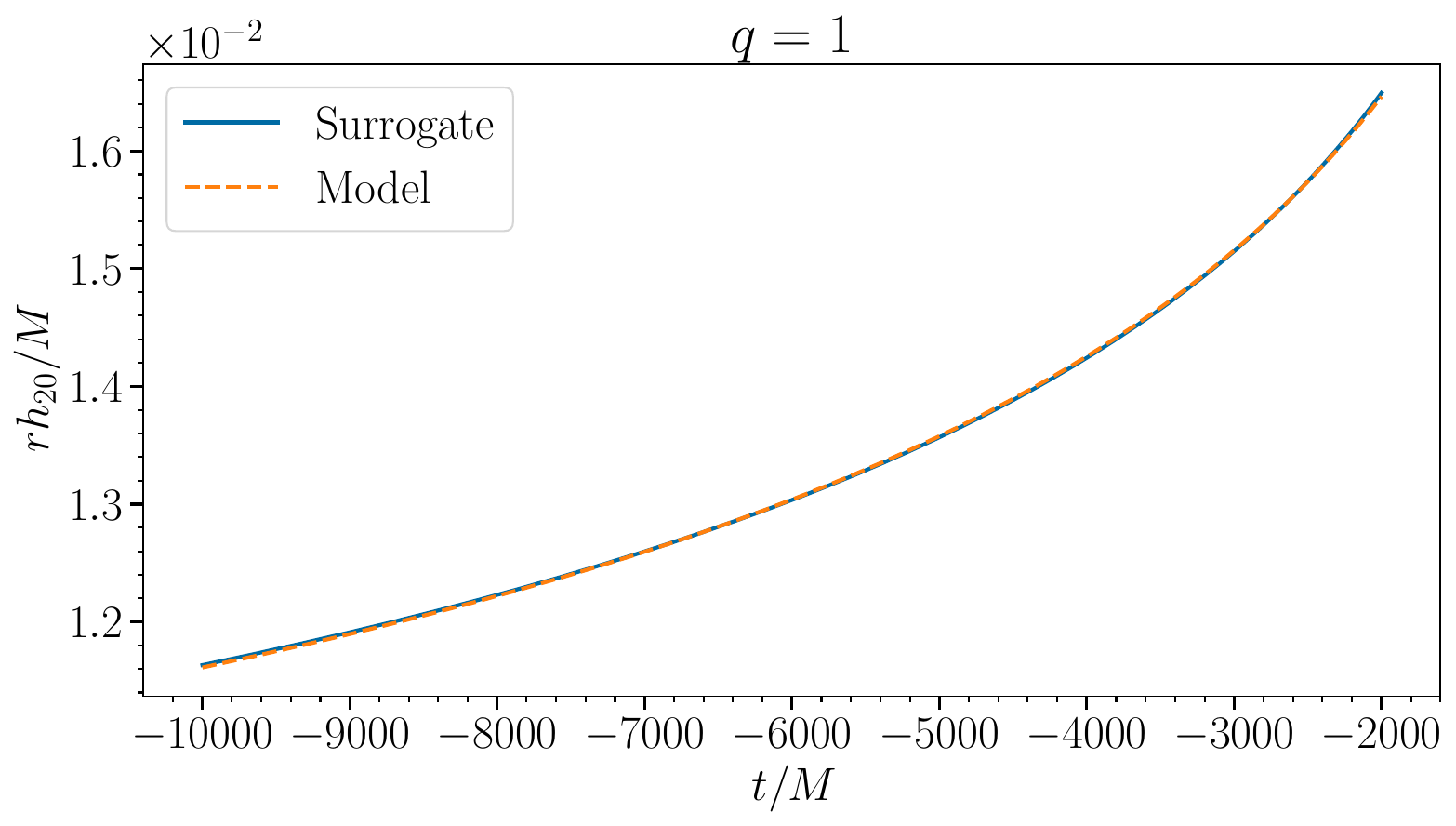}
    \includegraphics[width=0.48\textwidth]{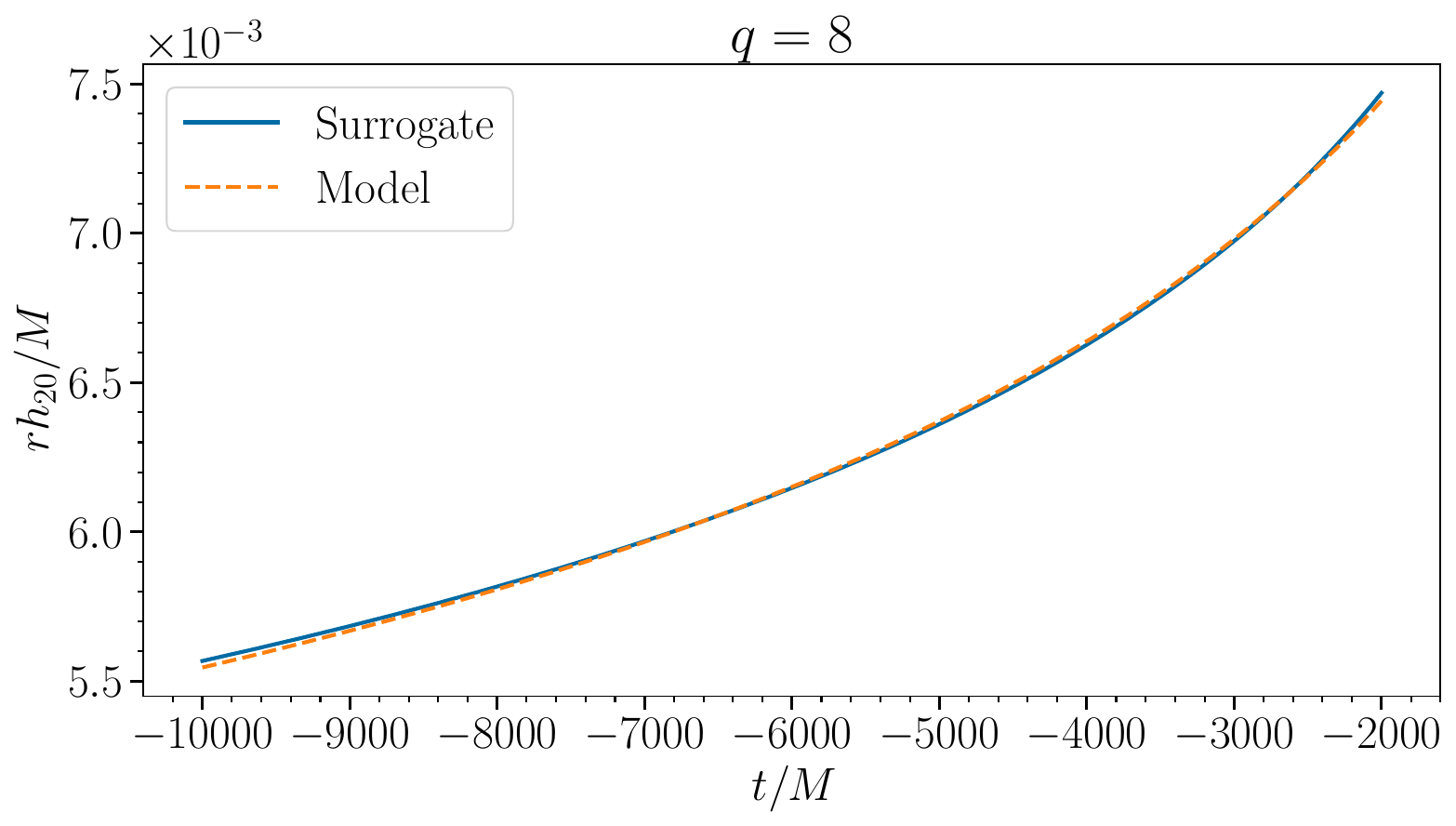}
    \includegraphics[width=0.48\textwidth]{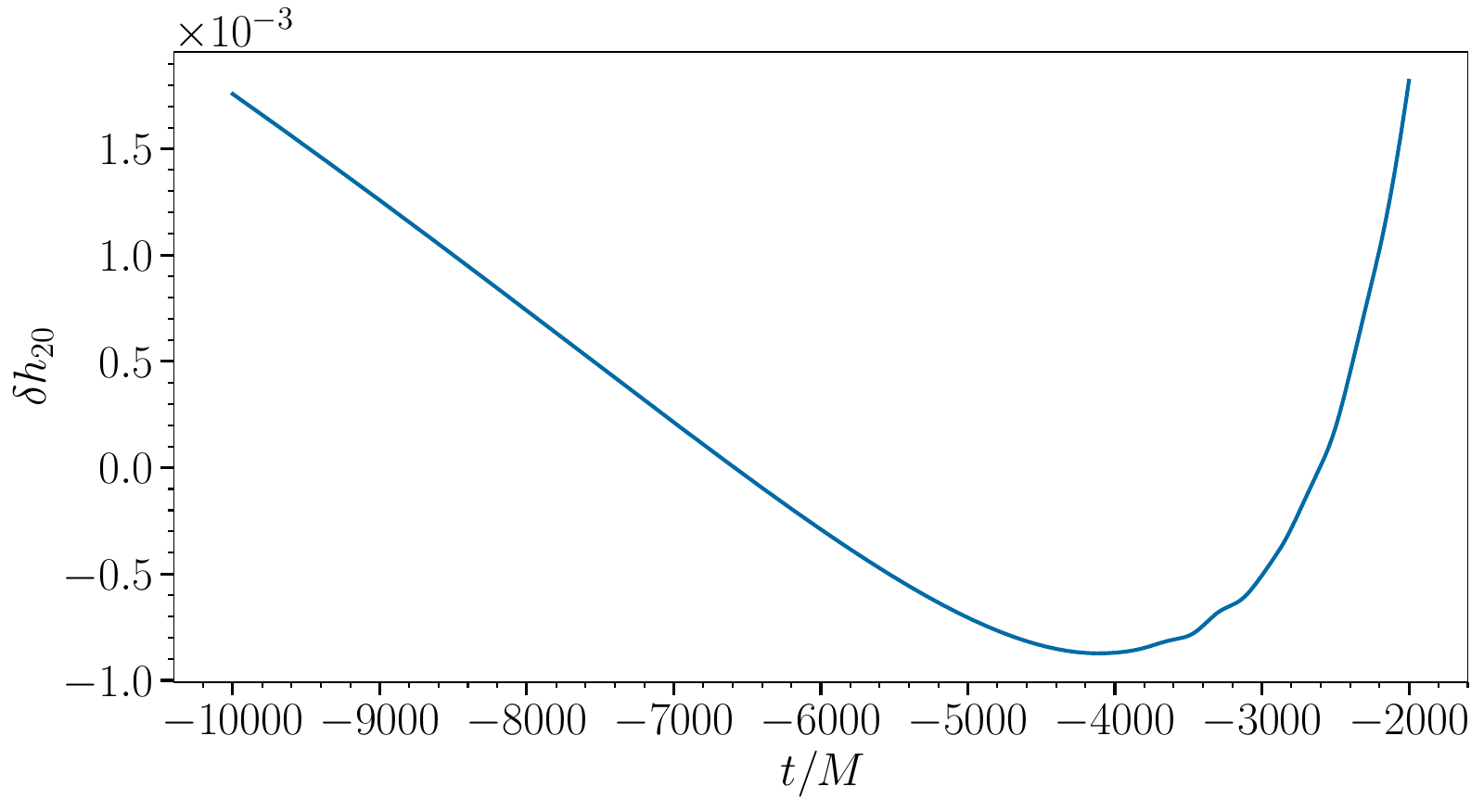}
    \includegraphics[width=0.48\textwidth]{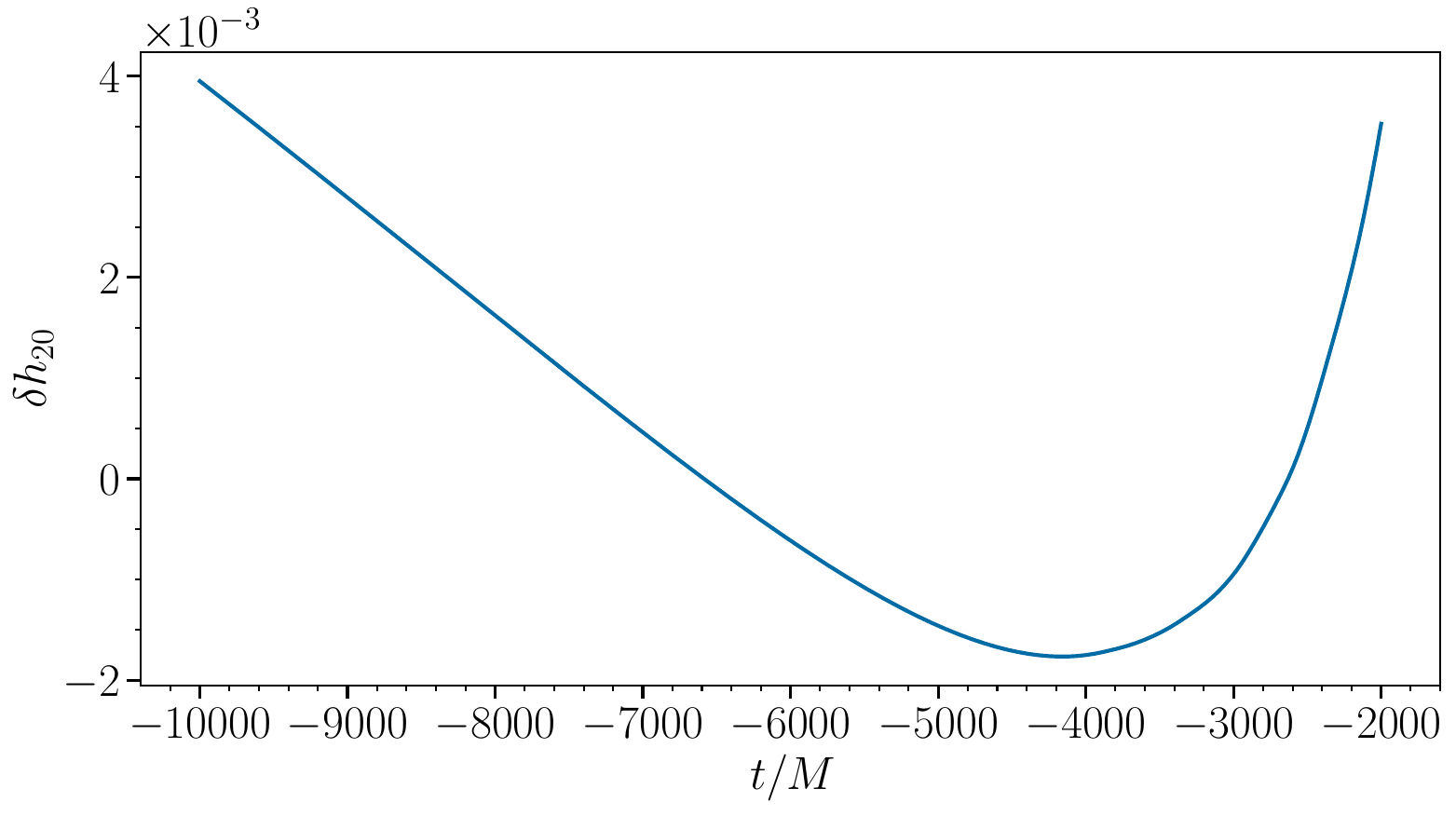}
    \caption{\textbf{Inspiral memory model and its relative error versus time}:
    \emph{Top}: The hybridized surrogate memory signal computed from the NRHybSur3dq8 surrogate waveform modes (solid, blue curve) and the inspiral time-domain memory model (dashed, orange curve) for an equal-mass non-spinning BBH merger (left) and for a mass ratio $q=8$ (right). 
    \emph{Bottom}: The relative error $\delta h_{20}$ in Eq.~\eqref{eq:relative_error}  the inspiral memory model for each of the corresponding mass ratios shown above.}
    \label{fig:inspiral_memory}
\end{figure*}

The inspiral memory signal model is shown in the top row of Fig.~\ref{fig:inspiral_memory} for two BBH systems with mass ratios $q=1$ (left) and $q=8$ (right).
The solid blue curve shows the memory computed from the surrogate model, and the dashed orange curve shows the memory computed from the inspiral model in Eq.~\eqref{eq:PN_memory}.
The accuracy of the model is measured by computing the relative error between the inspiral memory model and the inspiral part of the hybridized surrogate memory
\begin{equation}\label{eq:relative_error}
    \delta h_{20} = \frac{|h_{20}^{\surr}-h_{20}^{\model}|}{h_{20}^{\surr}} \, .
\end{equation}
The relative error at the two smallest and largest mass ratios in our mode is representative of the results for all mass ratios in the model ($q\in[1,8]$).
Namely, the relative error is the largest positive value near the times $t_1$ and $t_2$, and is most negative near $-4000M$, which was the $t_2$ value over which the surrogate was hybridized with the 3.5PN waveform.
For larger $q$ values, the relative error increases, because the PN approximation becomes less accurate at a fixed time from merger.

\subsection{Ringdown memory signal modeling} \label{subsec:ringdown}

During the ringdown phase, recent work has showed that a superposition of QNMs of different overtones provides a good fit to the waveform $(l,m)$ modes after the peak amplitude of the mode (see, e.g.,~\cite{Giesler:2019uxc,Cook:2020otn,MaganaZertuche:2021syq}). 
We will use this approach to model the ringdown phase of the oscillatory modes that are the inputs to our calculation of the nonlinear memory effect.
We will ignore the nonlinearities in the oscillatory ringdown modes~\cite{Mitman:2020bjf}, because they will be a further nonlinear correction to the nonlinear memory effect.
To determine the QNM frequencies, we use the \texttt{qnm} package~\cite{Stein:2019mop}.\footnote{This package uses a spectral method for solving the angular Teukolsky equation by Cook and Zalutskiy~\cite{Cook:2014cta} (which has some advantages over Leaver's original method~\cite{Leaver:1985ax}). 
In addition to the QNM frequencies it also gives the angular separation constant and the spherical-spheroidal mixing functions (discussed below) for any dimensionless spin parameter $\chi$ that satisfies $0\leq \chi<1$.}

The ringdown QNM waveforms are often expressed in terms of spin-weighted spheroidal harmonics $_{s}{S}_{lm}(\theta,\phi;a\omega)$.
The spheroidal harmonics can be expanded in terms of the spin-weighted spherical harmonics $_{s}{Y}_{lm}(\theta,\phi)$, which are the basis used for the NR surrogate waveform modes, as follows:
\begin{equation}\label{eq:Slm_expansion_Ylm}
    _{s}{S}_{lm}(\theta,\phi;a\omega) = \sum_{l'=l_\mathrm{min}}^{\infty} A_{l'lm}(a\omega) _{s}{Y}_{l'm}(\theta,\phi) \, .
\end{equation}
Here $l_\mathrm{min}$ is given by $l_{\mathrm{min}}=\mathrm{max}\left(|m|,|s|\right)$ and the coefficients $A_{l'lm}(a\omega)$ are complex-valued coefficients that are referred to as ``mixing'' coefficients.
The QNMs are functions of the final mass $M_f$ and final spin $\chi_f$ of the remnant BH, which are nontrivial functions of the initial masses $m_1$ and $m_2$ that are used during the inspiral phase.
We use the \texttt{SurfinBH} Python package to compute the final mass and spin parameters from the individual mass parameters.

The QNM frequencies are parametrized by three integers: the spheroidal harmonic indices $(l,m)$ and the overtone number $n=0,1,\dots$.
We will typically denote them by $\omega_{lmn}$.
%
%
They generically have nonzero real and imaginary parts.
The real part $\Re[\omega_{lmn}]$ is the angular frequencies and minus the imaginary part $-\Im[\omega_{lmn}]$ is the inverse of the damping time (we use the convention that the imaginary part is negative).
Higher overtones have a faster damping rate, so the fundamental mode ($n=0$) is longer lived than the $n>0$ overtones.
For each $(l,m,n)$ triple, there is a QNM with a positive and a negative real part, which are referred to as the ``ordinary'' ($\Re[\omega_{lmn}]>0$) and the ``mirror'' ( $\Re[\omega_{lmn}]<0$) modes.
For rotating black holes, for each $(l,m,n)$ value with $m\neq 0$, there are two modes, the ``prograde'' and the ``retrograde'' modes, for which $m$ has the same or opposite sign, respectively, as the sign of the energy.
There is a discrete symmetry of the modes that relates ordinary, prograde modes with the mirror, retrograde modes (and a similar symmetry for the ordinary, retrograde and mirror, prograde modes).
The \texttt{qnm} package computes the ordinary modes and thus prograde modes for positive $m$ and retrograde modes for negative $m$; the other cases can be obtained through the discrete symmetry described above.

The remnant black hole formed from a BBH merger is spinning, and the ringdown waves generically contain both prograde and retrograde modes. 
Prior work (see, e.g.,~\cite{Giesler:2019uxc}), however, has shown that when the remnant black hole is formed from a nonspinning BBH merger (the case we are focusing on in this paper), the ringdown waveform $(l,m)$ modes can be fit well using just the prograde modes.
For this reason, we also fit the ringdown waves using just prograde modes.

\subsubsection{Multimode ringdown fitting} \label{subsubsec:multimode}

Equation~\ref{eq:Slm_expansion_Ylm} implies that a single spheroidal-harmonic QNM mode can contribute to multiple spherical-harmonic strain modes.
As is described in~\cite{Cook:2020otn}, this implies that to fit for the QNM amplitudes from the spherical-harmonic modes of the NR surrogate waveform, then one would generically need to fit for multiple amplitudes simultaneously in the different spherical-harmonic modes rather than fitting sequentially in each mode.
For a black hole of a known final mass and spin, the frequencies are known, and the fitting problem is  linear in the complex amplitudes of the modes. 

We will use the ``eigenvalue method'' introduced in~\cite{Cook:2020otn} to fit the strain modes.
This method involves constructing overlap integrals between the NR-surrogate spherical-harmonic strain modes and a model of the ringdown formed from a superposition of QNM overtones written in terms of spheroidal harmonics.
The least-squares solution can be determined from maximizing an overlap between the normalized surrogate model and the waveform constructed from the superposition on QNMs with different overtones.
We describe the procedure more quantitatively below.

First, we write the ringdown GW signal as a linear combination of QNMs by
\begin{equation}\label{eq:h_model_ringdown}
    h^{\rd}_{\model} = \frac{M}{r}\sum_{l,m,n} C_{lmn} e^{-i \omega_{lmn} (t-t_0)}
    {}_{-2}{S}_{lm}(\theta,\phi;a\omega) .
\end{equation}
Here the coefficients $C_{lmn}$ are the complex amplitudes of the different QNMs and $t_0$ is the start time of the QNM mode, which is chosen to be some time after the peak amplitude of the $l=2$, $m=2$ waveform mode.\footnote{Note that we will use as our time, $t$, the time used in the NR surrogate model, which is measured in terms of the initial total mass of the system, $M$.
The QNM frequencies are more naturally expressed in terms of the mass of the final black hole, $M_f$, which differs from $M$. 
Thus, one should rescale the frequencies $\omega_{lmn}$ by $M/M_f$ so that they are expressed in units consistent with the time used in the NR surrogate.
To avoid introducing factors of $M/M_f$, we absorb this factor in the values of the frequencies $\omega_{lmn}$ (i.e., we assume that $t$ and $\omega_{lmn}$ are scaled to be in the same units).}
We will set the maximum overtone index $n$ in the sum to be $7$, based on the findings of prior studies~\cite{Giesler:2019uxc,Cook:2020otn}.
Following the notation in~\cite{Cook:2020otn}, we will rewrite the superposition of QNMs in $h_{\model}^{\rd}$ as
\begin{equation}\label{eq:h_model_ringdown_hk}
    h^{\rd}_{\model} =  \frac{M}{r} \sum_{K\in\{l,m,n\}} \! C_K h_K \, .
\end{equation}
Here the index $K$ is a short-hand for the triple index $lmn$ of a QNM.
The corresponding waveform for a QNM is
\begin{equation}\label{eq:hk_defined}
    h_K \equiv e^{-i\omega_{lmn} (t-t_0)}
    {}_{-2}{S}_{lm}(\theta,\phi; a\omega) \, .
\end{equation}
%
%
%

The eigenvalue method in~\cite{Cook:2020otn} is the solution that arises from maximizing the normalized overlap between the ringdown model and the surrogate waveform.
The overlap is defined by
\begin{equation}\label{eq:overlap_function}
    \rho^2 = \frac{|\langle h^{\rd}_{\model}|h_{\surr}\rangle|^2}{\langle h_{\surr}|h_{\surr}\rangle \langle h^{\rd}_{\model}|h^{\rd}_{\model}\rangle } \, ,
\end{equation}
where the inner product of two complex functions is given by
\begin{equation}
    \langle h_1|h_2 \rangle \equiv \int_{t_i}^{t_f} dt \int d^2\Omega \, \bar h_1(t,\Omega) h_2(t,\Omega) \, .
\end{equation}
Substituting $h^{\rd}_{\model}$ in Eq.~\eqref{eq:h_model_ringdown_hk} into the overlap function in Eq.~\eqref{eq:overlap_function} gives for $\rho^2$
\begin{equation}
    \rho^2 = \frac{\sum_K|\bar{C}_K A_K|^2}{\langle h_{\surr}|h_{\surr}\rangle \sum_{I,J} \bar{C}_I B_{IJ} C_J } \, .
\end{equation}
Here, we introduced the notation $A_k$ and $B_{ij}$ to represent the following inner products:
\begin{subequations}
\begin{align}
    A_K \equiv {} & \langle h_K| h_{\surr} \rangle \label{eq:Ak_defined} \, ,\\
    B_{IJ} \equiv {} & \langle h_I|h_J\rangle \, . \label{eq:Bij_defined}
\end{align}
\end{subequations}
The indices $I$ and $J$, like $K$, denote a triple of $lmn$ indices for a QNM.
We will introduce a matrix-vector notation, $\mathbf A_\eta$ and $\mathbb B_\eta$, to denote these collections of inner products for a specific value of $\eta$.
By maximizing the overlap function $\rho$ with respect to the unknown coefficients $\mathbf C_\eta$, the solution is given by~\cite{Cook:2020otn}
\begin{equation}\label{eq:QNMs_coefficients}
    \boldsymbol{C}_\eta = \mathbb{B}_\eta^{-1} \cdot \boldsymbol{A}_\eta \, .
\end{equation}
The $\eta$ subscript is used here to represent solving this fitting problem for a specific BBH system with symmetric mass ratio $\eta$.

We next give more explicit expressions for the components of $\boldsymbol{A}_\eta$ by substituting $h_K$ in Eq.~\eqref{eq:hk_defined} and the spin-weighted spherical harmonic expansion of $h_{\surr}$ into the definition of $A_K$ in Eq.~\eqref{eq:Ak_defined}.
\begin{align}
    A_K = {} & \sum_{l'm'\in\{\surr\}} \int dt \, e^{i\bar{\omega}_{lmn}(t-t_0)} h^{\surr}_{l'm'} \nonumber\\
    &\times  \int d^2\Omega {}_{-2}{\bar{S}}_{lm}(\theta,\phi;a\omega_{lmn}) {}_{-2}{Y}_{l'm'}(\theta,\phi) \, .
\end{align}
The notation $l',m'\in \{\surr\}$ in the sum means that the sum takes place over (a subset of) the modes that are included in the surrogate waveform.
Using the expansion of the spin-weighted spheroidal harmonics in terms of the spin-weighted spherical harmonics in Eq.~\eqref{eq:Slm_expansion_Ylm}, and the orthogonality of the spin-weighted spherical harmonics in Eq.~\eqref{eq:Ylm_orthogonality} gives
\begin{align}\label{eq:A_components}
    A_K = {} & \sum_{l'} \bar{A}_{l'lm}(a\omega_{lmn}) \int dt \, e^{i\bar{\omega}_{lmn}(t-t_0)}  h_{l'm}^{\surr}(t) \, .
\end{align}
The quantities $A_{l'lm}$ are the mixing coefficients that arise from expanding the spin-weighted spheroidal harmonics in Eq.~\eqref{eq:Slm_expansion_Ylm} in terms of spherical harmonics.
Similarly, the components of $\mathbb{B}_\eta$ are given by
\begin{align}
     B_{IJ} ={}& \int dt \int d^2 \Omega \, e^{i(\bar{\omega}_{lmn}-\omega_{l^\prime \! m^\prime \! n^\prime})(t-t_0)}  \nonumber\\
    \times {}& {}_{-2}{\bar{S}}_{lm}(\theta,\phi;a\omega_{lmn}) {}_{-2}{S}_{l'm'}(\theta,\phi;a\omega_{l'm'n'}) \, .
\end{align}
Using the expansion of the spin-weighted spheroidal harmonics in Eq.~\eqref{eq:Slm_expansion_Ylm} and using the orthogonality of the spin-weighted spherical harmonics, the result can be written as
\begin{align}\label{eq:B_components}
    B_{IJ} =  {} & \int dt \, e^{i(\bar{\omega}_{lmn}-\omega_{l^\prime \! m \! n^\prime})(t-t_0)} \nonumber \\
    \times {}&\sum_{l''}\bar{A}_{l''lm}(a\omega_{lmn})A_{l''l'm}(a\omega_{l'mn'}) \, .
\end{align}
The sum over $l''$ 
runs over all allowed values of $l''$ above the larger of $l_\mathrm{min}$ and $l'_\mathrm{min}$.
In practice, we must truncate the sum at a finite upper value of $l''$, for which the coefficients $A_{l''lm}$ are sufficiently small.

We compute the components of $\boldsymbol{A}_\eta$ and $\mathbb{B}_\eta$ by evaluating the time integrals in Eq.~\eqref{eq:A_components} numerically and those in~\eqref{eq:B_components} analytically. 
We integrate over the time interval $[t_i,t_f]=[0,130M]$.
We choose the start of the QNM modes $t_0$ to be at the merger time, which we define to be at the peak time of the $l=2$, $m=2$ mode of the waveform (i.e., $t_0=0$).
After constructing $\boldsymbol{A}_\eta$ and $\mathbb{B}_\eta$, we can solve Eq.~\eqref{eq:QNMs_coefficients} for the QNM amplitudes $\mathbf C_\eta$.

\subsubsection{Multiple mass-ratio, multimode fitting} \label{subsubsec:multi-mode-mass}

As with the inspiral model, we will calculate a fit for the ringdown amplitudes at a particular mass ratio that can be computed from a fitting function in $\eta$, so that it is not necessary to solve the multimode fitting problem in Eq.~\eqref{eq:QNMs_coefficients} at each mass ratio.
To make this fitting problem more tractable, we restrict the number of waveform modes that we include in the spherical-harmonic expansion of the surrogate model.
Specifically, we fit the $(l,m)$ modes $(2,\pm 1)$, $(2,\pm 2)$, and $(3,\pm 2)$ given in the first row of Table~\ref{tab:lm-modes-sur}.
We henceforth refer to just the positive $m$ values, because the negative $m$ values can be obtained through complex conjugation.

For our fitting ansatz, we expand the QNM amplitude coefficients as quadratic polynomials in $\eta$ (the form of these quadratic fits is given in Eq.~\eqref{eq:Clmn_eta_expansion}).
To account for the vanishing of the $l=2$, $m=1$ mode for equal-mass binaries (with mass ratio $q=1$ and symmetric mass ratio $\eta=1/4$), we multiply the coefficients of the $(2,1)$ mode by a factor of $\sqrt{1-4\eta}$:
\begin{subequations}\label{eq:Clmn_eta_expansion}
    \begin{align}
        C_{22n} = {} & \sum_{j=0}^{2} C_{22nj} \, \eta^j \, , \\
        C_{32n} = {} & \sum_{j=0}^{2} C_{32nj} \, \eta^j \, , \\
        C_{21n} = {} & \sqrt{1-4\eta} \, \sum_{j=0}^{2} C_{21nj} \, \eta^j \, .
    \end{align}
\end{subequations}

Because we fit for 24 mode amplitudes---three $(l, m)$ mode pairs with the fundamental mode and seven overtones---our fitting problem contains 72 undetermined parameters.
We determine these coefficients by fitting for these coefficients using three mass ratios.
For the $(2,2)$ and $(3,2)$ modes we include the smallest and biggest mass ratios available plus one in the middle of the range (specifically $q=1, 5, 8$).
However, because the $h_{21}$ mode vanishes for equal mass ratio binaries, we fit over the three mass ratios $q=2, 5, 8$.
Because there is no mode mixing between the $m=1$ and $m=2$ modes, these two fitting problems decouple.
Thus, we solve two fully determined problems for the 24 coefficients $C_{21nj}$ and for the 48 coefficients $C_{22nj}$ and $C_{32nj}$.
We introduce $N$ to denote the number of QNMs in the multimode fitting (i.e., 8 for the $l=2$, $m=1$ case and 16 for the joint $l=2,3$, $m=2$ case).

The eigenvalue method discussed in Sec.~\ref{subsubsec:multimode} requires some modifications to obtain the coefficients $C_{Kj}$ rather than $\mathbf C_\eta$. 
The problem is still linear, and the solution can be written as the solution to a linear system of equations, 
\begin{equation}\label{eq:expanded_QNMs_coefficients}
    \boldsymbol{C} = \mathbb{B}^{-1} \cdot \boldsymbol{A} \, .
\end{equation}
Here $\boldsymbol{A}$ is constructed by combining three vectors $\boldsymbol{A}_\eta$ for the three different mass ratios used in the fitting procedure of a specific $(l,m)$ mode.
\begin{equation}
    \boldsymbol{A} = 
        \begin{pmatrix}
        \boldsymbol{A}_{\eta_1} \\
        \boldsymbol{A}_{\eta_2} \\
        \boldsymbol{A}_{\eta_3}
        \end{pmatrix}_{3N \times 1} \, .
\end{equation}
The components of $\boldsymbol{A}_{\eta_i}$ are computed from Eq.~\eqref{eq:A_components} for the $i^\mathrm{th}$ mass ratio.
The matrix $\mathbb{B}$ can be written in terms of the product of two matrices.
The first one is a block diagonal matrix constructed from the matrices $\mathbb{B}_\eta$ for each of the three mass ratios.
The second matrix is similar to a Vandermonde matrix, but it differs in that it includes the terms $\eta^i$ that appear in the expansion of the coefficients in terms of the symmetric mass ratio in Eq.~\eqref{eq:Clmn_eta_expansion}.
For fitting the $l=2$, $m=2$ and $l=3$, $m=2$ modes, the $\mathbb{B}$ matrix is defined as
\begin{equation}
\label{eq:B_matrix_m2}
    \mathbb{B} = 
    \begin{pmatrix}
        \mathbb{B}_{\eta_1} & 0 & 0 \\
        0 & \mathbb{B}_{\eta_2} & 0 \\
        0 & 0 & \mathbb{B}_{\eta_3}
    \end{pmatrix}_{3N\times 3N}
    \cdot
    \begin{pmatrix}
        \mathbb{I} & \mathbb{S}_{\eta_1} & \mathbb{S}_{\eta_1}^2\\
            \mathbb{I} & \mathbb{S}_{\eta_2} & \mathbb{S}_{\eta_2}^2 \\
            \mathbb{I} & \mathbb{S}_{\eta_3} & \mathbb{S}_{\eta_3}^2
    \end{pmatrix}_{3N\times 3N} \, ,
    \end{equation}
where $\eta_i$ represents the $i^\mathrm{th}$ mass ratio over which the fitting is done.
The matrix $\mathbb{S}_j$ is a diagonal matrix with the symmetric mass ratio as its diagonal elements, defined for the $i^\mathrm{th}$ mass ratio as
\begin{equation}
\label{eq:S_matrix}
    \mathbb{S}_{\eta_i} = \mathrm{diag}(\eta_i, \ldots, \eta_i)_{N\times N} \, .
\end{equation}
For the $l=2$, $m=1$ mode, the factor $\sqrt{1-4 \eta_i}$ can be absorbed into the matrix $\mathbb{B}$ by adding another block-diagonal matrix
\begin{equation}
\label{eq:B_matrix_m1}
    \mathbb{B} = 
    \begin{pmatrix}
        \mathbb{B}_{\eta_1} & 0 & 0 \\
        0 & \mathbb{B}_{\eta_2} & 0 \\
        0 & 0 & \mathbb{B}_{\eta_3}
    \end{pmatrix}
    \cdot
    \begin{pmatrix}
        \mathbb{F}_{\eta_1} & 0 & 0 \\
        0 & \mathbb{F}_{\eta_2} & 0 \\
        0 & 0 & \mathbb{F}_{\eta_3}
    \end{pmatrix}
    \cdot
    \begin{pmatrix}
        \mathbb{I} & \mathbb{S}_{\eta_1} & \mathbb{S}_{\eta_1}^2\\
            \mathbb{I} & \mathbb{S}_{\eta_2} & \mathbb{S}_{\eta_2}^2 \\
            \mathbb{I} & \mathbb{S}_{\eta_3} & \mathbb{S}_{\eta_3}^2
    \end{pmatrix} \, ,
\end{equation}
where the matrix of additional factors $\mathbb{F}_{\eta_i}$ is defined to be
\begin{equation}
\label{eq:SF_matrix}
    \mathbb{F}_{\eta_i} = \mathrm{diag}(\sqrt{1-4\eta_i},\ldots,\sqrt{1-4\eta_i})_{N\times N} \, .
\end{equation}
The vector of QNM amplitude coefficients $\boldsymbol{C}$ has the form 
\begin{equation}
    \boldsymbol{C} = 
    \begin{pmatrix}
        C_{K0} \\
        C_{K1} \\
        C_{K2} \\
    \end{pmatrix}_{3N \times 1} ,
\end{equation}
where $K$ runs over all QNM indices $lmn$ included in the multimode fit.
The coefficients $C_{Kj}$ were defined in Eq.~\eqref{eq:Clmn_eta_expansion}.

We solve the system of linear equations in Eq.~\eqref{eq:expanded_QNMs_coefficients} by computing the inverse of the constructed matrix $\mathbb{B}$ using its singular value decomposition.
The components in the vector of coefficients $\boldsymbol{C}$ in Eq.~\eqref{eq:expanded_QNMs_coefficients} are written in terms of their amplitude and phase in Table~\ref{tab:QNM-coeffs} in Appendix~\ref{app:rd-coeffs}.

The model for the GW strain modes can be computed from Eq.~\eqref{eq:h_model_ringdown} with the coefficients $C_{Kj}$, the spherical-spheroidal mixing coefficients $A_{l'lm}(a\omega)$, and the QNM frequencies $\omega_{lmn}$.
Substituting the expansion of the spin-weighted spheroidal harmonics in Eq.~\eqref{eq:Slm_expansion_Ylm} into the strain model $h_{\model}^\rd$ in Eq.~\eqref{eq:h_model_ringdown} gives
\begin{align}
    h_{\model}^\rd = {} & \frac{M}{r} \sum_{l,m,n,j} C_{lmnj} \eta^j e^{-i\omega_{lmn}(t-t_0)}   \nonumber \\
    & \times \sum_{l'} A_{l'lm}(a\omega_{l'mn}) {}_{-2}Y_{l'm} \, .
\end{align}
It then follows that the $(l,m)$ mode ringdown strain expanded in spherical harmonics can be written as
\begin{equation}\label{eq:hlm_ringdown_model}
    h_{lm}^\rd = \frac{M}{r} \sum_{l',n,j} C_{l'mnj} \eta^j A_{ll'm}(a\omega_{lmn}) e^{-i\omega_{l'mn}(t-t_0)} \, .
\end{equation}
It can be seen from the form of the ringdown strain modes in Eq.~\eqref{eq:hlm_ringdown_model} that the mixing of different modes arises for different $l$ with the same $m$, because of the coefficients $A_{ll'm}$.

\begin{figure}
    \centering
    \includegraphics[width=0.48\textwidth]{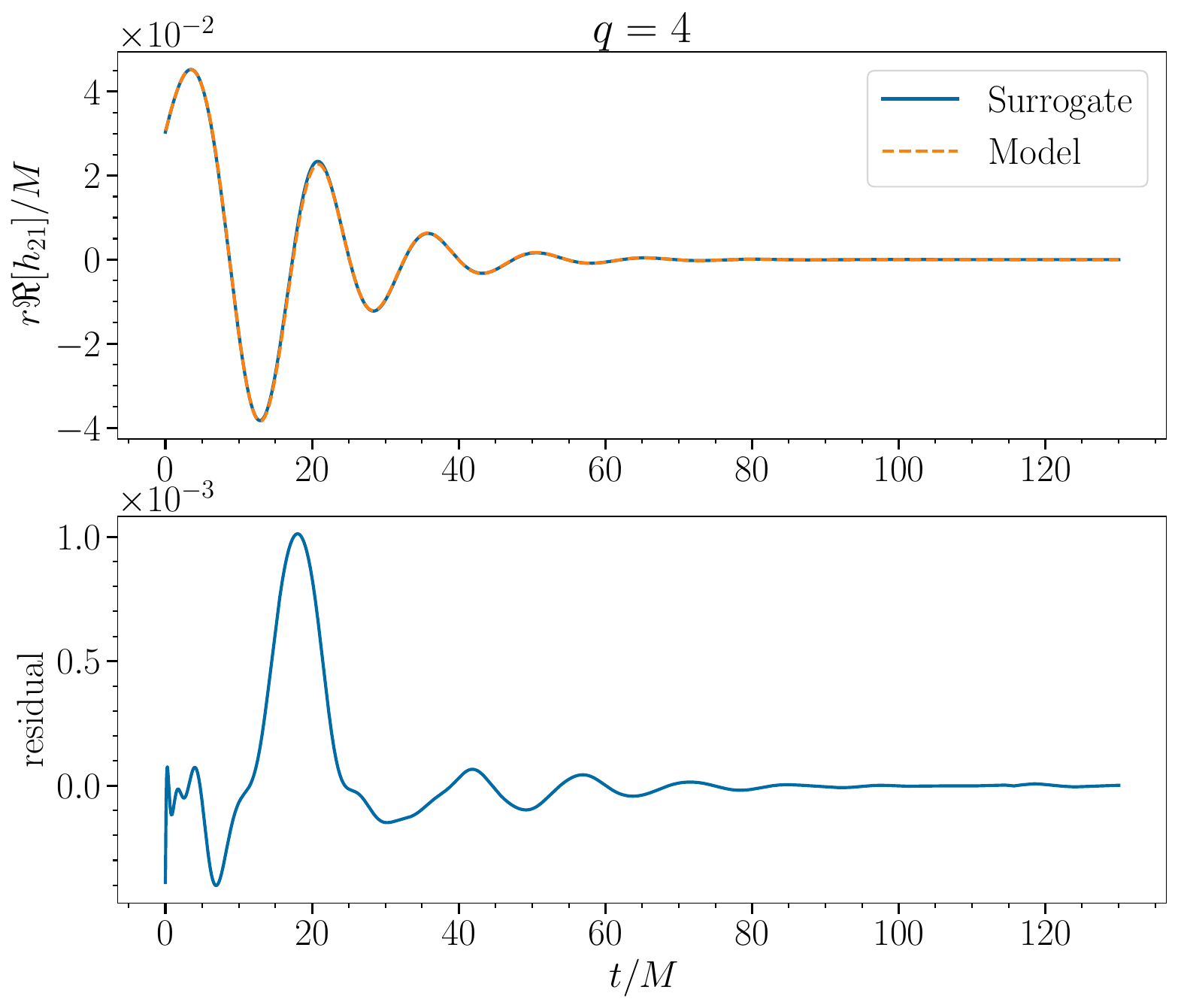}
    \caption{\textbf{Waveform for the $h_{21}$ mode}:
    \emph{Top}: The real part of the $h_{21}$ mode computed from the surrogate model shown as a solid, blue curve, and the same mode computed from our ringdown model in Eq.~\eqref{eq:hlm_ringdown_model} is shown as an orange, dashed curve.
    \emph{Bottom}: The residuals between the $h_{21}$ waveform evaluated from the surrogate model and our ringdown model waveform.}
    \label{fig:h21_model}
\end{figure}
We first show the results of the fitting procedure for the $l=2$, $m=1$ mode in Fig.~\ref{fig:h21_model}.
We illustrate just the mass ratio $q=4$, as a representative mass ratio which we did not use in constructing the ringdown fit.
The top panel shows the surrogate model as a solid, blue curve.
The ringdown model of the mode is the orange, dashed curve.
The residual (difference between the QNM ringdown model and the NR surrogate) is shown in the bottom panel.
For the $l=2$, $m=1$ mode, the difference between the ringdown model and the surrogate is at most a few percent.

\begin{figure*} 
    \centering
    \includegraphics[width=0.48\textwidth]{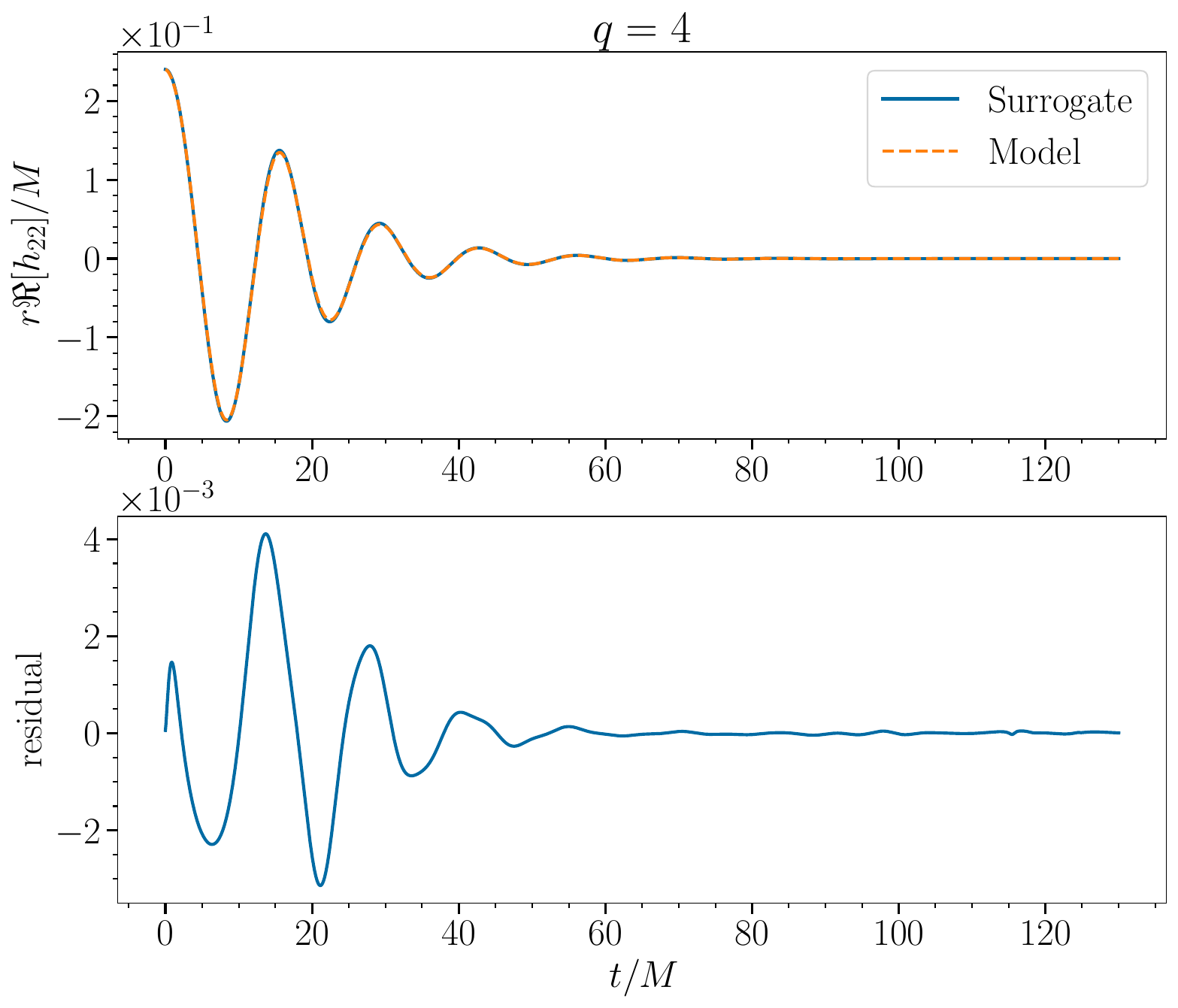}
    \includegraphics[width=0.48\textwidth]{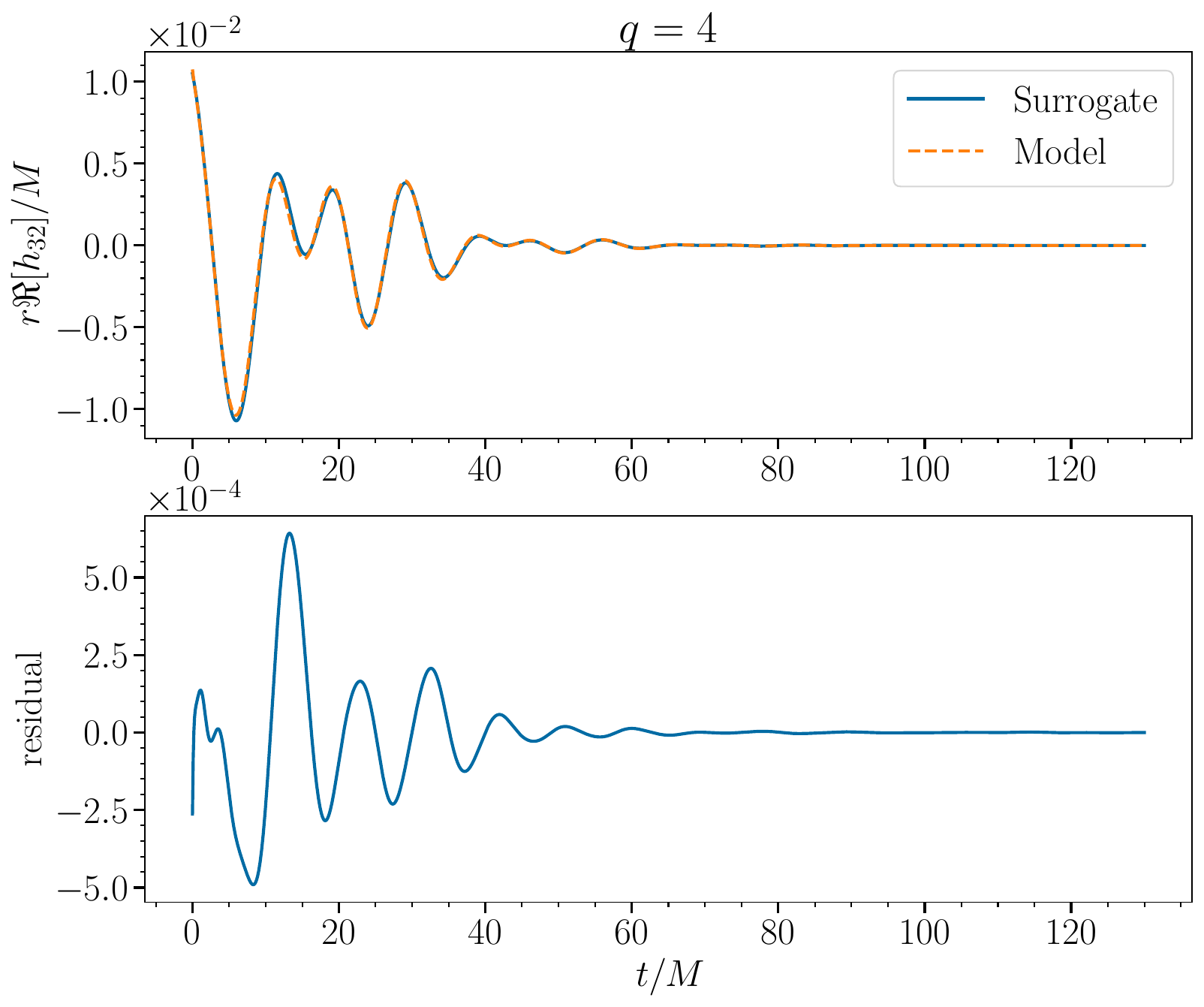}
    \caption{\textbf{Waveform for the $h_{22}$ and $h_{32}$ modes}:
    \emph{Top}: The real part of the $h_{22}$ mode (left) from the surrogate model shown in solid, blue curve.
    The same mode computed from our ringdown model is the orange, dashed curve.
    On the right are the equivalent results for the $h_{32}$ mode.
    \emph{Bottom}: The residuals between the $h_{22}$ waveform evaluated from the surrogate model and our ringdown model waveform (left) and the same quantity for the $h_{32}$ mode (right).}
    \label{fig:h22_32_model}
\end{figure*}
Next, we show the similar results for the $l=2,3$ and $m=2$ modes in Fig.~\ref{fig:h22_32_model} for the mass ratio $q=4$.
The results for $l=2$ are shown on the left and $l=3$ are on the right.
The depiction of the curves and the residuals are completely analogous to those in Fig.~\ref{fig:h21_model} for the $l=2$, $m=1$ mode.
The $l=2$, $m=2$ mode is the largest of the three (more than an order of magnitude larger than the $l=3$, $m=2$ mode and a factor of five larger than the $l=2$, $m=1$ mode).
The multimode ringdown model again reproduces the NR surrogate results with an accuracy of roughly a few percent.
The $l=3$, $m=2$ mode of the ringdown has a more complicated oscillatory structure than the $l=2$, $m=2$ mode, because the $l=2$, $m=2$ QNMs mix into the $l=3$ and $m=2$ harmonic with a larger relative amplitude than the mixing of the $l=3$, $m=2$ modes into the $l=2$, $m=2$ harmonic.

Note that we did not fit for the QNM frequencies $\omega_{lmn}$ or the mixing coefficients $A_{l'lm}$ for different mass ratios.
To evaluate the model for the ringdown memory modes in Eq.~\eqref{eq:hlm_ringdown_model}, we use the \emph{Python} package \texttt{qnm} for a certain final mass $M_f$ and spin parameter $\chi_f$ for a given mass ratio.
We obtain the values of the final mass and spin from the \texttt{SurfinBH} package rather than refitting those values.

\subsubsection{The ringdown memory model} \label{subsubsec:ringdown-mem}

We now compute the model for the ringdown memory strain signal using the modes $h_{lm}^{\rd}$ in Eq.~\eqref{eq:hlm_ringdown_model}.
To obtain the $h_{20}$ memory mode, we substitute the ringdown strain modes computed in Eq.~\eqref{eq:hlm_ringdown_model} into Eq.~\eqref{eq:h20Memory_hlm_nonprecessing}.
As we described near Eq.~\eqref{eq:h20_insp_int_rd}, during the ringdown, we compute the memory signal for some $t \geq t_0$ by integrating from late times ($t\rightarrow\infty$) back to a time $t$ and adding the result from the final memory offset.
This is equivalent to subtracting the integral from this time $t \geq t_0$ to infinity from $\Delta h_{20}$.
We can explicitly evaluate this integral by using the form of the ringdown modes in Eq.~\eqref{eq:hlm_ringdown_model}.
We find that
\begin{align} \label{eq:h_rd-int}
    \int_t^\infty dt' \,{}& \dot{h}_{l'm'}(t')  \dot{\bar h}_{l''m'}(t') = \nonumber \\
    & \frac{M^2}{r^2}\sum_{\bar{l},\bar{\bar{l}},\bar{n},\bar{\bar{n}},j,j'} \bigg( \frac{i\omega_{\bar{l}m'\bar{n}} \bar \omega_{\bar{\bar{l}}m'\bar{\bar{n}}}}{\omega_{\bar{l}m'\bar{n}} - \bar \omega_{\bar{\bar{l}}m'\bar{\bar{n}}}} \bigg) C_{\bar{l}m'\bar{n}j'} \bar C_{\bar{\bar{l}}m'\bar{\bar{n}}j} \eta^{j+j'}   \nonumber \\
    & \times A_{l'\bar{l}m'} \bar A_{l''\bar{\bar{l}}m'} e^{-i(\omega_{\bar{l}m'\bar{n}} - \bar\omega_{\bar{\bar{l}}m'\bar{\bar{n}}})(t-t_0)} \, .
\end{align}
Because the imaginary part of the QNM frequency is negative (i.e. $\Im[\omega_{lmn}]<0$), the exponential term vanishes at infinity.
We have also used the shorter notation of $A_{\bar l lm}(a\omega_{lmn})\equiv A_{\bar l lm}$ for conciseness.
By substituting Eq.~\eqref{eq:h_rd-int} into the expression for the memory signal Eq.~\eqref{eq:h20Memory_hlm_nonprecessing}, we get the following form of the ringdown memory model for nonprecessing BBH mergers
\begin{align} \label{eq:h20_ringdown_model}
    h_{20}^{\rd} {}& (t) =   \nonumber\\
    & \frac{M^2}{\sqrt{6}r} \sum_{l',l''}\sum_{m'=1}^{l'} (-1)^{m'} C_l(-2,l',m';2,l'',-m') \nonumber\\
    & \times \sum_{\bar{l},\bar{\bar{l}},\bar{n},\bar{\bar{n}},j,j'} \Im \bigg[\bigg( \frac{\omega_{\bar{l}m'\bar{n}} \bar \omega_{\bar{\bar{l}}m'\bar{\bar{n}}}}{\omega_{\bar{l}m'\bar{n}} - \bar \omega_{\bar{\bar{l}}m'\bar{\bar{n}}}} \bigg) C_{\bar{l}m'\bar{n}j} \bar C_{\bar{\bar{l}}m'\bar{\bar{n}}j'} \eta^{j+j'}  \nonumber \\
    & \times A_{l'\bar{l}m'} \bar A_{l''\bar{\bar{l}}m'} e^{-i(\omega_{\bar{l}m'\bar{n}} - \bar\omega_{\bar{\bar{l}}m'\bar{\bar{n}}})(t-t_0)} \bigg] \, .
\end{align}

Equation~\eqref{eq:h20_ringdown_model} can be used to compute the ringdown memory signal, because we obtain the QNM frequencies $\omega_{lmn}$ and spherical-spheroidal mixing coefficients $A_{l'lm}(a\omega_{lmn})$ from the \texttt{qnm} package (using the final mass and spin from the \texttt{SurfinBH} package), and we have fit for the polynomial coefficients of the QNM amplitudes $C_{lmnj}$ (see Table~\ref{tab:QNM-coeffs}).
Given that our modeling of the oscillatory ringdown modes contains the three largest $(l,m)$ modes $[(2,2), (3,2), (2,1)]$, we compute the memory model for the $(2,0)$ mode from Eq.~\eqref{eq:h20_ringdown_model} from these three modes.

\begin{figure*}
    \centering
    \includegraphics[width=0.48\textwidth]{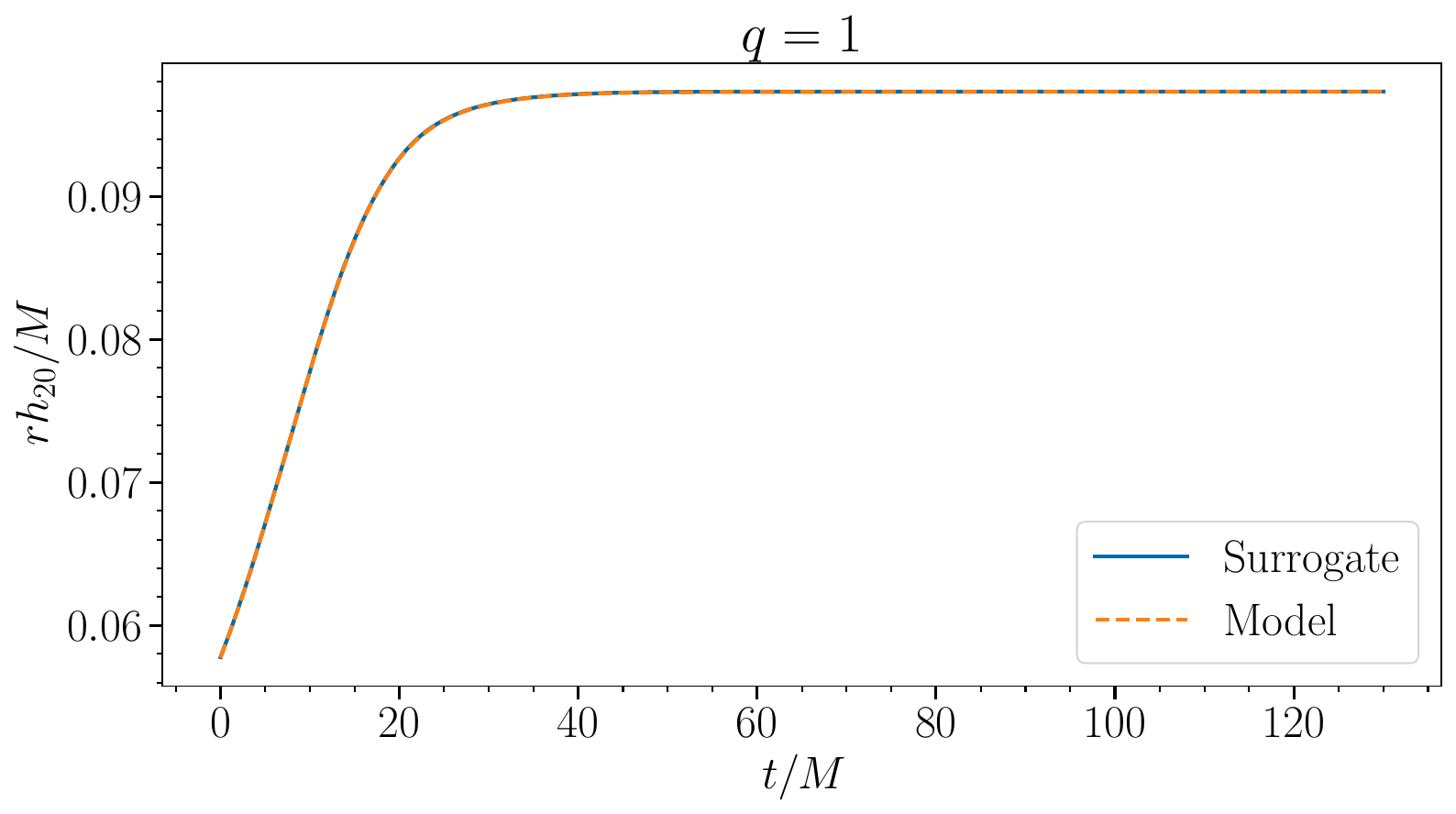}
    \includegraphics[width=0.48\textwidth]{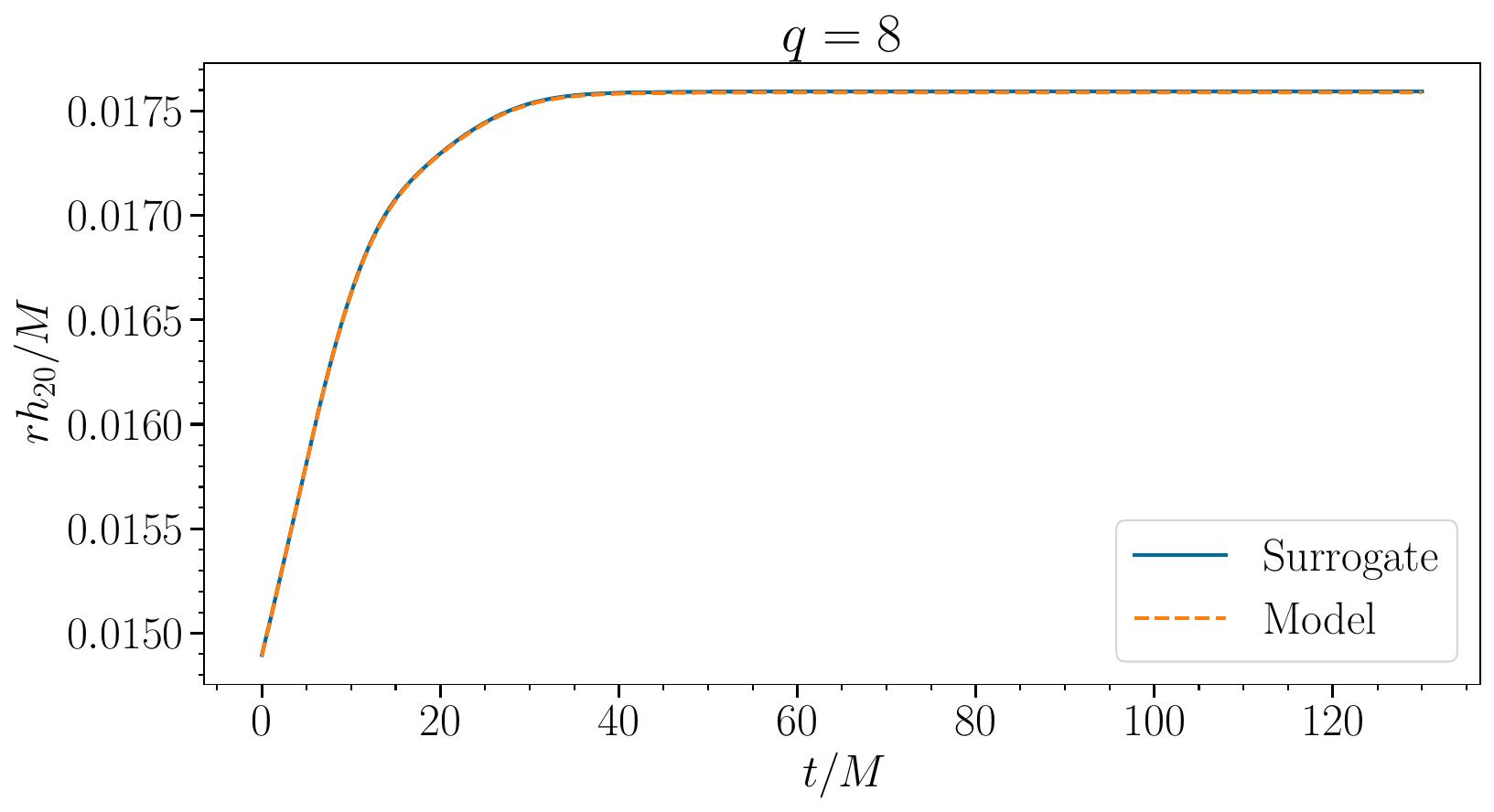}
    \includegraphics[width=0.48\textwidth]{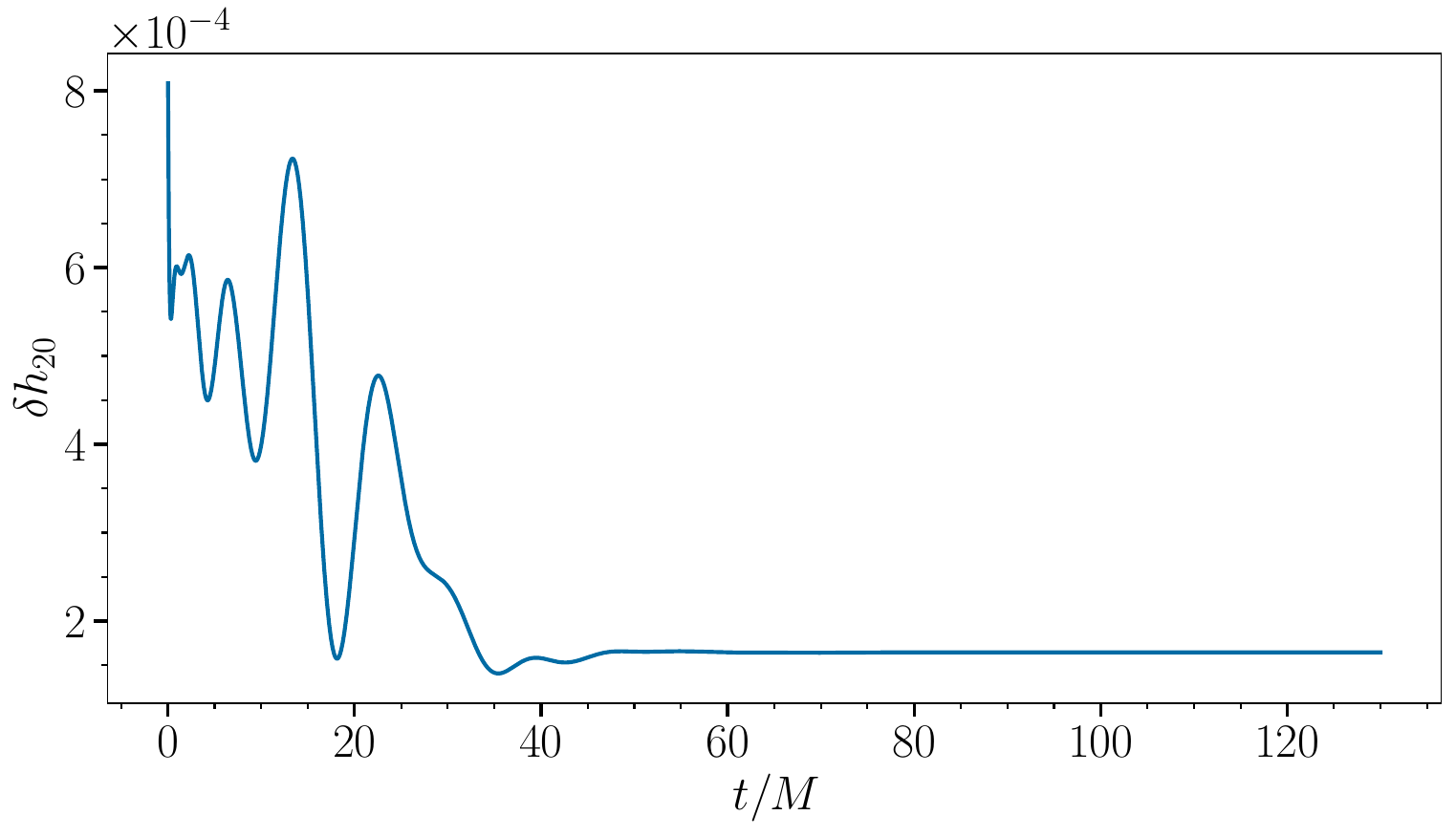}
    \includegraphics[width=0.48\textwidth]{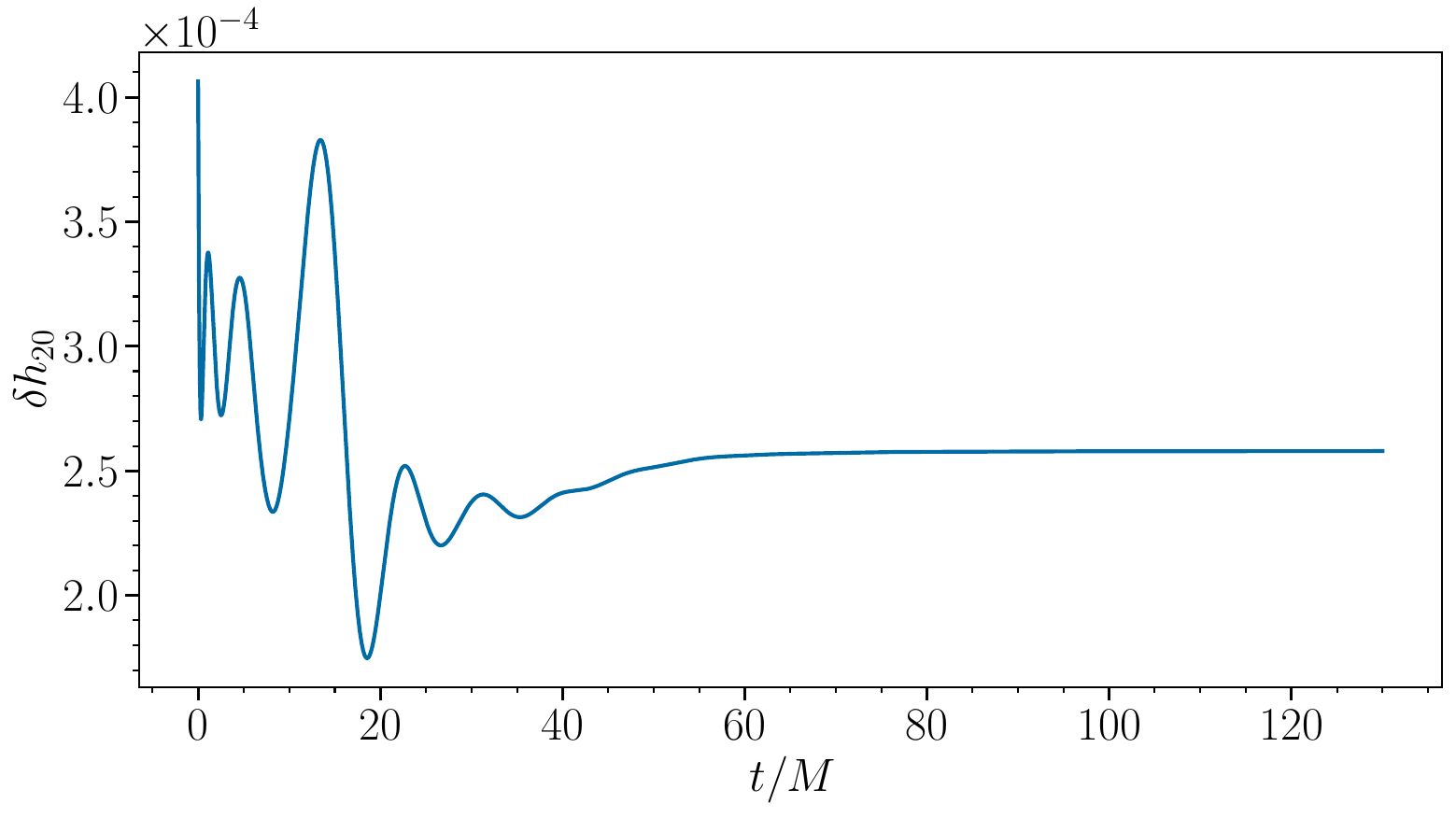}
    \caption{\textbf{Ringdown memory model and relative error versus time}:
    \emph{Top}: The hybridized surrogate memory signal computed from the NRHybSur3dq8 surrogate waveform modes (solid, blue curve) and the ringdown memory model (dashed, orange curve) for non-spinning BBH mergers.
    The left panel contains the results for the mass ratio $q=1$ and the right shows the mass ratio $q=8$. 
    \emph{Bottom}: The relative error of the ringdown memory model and the memory signal computed directly from the NR surrogate for the corresponding mass ratios in the panels above.}
    \label{fig:ringdown_memory}
\end{figure*}
We show the full ringdown memory signal in Fig.~\ref{fig:ringdown_memory}.
We show the extreme two mass ratios $q=1,8$ for our model, along with the relative error between our ringdown model and the ringdown memory computed from the NR surrogate waveforms.
The left panels are for the mass ratio $q=1$ and the right panels are for $q=8$.
Note that the relative error is at least an order of magnitude smaller at mass ratios $q=1$ and $q=8$ than the corresponding relative error associated with the ringdown modes for $q=4$ (described in Sec.~\ref{subsubsec:multi-mode-mass}).
The error is significantly smaller for these mass ratios because these were two of the mass ratios used to perform the multimode fitting for the $l=2,3$ and $m=2$ oscillatory ringdown modes.
The relative error for the memory ringdown signal for $q=4$ is comparable to the relative error in the modes depicted in Figs.~\ref{fig:h21_model} and~\ref{fig:h22_32_model}.

\subsection{Intermediate-time memory model}

Analytical models for the GW strain of BBH mergers during the late-inspiral and merger phases often have some phenomenological component, because PN theory and black-hole perturbation theory alone do not have sufficient accuracy to be a robust model of these phases.
For this reason, we will take a purely phenomenological modeling approach to model the memory during this interval. 

We model the intermediate memory as a sum of exponential functions (and a constant)
\begin{equation}
\label{eq:h_model_intermediate}
    h_{20}^{\intr}(t) = \frac Mr\sum_{j=0}^{6} c_j e^{p_j t} .
\end{equation}
We denote the amplitudes of the exponential functions by $c_j$ and the exponents by $p_j$, with the convention that $p_0=0$ (so that the first term is a constant).
We require that the intermediate memory is continuous with inspiral and ringdown memory at times $t_\mathrm{int}=-2000M$ and $t_\rd=2M$, respectively.
Similarly, we enforce continuity of the first and second time derivatives of the memory signal at these times.
The choice of starting the ringdown model at $t_\rd=2M$ was made empirically by observing that it produced a better matching of the time-derivatives of the memory signal than the start time of the ringdown memory signal ($t=t_0=0$).

The choice of exponential functions as the fitting ansatz (though ultimately an arbitrary choice) was made for several reasons.
First, this model has a simple analytical Fourier transform, which will make it easier to construct a frequency-domain model directly from this time-domain ansatz.
Second, other simple fitting functions (such as polynomials, rational functions, or their superposition) we found fit less accurately.
Third, the more complicated ansatz used in the IMRPhenomT waveform models~\cite{Estelles:2020osj} for the amplitude of the dominant quadrupole mode does not match the morphology of the memory waveform signal well.
However, the spirit of finding a phenomenological fitting function for the late inspiral and merger is similar.\footnote{Our sum of exponential functions in Eq.~\eqref{eq:h_model_intermediate} is more similar to the model of a linear function plus an exponential function (with a base not equal to Euler's constant) that was used in~\cite{Rossello-Sastre:2024zlr}. Given that our intermediate region spans a longer time interval than the late-inspiral and merger model of~\cite{Rossello-Sastre:2024zlr}, it is reasonable that we require more terms in our model to fit the memory signal over these times.}
Although the exponential model may not be a unique (or obvious) ansatz, it is able to fit the numerical-relativity results over the intermediate time interval $[t_\mathrm{int},t_\rd]$.

To construct a memory signal that can be evaluated for any mass ratio in the range $q\in[1,8]$, we expand the parameters $p_j$ (for $j\neq 0$) as linear functions of the symmetric mass ratio $\eta$:
\begin{equation}
    p_j = p_{j0} + p_{j1} \eta \, .
\end{equation}
The intermediate model nominally has nineteen free parameters: seven amplitude coefficients $c_j$ and twelve exponent coefficients $p_{ji}$.
By matching the value of the memory model and its first and second derivatives at the times $t_\mathrm{int}$ and $t_\mathrm{rd}$, we fix six of the free parameters, leaving thirteen in our model.
We choose to solve linear equations for the coefficients $c_j$ for $j=0,\ldots,5$ in terms of the other parameters, which leaves $c_6$ and $p_{10}$, $p_{11},\ldots, p_{61}$ as the thirteen free parameters.

The system of linear equations for the six coefficients $c_0, \ldots, c_5$ that we solve can be written in matrix form as 
\begin{equation}\label{eq:h_model_int_lineq}
    \mathbb{M} \cdot \mathbf c = \mathbf h \, .
\end{equation}
Here $\mathbb{M}$ is the model in Eq.~\eqref{eq:h_model_intermediate} evaluated at the initial and final times, $\mathbf c$ is a vector of the coefficients $c_0, \ldots, c_5$, and $\mathbf h$ is the inspiral and ringdown memory models evaluated at the initial and final times, respectively.
The components of $\mathbb{M}$ are constructed by evaluating the exponential terms in the memory model (and its first and second derivatives) in Eq.~\eqref{eq:h_model_intermediate} at times $t_\mathrm{int}$ and $t_\rd$
\begin{equation}
    \mathbb{M} = \begin{pmatrix}
        1 & e^{p_1 t_\mathrm{int}} & \ldots & e^{p_5 t_\mathrm{int}} \\
        1 &  e^{p_1 t_\rd} & \ldots & e^{p_5 t_\rd} \\
        0 & p_1e^{p_1 t_\mathrm{int}} & \ldots & p_5e^{p_5 t_\mathrm{int}} \\
        0 & p_1e^{p_1 t_\rd} & \ldots & p_5e^{p_5 t_\rd} \\
        0 & p_1^2 e^{p_1 t_\mathrm{int}} & \ldots & p_5^2e^{p_5 t_\mathrm{int}} \\
        0 & p_1^2 e^{p_1 t_\rd} & \ldots & p_5^2e^{p_5 t_\rd} \\
    \end{pmatrix} \, .
\end{equation}
The vector $\mathbf c$ is
\begin{equation}
    \boldsymbol{c} = \begin{pmatrix}
        c_0 \\ \vdots \\ c_5
    \end{pmatrix} \, .
\end{equation}
Finally, the vector $\mathbf h$ is constructed from the values of the inspiral and ringdown models (and their first and second derivatives) evaluated at the times $t_\mathrm{int}$ and $t_\rd$, respectively
\begin{equation}
    \mathbf h = \begin{pmatrix}
        h_{20}^{\insp}(t_\mathrm{int}) \\ h_{20}^{\rd}(t_\rd) \\ \dot h_{20}^{\insp}(t_\mathrm{int}) \\ \dot h_{20}^{\rd}(t_\rd) \\ \ddot h_{20}^{\insp}(t_\mathrm{int}) \\ \ddot h_{20}^{\rd}(t_\rd)
    \end{pmatrix} \, .
\end{equation}
Solving the system of linear equations in Eq.~\eqref{eq:h_model_int_lineq} gives the values of the coefficients ($c_0, \ldots, c_5$) that matches the intermediate memory model (and its first and second time-derivatives) to the inspiral and ringdown models, for a given choice of the remaining free parameters ($c_6$ and $p_{10}$, $p_{11},\ldots, p_{61}$).

We determine values of the remaining thirteen parameters by performing a nonlinear least-squares fit by minimizing the cost function $C[h_{20}^{\surr},h_{20}^{\intr}]$:
\begin{equation}\label{eq:cost_function_intermediate}
    C[h_{20}^{\surr},h_{20}^{\intr}] = \sum_{q=1}^8 C_q [h_{20}^{\surr},h_{20}^{\intr}] \, .
\end{equation}
The cost function is defined analogously to that in Eq.~\eqref{eq:cost_function_inspiral}: namely, it is the sum over eight mass ratios $q=[1,\ldots,8]$, but now it is evaluated using the intermediate model over the time interval $[t_\mathrm{int},t_\rd]=[-2000M,2M]$.
The cost function is minimized using the \texttt{SciPy} \emph{minimize} function.
The resulting coefficients that minimize the cost function are listed in Table~\ref{tab:pij_coefficients}.

\begin{table}[t]
    \centering
    \caption{The coefficients of the intermediate memory model in Eq.~\eqref{eq:h_model_intermediate}. The remaining coefficients $c_0,\ldots,c_5$ are obtained from solving the linear system of equations in~\eqref{eq:h_model_int_lineq}.}
    \begin{tabular}{c|l}
    \hline
    \hline
    
       Coefficient & Numerical Value \\
       \hline
       
    $p_{10}$ & $8.49306 \times 10^{-1}$  \\
    $p_{11}$ & $1.02024 \times 10^{1}$   \\
    $p_{20}$ & $4.29098 \times 10^{-2}$   \\
    $p_{21}$ & $2.36601 \times 10^{-1}$  \\
    $p_{30}$ & $8.37061 \times 10^{-3}$   \\
    $p_{31}$ & $1.59611 \times 10^{-2}$  \\
    $p_{40}$ & $5.97250 \times 10^{-4}$  \\
    $p_{41}$ & $1.10423 \times 10^{-3}$  \\
    $p_{50}$ & $8.23572 \times 10^{-1}$  \\
    $p_{51}$ & $1.01875 \times 10^{0}$   \\
    $p_{60}$ & $9.40991 \times 10^{-4}$  \\
    $p_{61}$ & $6.22363 \times 10^{-3}$  \\
    $c_{6}$  & $1.19119 \times 10^{-3}$ \\
    \hline
    \hline
    \end{tabular}
    \label{tab:pij_coefficients}
\end{table}

\subsection{Full time-domain memory signal model}

\begin{figure*}
    \centering
    \includegraphics[width=0.48\textwidth]{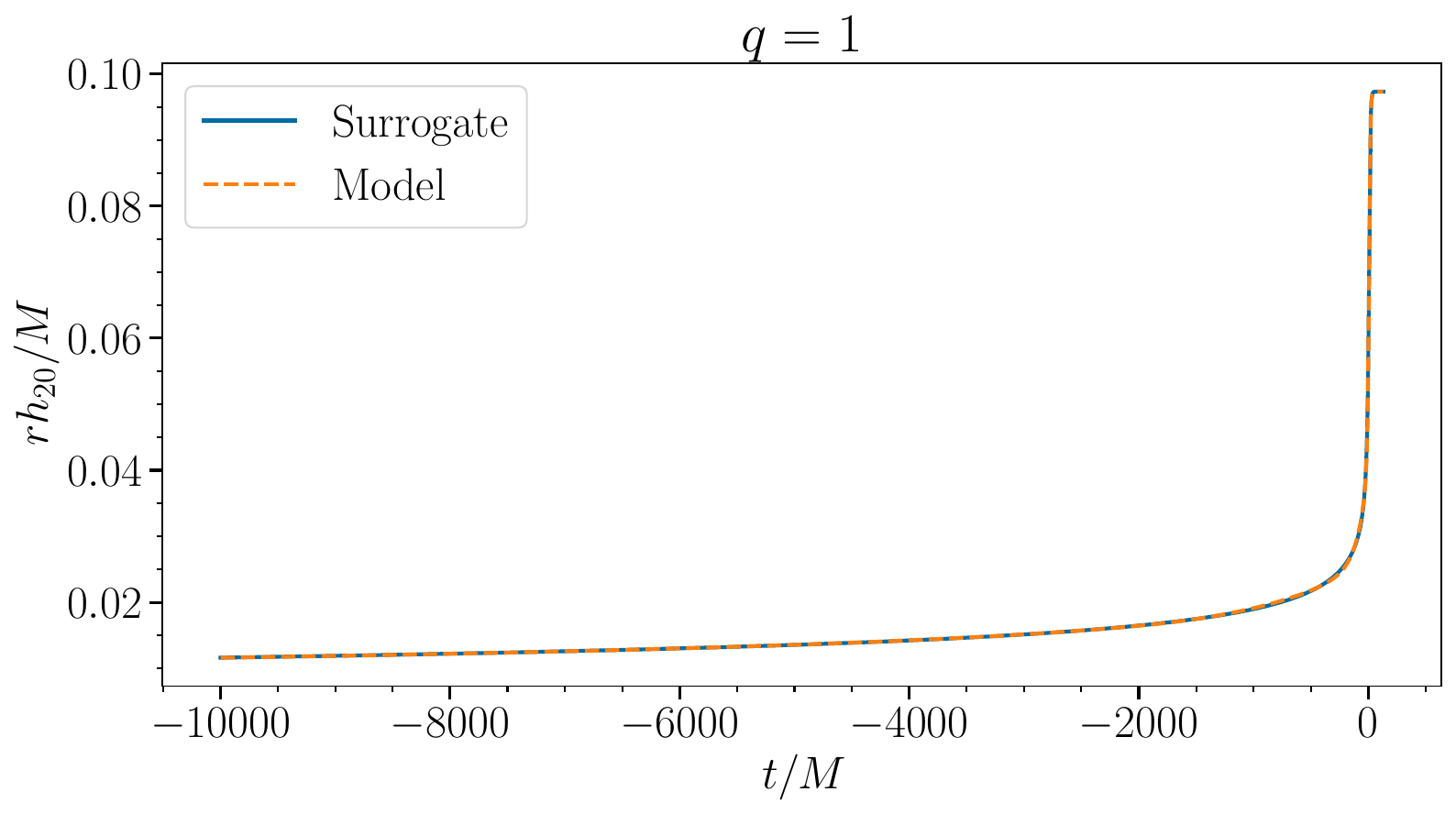}
    \includegraphics[width=0.48\textwidth]{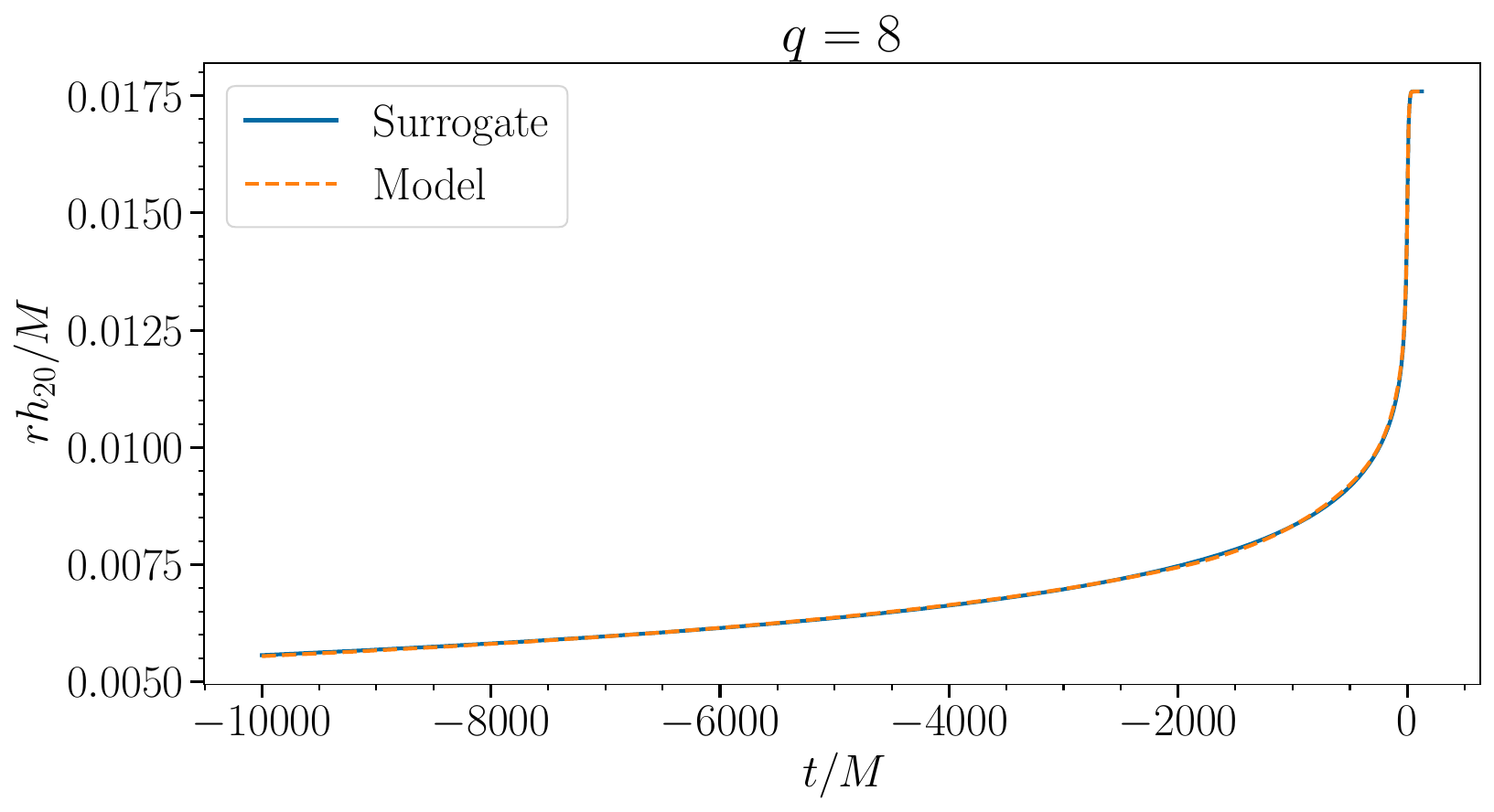}
    \includegraphics[width=0.48\textwidth]{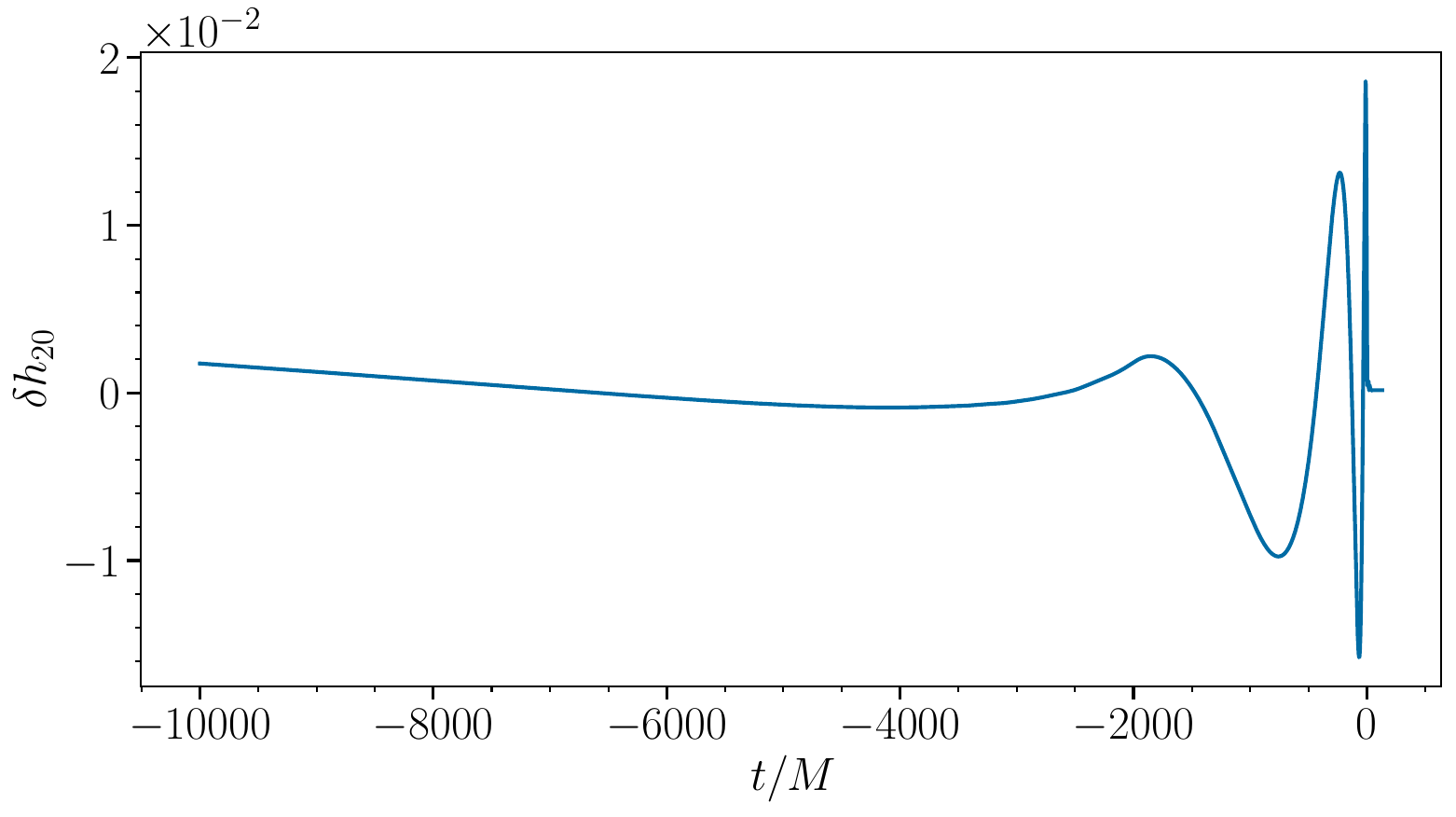}
    \includegraphics[width=0.48\textwidth]{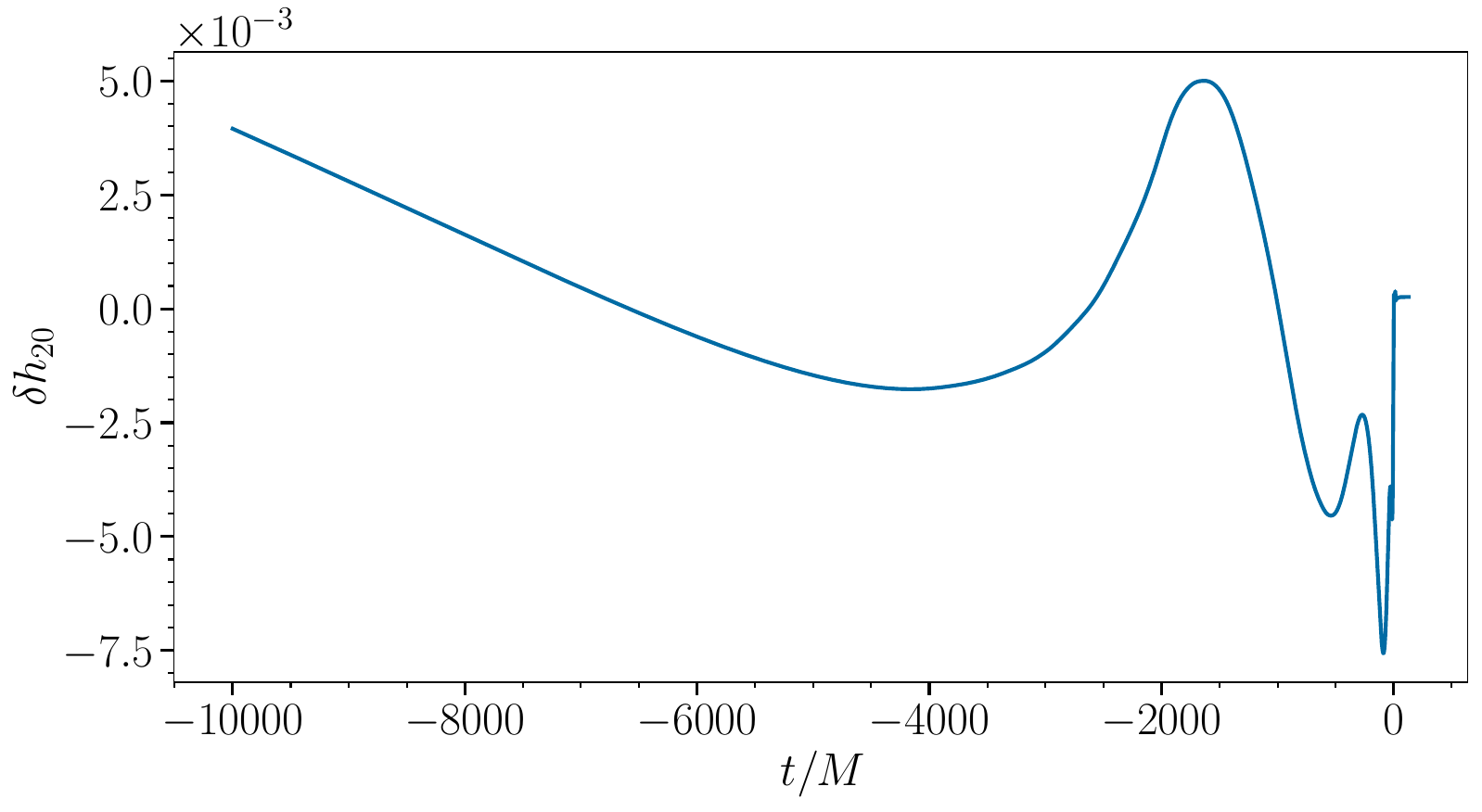}
    \caption{\textbf{Memory model and relative error versus time}:
    \emph{Top}: The GW memory signal computed directly from the hybridized surrogate (solid blue) and from the time-domain model (dashed orange).
    The left panel is for an equal mass binary ($q=1$), whereas the right panel is for the largest mass ratio of $q=8$ that we model.
    \emph{Bottom}: The relative error between the surrogate and the time-domain model for the two mass ratios.}
    \label{fig:memory_model}
\end{figure*}

Now that we have described how we construct all three parts of the model, we can now show the complete time-domain model.
A Python package, \texttt{GWMemoryModel} that implements the time-domain model, is available on GitHub~\cite{code:GWMemoryModel}.
We do so in Fig.~\ref{fig:memory_model}, which displays the scaled memory signal $(r/M)h_{20}$ as a function of time.
In both the left and right panels on top, the solid, blue curve is the memory signal computed directly from the NR hybrid surrogate model.
The dashed, orange curve in these panels is the memory signal computed using our model.
The left panel is the memory signal from an equal mass ($q=1$) BBH system, whereas the right panel is that for a mass ratio of $q=8$.
The relative error between the two signals in the top panels (as defined in Eq.~\eqref{eq:relative_error}) is depicted in the bottom row of Fig.~\ref{fig:memory_model} for the corresponding mass ratios illustrated above them.

For the equal mass case, the relative error during the inspiral and ringdown portions of the waveform are smaller than that during the times when the phenomenological intermediate model is used.
Nevertheless, there is good agreement between the surrogate (namely, relative errors of at most a few percent).
At the larger mass ratio of $q=8$, the magnitude of the relative error is smaller than in the $q=1$ case (specifically, it is less than one percent).
This makes the relative error during the PN inspiral comparable to the error during the intermediate stage (the error during the ringdown is smaller, because the mass ratio $q=8$ was a value used to fit the coefficients in the memory model, as discussed in Sec.~\ref{subsubsec:ringdown-mem}).

To obtain more intuition about how significant a relative error of a few percent is for analyzing GW memory signals, it will be useful to compute the mismatch between the NR surrogate and our GW memory signal model.
To do so, however, it is advantageous to analyze the signals in the frequency domain, in which the mismatch is typically computed.
We thus turn in the next section to transforming our time-domain memory model to the frequency domain.

\section{Frequency-domain memory signal} \label{sec:freq-domain}

In this section, we compute and discuss the GW memory signal in the frequency domain.
In the first subsection, we derive an analytic expressions for the frequency-domain model by taking the (continuous) Fourier transform (FT) of the time-domain model.
Next, we discuss how we compute the (discrete) fast Fourier transform (FFT) of the time-domain memory model.
We compare both results with the FFT of the time-domain signal computed from the NR hybrid surrogate model.
In certain parts of the discussion of the frequency-domain memory signal below, we will drop the delta function term at zero-frequency, because most GW detectors cannot measure the zero-frequency component of the signal.
We will note the points at which we make this approximation.

\subsection{Analytic frequency-domain model}

Our convention for the (forward) Fourier transform is
\begin{equation}
    \tilde{H}(f) = \int_{-\infty}^{\infty} dt e^{-2\pi i f t} H(t).
\end{equation}
Given the piecewise nature of our time-domain GW memory model in Eq.~\eqref{eq:h20_insp_int_rd}, its continuous Fourier transform can be written as
\begin{align}
\label{eq:h20tilde_insp_int_rd}
    \tilde{h}_{20}(f) ={}& \int_{-\infty}^{t_\intr} dt \, e^{-2\pi i f t} h^{\insp}_{20}(t) \nonumber \\ 
    +{}& \int_{t_\intr}^{t_\rd} dt \, e^{-2\pi i f t} h^{\intr}_{20}(t) \nonumber\\
    +{}& \int_{t_\rd}^{\infty} dt \, e^{-2\pi i f t} \big(\Delta h_{20} -  h_{20}^{\rd}(t)\big) \, .
\end{align}
We discuss how we evaluate the three integrals on the right-hand side of Eq.~\eqref{eq:h20tilde_insp_int_rd} in the next three parts of this subsection.

\subsubsection{Fourier transform of the inspiral memory signal}

We write the inspiral memory model (3.5PN memory in Eq.~\eqref{eq:PN_memory}) as
\begin{align}
    h_{20}^{\insp}(t) \equiv {}& \frac{4M}{7r} \sqrt{\frac{5 \pi}{6}} \eta \sum_n C^{\PN}_{n} x^{1+n} \, ,
\end{align}
where $n$ runs over the PN orders ($n=[0,1, 2, 2.5, 3, 3.5]$).
For conciseness, we denote the coefficients that appear in the PN expansion in Eq.~\eqref{eq:PN_memory} as $C_n^{\PN}$.
Using the PN parameter $x(t)$ in Eq.~\eqref{eq:x-of-t}, we find that the Fourier transform of the inspiral memory signal requires evaluating the following integrals:
\begin{align}
    \tilde{h}_{20}^{\insp}(f) = {}& \frac{4M}{7r} \sqrt{\frac{5 \pi}{6}} \eta \sum_n C^{\PN}_{n} (C_{x})^{1+n} \nonumber\\
    \times {}&\int_{-\infty}^{t_\intr} dt \, e^{-2\pi ift}   (t_c - t)^{-(1+n)/4} \,.
\end{align}
We defined $C_x \equiv (1/4)(\eta/5)^{-1/4}$ to be the coefficient multiplying $(t_c-t)^{-1/4}$ in the definition of the $x$.
After a change of variables to $y = (t_\intr-t)/(t_c-t_\intr)$, we can write the integral as
\begin{align}
    \tilde{h}_{20}^{\insp}(f) = {}& \frac{4M}{7r} \sqrt{\frac{5 \pi}{6}} \eta \sum_n C^{\PN}_{n} C^{(1+n)}_{x} (t_c - t_\intr)^{(3-n)/4} \nonumber\\
    \times {}& e^{-2\pi ift_\intr} \int^{\infty}_{0} dy \, e^{2\pi if(t_c-t_\intr)y} (1+y)^{-(1+n)/4} \, .
\end{align}
As was noted by Favata~\cite{Favata:2009ii} for the Newtonian term, the result can be written in terms of Kummer's confluent hypergeometric function of the second kind, $U(a,b,z)$, which has the following integral representation (see, e.g.,~\cite{NIST:DLMF}):
\begin{equation}
    \Gamma(a) U(a,b,z) = \int_0^\infty dt \, e^{-zt} t^{a-1} (1+t)^{b-a-1} .
\end{equation}
Specifically, by choosing $a=1$, $b=(7-n)/4$, and $z\equiv-2\pi i f (t_c-t_\intr)$,
the inspiral memory can be written in terms of Kummer's function as
\begin{align} \label{eq:tilde-h-insp}
    \tilde{h}_{20}^{\insp}(f) = {}& \frac{4M}{7r} \sqrt{\frac{5 \pi}{6}} \eta \sum_n C^{\PN}_{n} C^{(1+n)}_{x} (t_c - t_\intr)^{(3-n)/4} \nonumber\\
    \times {}& e^{-2\pi ift_\intr} U\boldsymbol(1,(7-n)/4,-2\pi i f (t_c-t_\intr)\boldsymbol) \, .
\end{align}
We use the \texttt{scipy} Python package to compute the Kummer's confluent geometric functions in Eq.~\eqref{eq:tilde-h-insp}.\footnote{Although \texttt{scipy} has an implementation of $U(a,b,x)$, it is restricted to real values of $x$, whereas the third arguments of the Kummer functions in Eq.~\eqref{eq:tilde-h-insp} are evaluated on the imaginary axis. 
However, the \texttt{scipy} confluent hypergeometric function ${}_1F_1(a,b,z)$ and its implementation of the exponential integral $E_1(z)$ can be evaluated for complex $z$.
Thus, one can use Eq.~(13.2.42) of~\cite{NIST:DLMF} to rewrite $U(a,b,z)$ in terms of ${}_1F_1(a,b,z)$, when $b$ is not an integer (i.e., for $n\neq 3$).
When $n=3$, the Kummer function is $U(1,1,z)$, which from Eq.~(13.6.6) of~\cite{NIST:DLMF} is equal to $e^z E_1(z)$.}
%

\subsubsection{Fourier transform of the intermediate memory signal}

Taking the Fourier transform of the intermediate part of the memory model in Eq.~\eqref{eq:h_model_intermediate} is reasonably straightforward.
The relevant integrals that must be evaluated are
\begin{equation}
    \tilde{h}_{20}^{\intr}(f) = \frac M r\sum_{j=0}^6 \int_{t_\intr}^{t_\rd} dt \, e^{-2\pi i f t} c_j \, e^{p_j t} \, .
\end{equation}
A short calculation shows that the frequency-domain intermediate memory model is
\begin{equation} \label{eq:tilde-h-int}
    \tilde{h}_{20}^{\intr}(f) = \frac Mr\sum_{j=0}^6 \frac{c_j}{p_j-2\pi i f}\bigg[ e^{(p_j-2\pi i f)t_\rd} - e^{(p_j-2\pi i f)t_\intr} \bigg]
\end{equation}
(recall that $p_0=0$).
We dropped the zero-frequency Dirac delta-function contribution in Eq.~\eqref{eq:tilde-h-int}.

\subsubsection{Fourier transform of the ringdown memory signal}

The FT of the ringdown memory in Eq.~\eqref{eq:h20_ringdown_model} is a similar to that of the intermediate signal, because it is a superposition of complex (rather than real) exponential functions.
Given the somewhat involved form of the coefficients and the arguments of the complex exponentials, the result is rather lengthy:
\begin{widetext}
\begin{align} \label{eq:tilde-h-rd}
    \tilde{h}_{20}^{\rd} (f) = {}& \frac{-M^2}{2\sqrt{6}r}  \sum_{l',l''}\sum_{m'=1}^{l'} (-1)^{m'} C_l(-2,l',m';2,l'',-m') \nonumber\\
    {} & \times \sum_{\bar{l},\bar{\bar{l}},\bar{n},\bar{\bar{n}},j,j'} \eta^{j+j'} \bigg[ C_{\bar{l}m'\bar{n}j} \bar C_{\bar{\bar{l}}m'\bar{\bar{n}}j'} A_{l'\bar{l}m'} \bar A_{l''\bar{\bar{l}}m'} \bigg( \frac{\omega_{\bar{l}m'\bar{n}} \bar \omega_{\bar{\bar{l}}m'\bar{\bar{n}}}}{\omega_{\bar{l}m'\bar{n}} - \bar \omega_{\bar{\bar{l}}m'\bar{\bar{n}}}} \bigg) \frac{e^{-i((\omega_{\bar{l}m'\bar{n}} - \bar\omega_{\bar{\bar{l}}m'\bar{\bar{n}}})+2\pi f)t_\rd}}{(\omega_{\bar{l}m'\bar{n}} - \bar\omega_{\bar{\bar{l}}m'\bar{\bar{n}}})+2\pi f} \nonumber \\
    {}&- \bar C_{\bar{l}m'\bar{n}j} C_{\bar{\bar{l}}m'\bar{\bar{n}}j'} \bar A_{l'\bar{l}m'} A_{l''\bar{\bar{l}}m'} \bigg( \frac{\bar \omega_{\bar{l}m'\bar{n}} \omega_{\bar{\bar{l}}m'\bar{\bar{n}}}}{\bar \omega_{\bar{l}m'\bar{n}} - \omega_{\bar{\bar{l}}m'\bar{\bar{n}}}} \bigg) \frac{e^{-i(-(\bar\omega_{\bar{l}m'\bar{n}} - \omega_{\bar{\bar{l}}m'\bar{\bar{n}}})+2\pi f)t_\rd}}{(-(\bar\omega_{\bar{l}m'\bar{n}} - \omega_{\bar{\bar{l}}m'\bar{\bar{n}}})+2\pi f)}
    \bigg] \, .
\end{align}
\end{widetext}
We have used the fact that $\Im[\omega_{lmn}-\bar\omega_{l'm'n'}]<0$ to derive the above result.
The FT of the $\Delta h_{20}$ term in Eq.~\eqref{eq:h20tilde_insp_int_rd} is simply $\Delta h_{20}$ times the FT of the step function $\Theta(t-t_\rd)$.
We denote this contribution by
\begin{equation} \label{eq:tilde-h-Delta}
    \tilde h_{20}^\Delta(f) = \frac{\Delta h_{20}}2\left[\delta(f) + \frac{e^{-2\pi i f t_\rd}}{\pi i f}\right],
\end{equation}
where $\delta(f)$ is the Dirac delta function.

\subsubsection{Fourier transform of the full memory signal}

Although the time-domain memory signal is a piecewise function, each piece in the time domain has contributions at all frequencies.
The, Fourier transform is linear, so the total frequency-domain memory signal is given by the sum of the different time domain parts:
\begin{equation} \label{eq:h20_model_FT}
    \tilde h_{20}(f) = \tilde h_{20}^\insp(f) + \tilde h_{20}^\intr(f) +\tilde h_{20}^\Delta(f) - \tilde h_{20}^\rd(f) .
\end{equation}
The inspiral, intermediate, offset, and ringdown expressions are given in Eqs.~\eqref{eq:tilde-h-insp},~\eqref{eq:tilde-h-int},~\eqref{eq:tilde-h-Delta} and~\eqref{eq:tilde-h-rd}, respectively.

In some of the results below, we find it useful to compare our memory model with a step-function approximation of the memory, where the memory signal starts from zero and jumps to the final value computed from the final memory fit in Eq.~\eqref{eq:Deltahmem_fit}, at the peak time of the $l=2$, $m=2$ mode of the waveform $t/M = 0$:
\begin{align}
    h_{20}^{\step} = {}& \Delta h_{20} \Theta(t) . 
\end{align}
The FT of a step function was computed above, so we immediately write down the Fourier transform: 
\begin{equation}\label{eq:step_function_FT}
    \tilde{h}_{20}^{\step}(f) = \frac{\Delta h_{20}}2 \left[\delta(f) + \frac{1}{2\pi i f} \right] .
\end{equation}
Although we include the Dirac delta-function term in our analytical expression, when we compute the signal for any nonzero frequency range, it does not contribute to the expression (and we will often ignore it henceforth).
Thus, the step-function approximation to the memory is proportional to $1/f$ at all nonzero frequencies.

\subsection{Continuous and discrete Fourier transforms of the memory signal} \label{subsec:FTandFFT}

In this section, we compare our analytical frequency-domain model with the numerical FFT of the surrogate model, the FFT of the time-domain memory model, and the step-function approximation.
We first discuss some subtleties related to the computation of the FFT of the memory signal, which were also recently discussed in~\cite{Chen:2024ieh,Valencia:2024zhi}.
Our resolution to these subtleties was not discussed in either paper~\cite{Chen:2024ieh,Valencia:2024zhi}.

The main subtleties arise from the fact that the memory signal asymptotes to a nonzero value and is computed for a finite stretch of time.
When taking the discrete Fourier transform (or FFT) of this time series, the series is represented as a periodic function.
Thus, the offset between the earliest time and the final time in the time series behaves like a discontinuity when decomposed into the periodic basis of the FFT.
This discontinuity produces an artificial $1/f$ fall off at high frequencies associated with the difference between the initial and final values of the memory signal.

This data artifact in the FFT can be mitigated by windowing (or tapering) the signal. 
However, as was shown recently in~\cite{Chen:2024ieh,Valencia:2024zhi}, the choice of the window can produce other artifacts in the memory signal.
Methods were introduced to improve these effects, specifically the linear subtraction method of~\cite{Chen:2024ieh} and the symbolic sigmoid subtraction of~\cite{Valencia:2024zhi}.
We discuss a different method for mitigating the effects of the final offset when computing the FFT of the memory signal.

As we show in Appendix~\ref{app:hdot-calc}, one can compute the Fourier transform of the memory effect from the time derivative of the memory signal (namely, the integrand, which is proportional to the product of two $\dot h_{lm}$ modes).
Using the Fourier integral theorem, we find that the time derivative of the memory signal determines the Fourier transform, up to a term involving a delta function at zero frequency---see, Eq.~\eqref{eq:hdot-mem}.
Because we compute the memory signal for frequencies $f>0$, then we will use Eq.~\eqref{eq:hdot-mem} to compute the FFT, and we will neglect the zero-frequency contribution.
We will compare the result with our analytical expression for the frequency-domain waveform to show the efficacy of this procedure.

\begin{figure}
    \centering
    \includegraphics[width=0.48\textwidth]{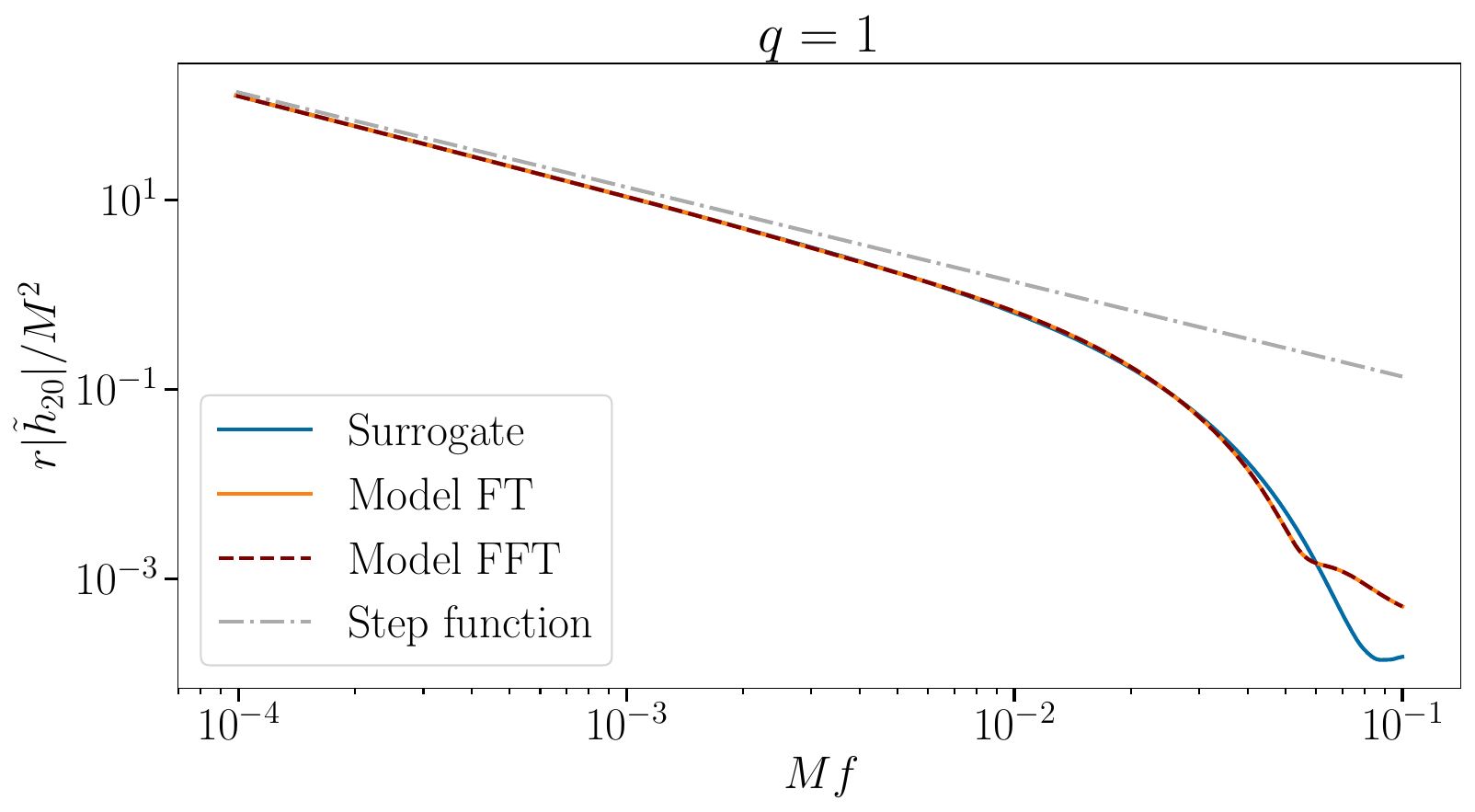}
    \includegraphics[width=0.48\textwidth]{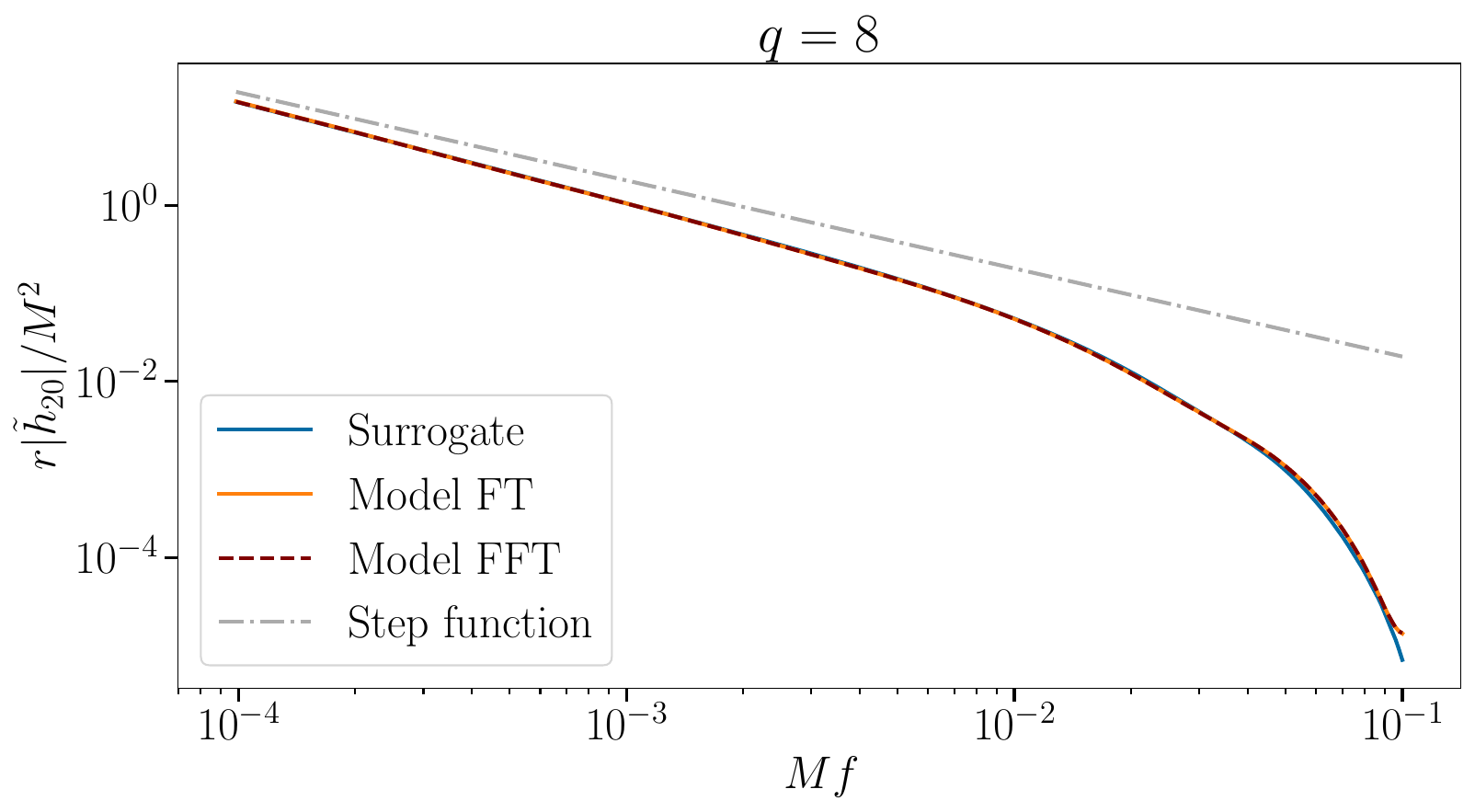}
    \caption{\textbf{Amplitude of the frequency-domain GW memory signal versus frequency}:
    The FFT of the surrogate (solid blue), the analytical FT of the time domain signal (solid orange), the FFT of the time-domain model (dashed maroon), and the step-function model (dashed-dotted gray) are shown for a mass ratio $q=1$ (top panel) and $q=8$ (bottom panel). 
    We discuss the qualitative features of the memory signals in the text of Sec.~\ref{subsec:FTandFFT}.}
    \label{fig:memory_fft}
\end{figure}
Specifically, Fig.~\ref{fig:memory_fft} shows the amplitude of the FFT of the surrogate memory signal (solid blue), the analytical FT (solid orange), the FFT of our time-domain model (dashed maroon), and the step-function approximation (dotted-dashed gray).
We show the two mass ratios $q=1$ (top panel) and $q=8$ (bottom panel), which are the largest and smallest mass ratios in our model.

At low frequencies, all four curves have the same $1/f$ behavior, which is related to the fact that the signal has the same final offset in all cases.
For $q=1$, the surrogate and the model converges more rapidly to the step function approximation at low frequencies than they do for $q=8$.
We suspect this occurs for a similar reason to why the inspiral memory signal in the time domain fits less well at $q=8$ than at $q=1$.
For a fixed frequency, the higher mass ratio system is more relativistic at this frequency, thereby making the PN inspiral contribution more important.
Thus, a larger fraction of the memory accumulates from frequencies lower than those depicted in the figure, thereby making the signal converge more slowly to the step-function approximation.

As the frequency increases, the amplitude of all the signals (aside from the step-function approximation) fall off faster than $1/f$.
This occurs because the surrogate and the memory models are smoother functions, which causes the frequency-domain representation of the functions to fall off more rapidly.
At frequencies above $Mf \sim 10^{-1}$, both the surrogate model and our memory model have high frequency artifacts.
For the memory model, the artifacts are related to the finite order of continuity of the derivatives of the time-domain signal at the times $t_\mathrm{int}$ and $t_\rd$.
We do not know the origin of the artifacts in the surrogate, but we suspect that they are related to the finite accuracy of the interpolant which forms the basis of the surrogate model.
Thus, we do not use the model above the dimensionless frequency of $0.1$.

\subsection{Mismatch results for advanced LIGO}

With the analytical frequency-domain waveform or the FFT of the time-domain waveform, we can compute the mismatch, so as to assess the performance of our memory waveform model.
The mismatch is defined to be
\begin{equation}
\label{eq:mismatch}
    \mathcal{M} = 1-\frac{\big<h_{\surr},h_{\model}\big>}{\sqrt{\big<h_{\surr},h_{\surr}\big>\big<h_{\model},h_{\model}\big>}} ,
\end{equation}
where the noise-weighted inner product between two signals $h_1$ and $h_2$ is given by
\begin{equation}
    \big<h_{1},h_{2}\big> = 4\mathcal{R}\Big[\int_{f_\mathrm{min}}^{f_\mathrm{max}}df \frac{\tilde{h}_1(f) \bar{\tilde{h}}_2(f)}{S_n(f)} \Big] .
\end{equation}
For the noise power-spectral density, $S_n(f)$, we use the ``low'' advanced LIGO design sensitivity curve for the fourth observing run, which was used in~\cite{KAGRA:2013rdx} and can be downloaded from~\cite{LIGOpsds}.

\begin{figure}
    \centering
    \includegraphics[width=0.48\textwidth]{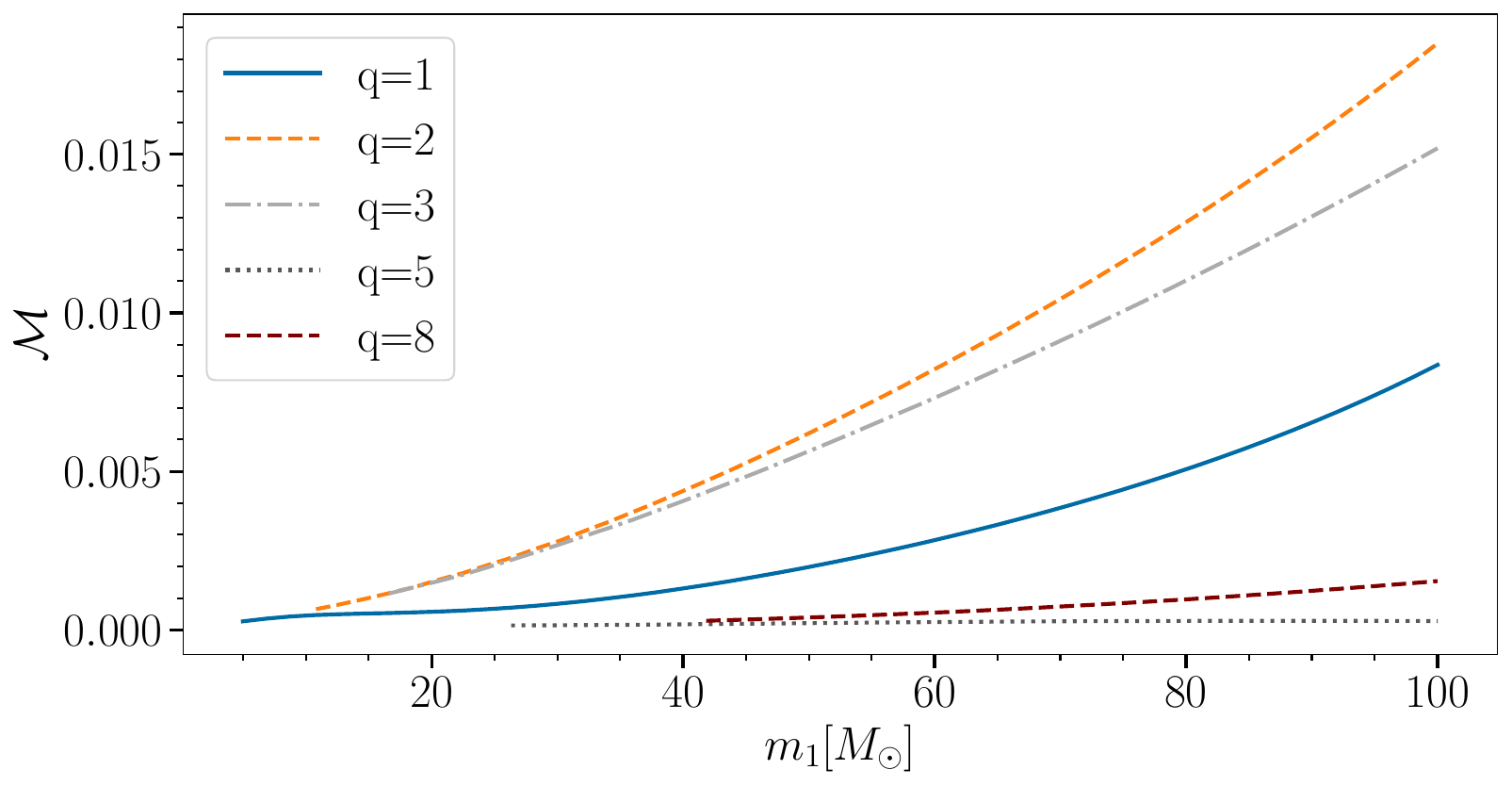}
    \includegraphics[width=0.48\textwidth]{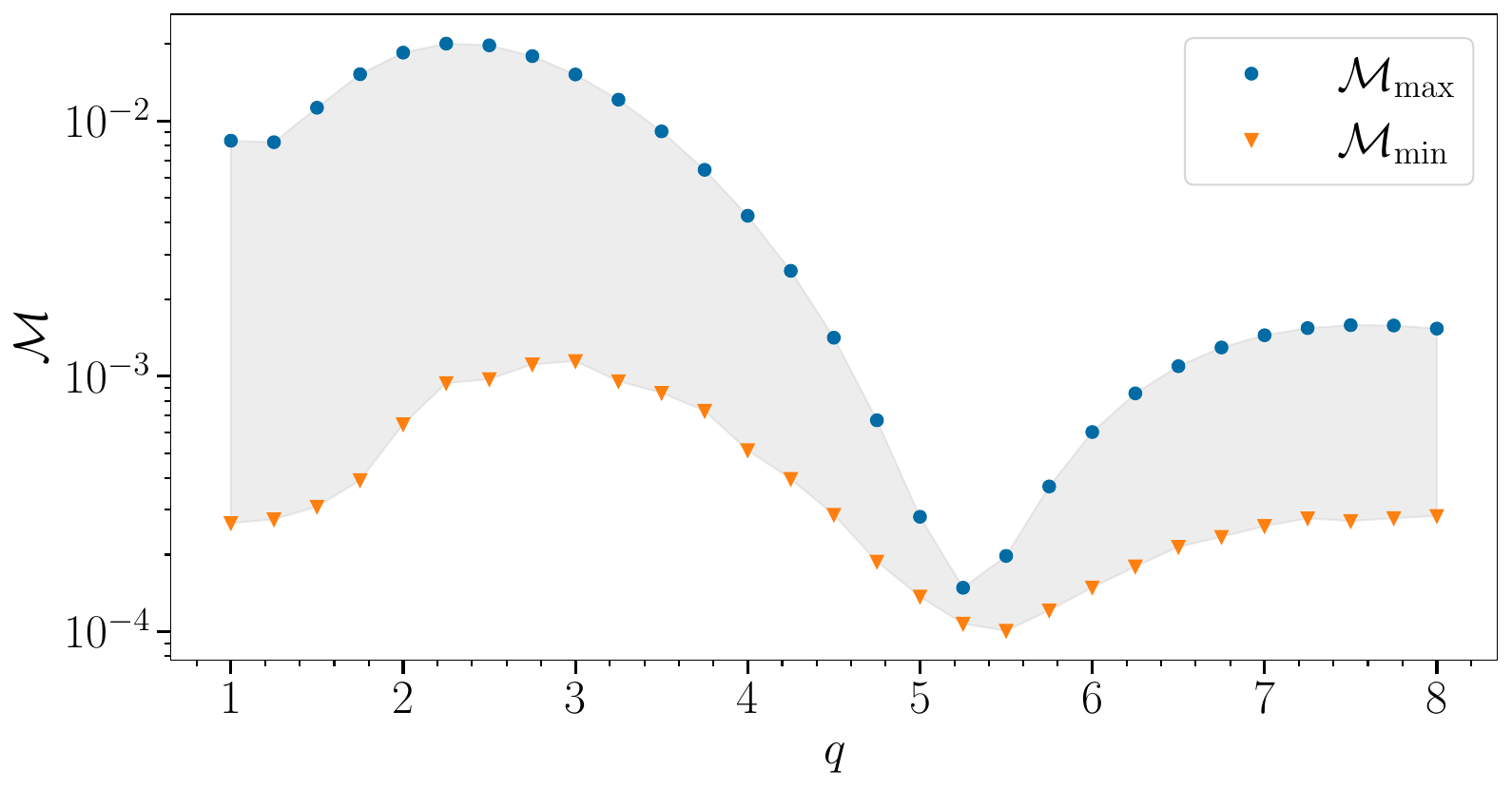}
    \caption{\textbf{Mismatch for different binary masses and mass ratios}:
    \emph{Top}: The mismatch $\mathcal{M}$ between our GW memory model and memory computed directly from the surrogate model versus the primary mass $m_1$ for BBH systems with different mass ratios $1\leq q\leq 8$.
    The specific mass ratios considered are indicated in the figure legend.
    We require that both the primary and secondary masses be greater than $5\, M_\odot$ but less than $100\,M_\odot$.
    \emph{Bottom}: The maximum and minimum mismatch versus mass ratio shown in the blue circles and orange triangles, respectively.
    The gray shaded region shows the range of mismatch values between the minimum and maximum.}
    \label{fig:mismatch}
\end{figure}
We compute the mismatch between the memory signal computed from the FFT of our time-domain model and the memory signal computed from the FFT of the surrogate, for different BBH systems with mass ratios $1\leq q \leq 8$.
The resulting mismatch for a selection of BBH systems with mass ratios ($q\in\{1,2,3,5,8\}$) is shown in the top panel of Fig.~\ref{fig:mismatch} as a function of the primary mass $m_1$.
We choose this mass to be in a range typical of the LIGO BBH detections, namely $m_1 \in [5,100] M_\odot$.
We also require that $m_2$ is in this same range, so the least massive primary mass increases as a function of increasing mass ratio.

The mismatch increases with the primary mass $m_1$ for all mass ratios.
This occurs because lower mass systems have more of the low-frequency $1/f$ behavior of the memory signal in the LIGO band, which is the range of frequencies at which our model and the surrogate model most closely agree.
At higher frequencies, the part of the signal that is in the LIGO band for more massive binaries, there is a larger disagreement between the two.
The bottom panel shows the maximum and minimum mismatch versus the mass ratio $q$.
The gray shaded region spans the same range of primary masses shown in the top panel; however the plot is displayed against the mass ratio on the horizontal axis.
The typical mismatch is of order $10^{-3}$, with a range that spans an order of magnitude.
This should be a sufficient accuracy for most analyses with LVK data.

Because the mismatch computes a normalized inner product between two GW signals, it is a measure of the alignment of the two signals, which depends primarily on the phase.
As we show in Appendix~\ref{app:memory_phase}, however, the phase of the complex frequency-domain memory signal does not evolve significantly over the frequency range of our model.
It would also be useful to determine how well the amplitudes of two waveforms agree.
For this purpose, we introduce a ``signal-to-noise ratio (SNR) mismatch'' which we define to be the following normalized difference of the optimal SNRs:
\begin{equation}
\label{eq:amplitude_mismatch}
    \mathcal{M}_{\rho} = \frac{\rho_{\surr}-\rho_{\model}}{\rho_{\surr}} .
\end{equation}
We use the notation $\rho=\sqrt{\left<h,h\right>}$ to denote the SNR.

\begin{figure}
    \centering
    \includegraphics[width=0.48\textwidth]{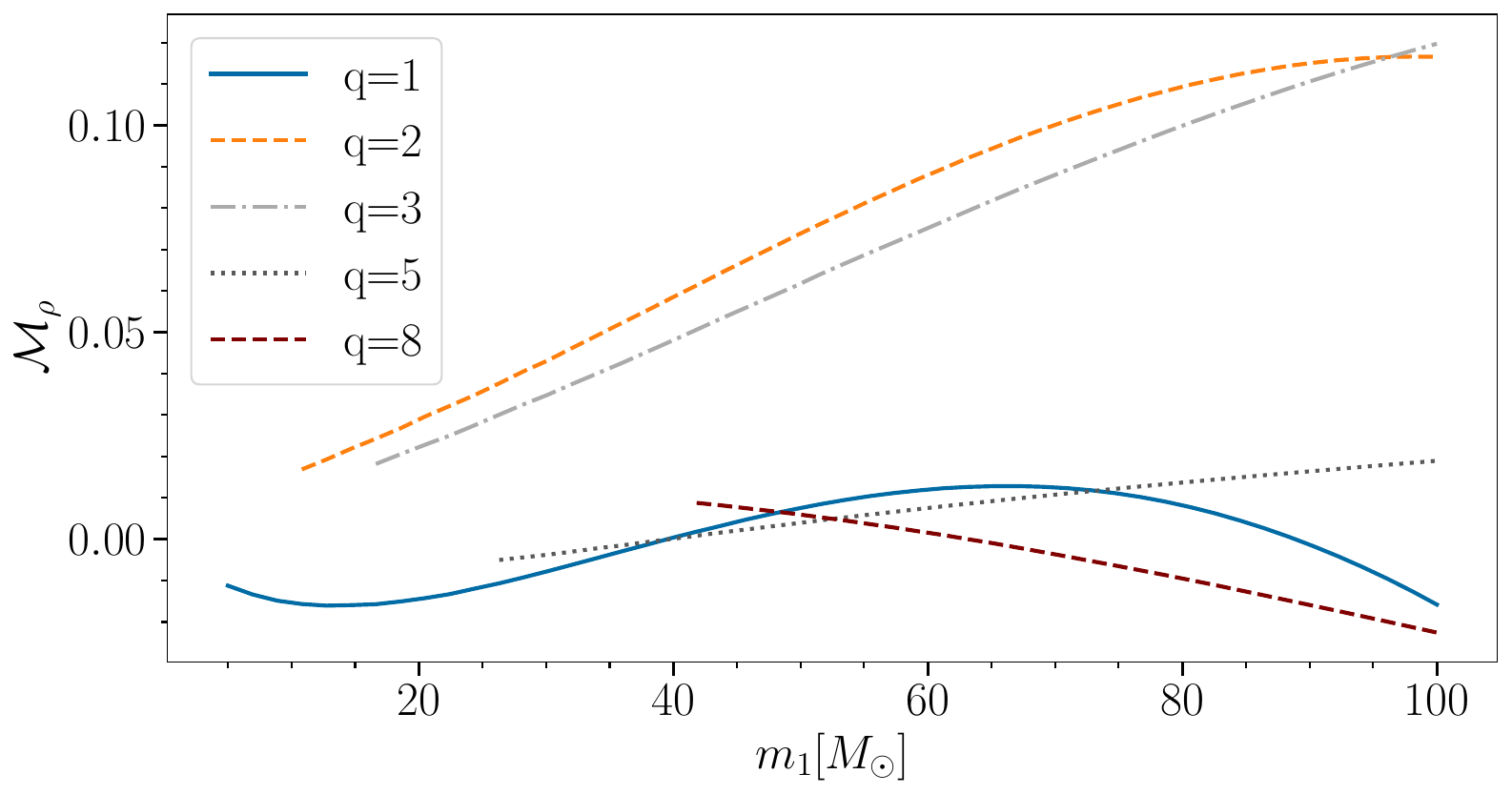}
    \includegraphics[width=0.48\textwidth]{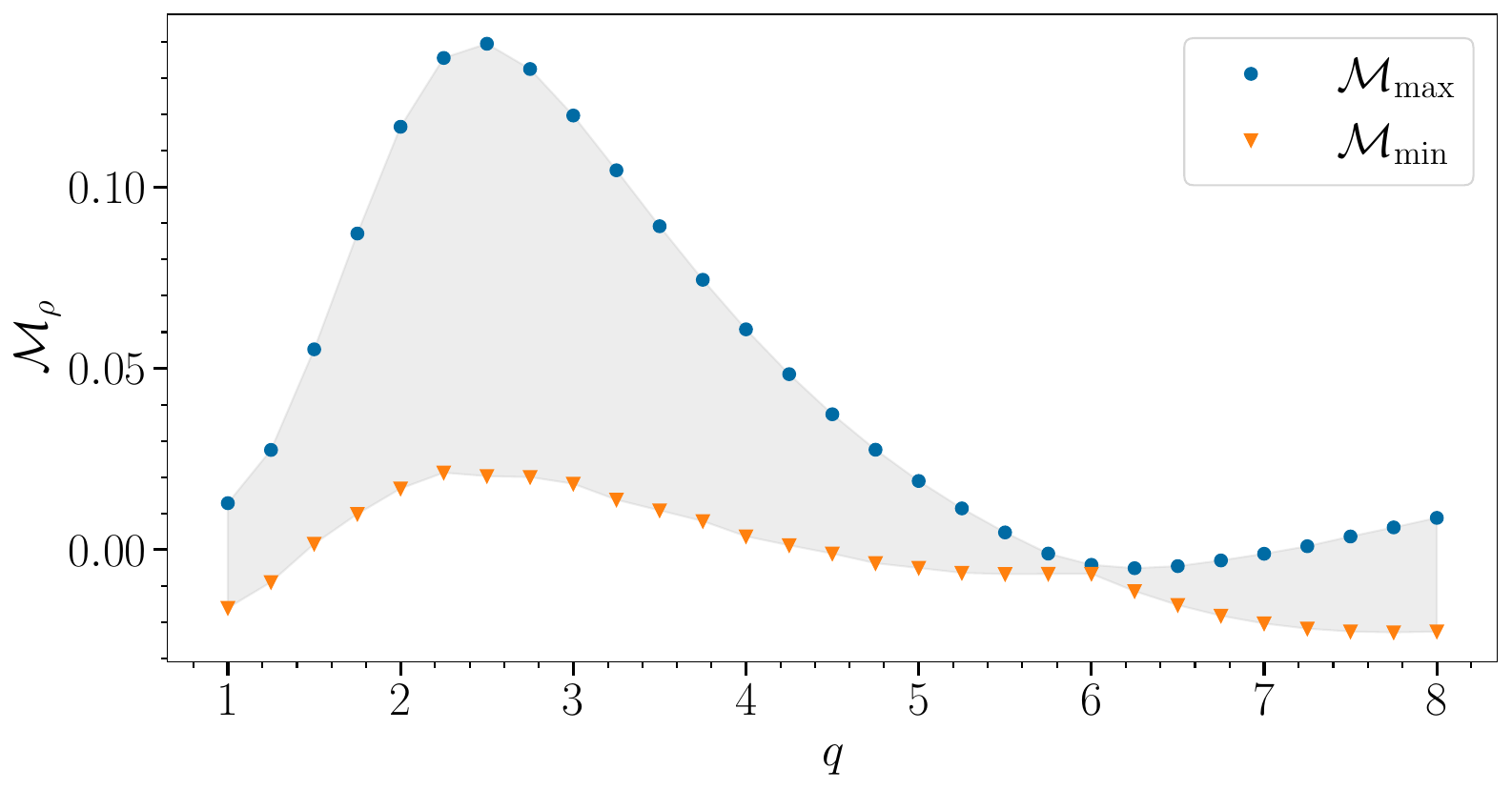}
    \caption{\textbf{SNR mismatch for different binary masses and mass ratios}:
    This figure is similar to Fig.~\ref{fig:mismatch} but it shows the SNR mismatch.
    \textit{Top}: The SNR mismatch $\mathcal{M}_\rho$ versus the primary mass $m_1$, which is computed for different BBH systems with different mass ratios $1\leq q\leq 8$ given in the legend of the figure.
    \textit{Bottom}: The maximum and minimum SNR mismatch versus mass ratio. 
    The orange triangles are again the minimum and the blue circles are the maximum. 
    The gray shaded region indicates the range of values between the extremes.}
    \label{fig:amplitude_mismatch}
\end{figure}
Figure~\ref{fig:amplitude_mismatch} shows the SNR mismatch for BBH systems with different mass ratios.
The top panel depicts the same mass ratios as in Fig~\ref{fig:mismatch}, though now for the SNR mismatch is plotted against $m_1$ on the vertical axis.
The bottom panel shows the range of SNR mismatches against the mass ratio.
Note that the SNR mismatch is no longer a monotonic function of the primary mass, and can be either positive or negative (namely, there are mass ratios for which the model can either underestimate or overestimate the SNR).
However, there is a trend that at smaller mass ratios, the SNR mismatch increases with increasing primary mass.
This likely occurs for reasons similar to those described above for the mismatch.

\section{Conclusions} \label{sec:conclusions}

In this paper, we continued our development of waveform models of the gravitational-wave memory effect, which was initiated in Paper I~\cite{Elhashash:2024thm}.
We produced a time-domain waveform model for the memory effect, which covered the inspiral, merger and ringdown stages of the waveform.
During the inspiral, we used an existing 3.5PN memory signal, which we calibrated by hybridizing the result to the NR hybrid surrogate model.
During the ringdown, we performed multimode QNM fitting for three oscillatory modes, and used the multimode fits to compute the corresponding memory signal.
Finally, we used a phenomenological ansatz for an intermediate temporal region of the memory waveform between the inspiral and merger-ringdown phases.
The memory signal over the three regions is continuous and had continuous first and second derivatives.
The model was calibrated to the NR surrogate, and it spanned a parameter space of nonspinning binary black holes with mass ratios from one to eight.

The Fourier transform of the full time-domain signal (inspiral, intermediate and ringdown stages) was computed analytically.
We also computed the FFT of the time-domain signal, which agreed with the analytical model.
We assessed the performance of the model by computing the mismatch between the memory computed from the surrogate and from our time-domain model.
The mismatch, at its largest, was of order $10^{-2}$, but the typical value was of order $10^{-3}$.
We also introduced an SNR mismatch to better determine how well the amplitudes of the two signals agreed or disagreed.
The SNR mismatch was an order of magnitude larger than the usual mismatch.

In future work, we would like to generalize this model to include the effects of black-hole spins.
We would start with spins aligned or anti-aligned with the orbital angular momentum.
Precession could also be added later. 
Given the relatively simple form of the frequency-domain amplitude and phase of the memory signal, we also plan to investigate purely phenomenological frequency-domain waveforms that directly model the signal in the Fourier domain.
Covering a larger region of the BBH parameter space is important for being able to use the model to analyze GW data from the LVK collaboration.
This will be the primary application of this and future iterations of our GW memory signal model.

\acknowledgments

A.E.\ and D.A.N.\ acknowledge support from the NSF grants No.\ PHY-2011784 and No.\ PHY-2309021. D.A.N.\ also acknowledges support from the NSF CAREER Award PHY-2439893.

\appendix

\section{Phase of the frequency-domain GW memory signals}
\label{app:memory_phase}

In this appendix, we discuss the phase of the complex frequency-domain GW memory signal.
To help interpret the phase, it is first helpful to consider the phase of the analytical FT of the step-function approximation to the memory signal in~Eq.~\eqref{eq:step_function_FT}.
The phase is the constant value of $-\pi/2$ for all frequencies $f > 0$, because the $1/f$ part of the solution has a negative, purely imaginary constant multiplying the $1/f$ term.
Thus, we expect our model to approach a phase of $-\pi/2$ at low frequencies.

\begin{figure}
    \centering
    \includegraphics[width=0.48\textwidth]{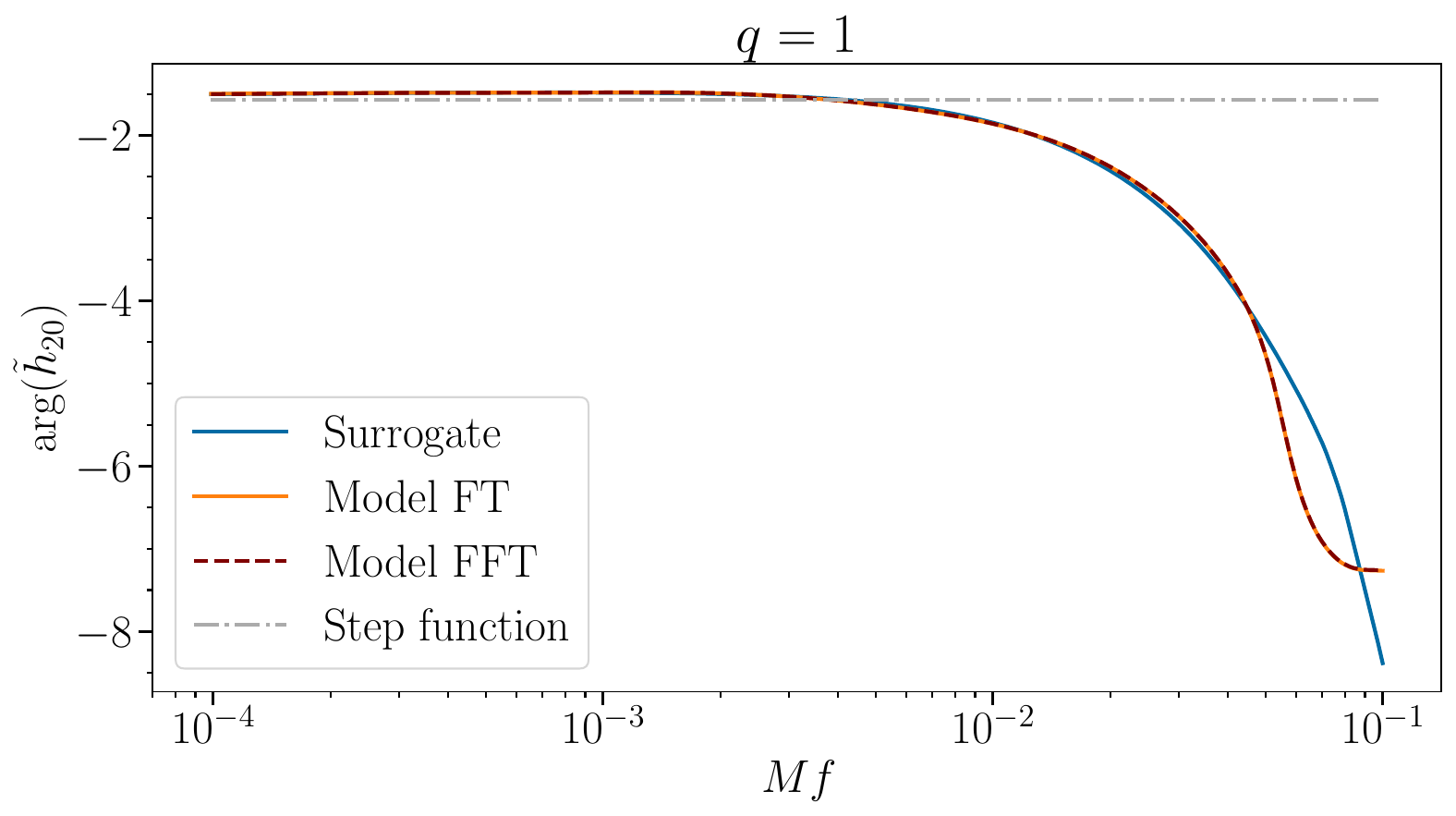}
    \includegraphics[width=0.48\textwidth]{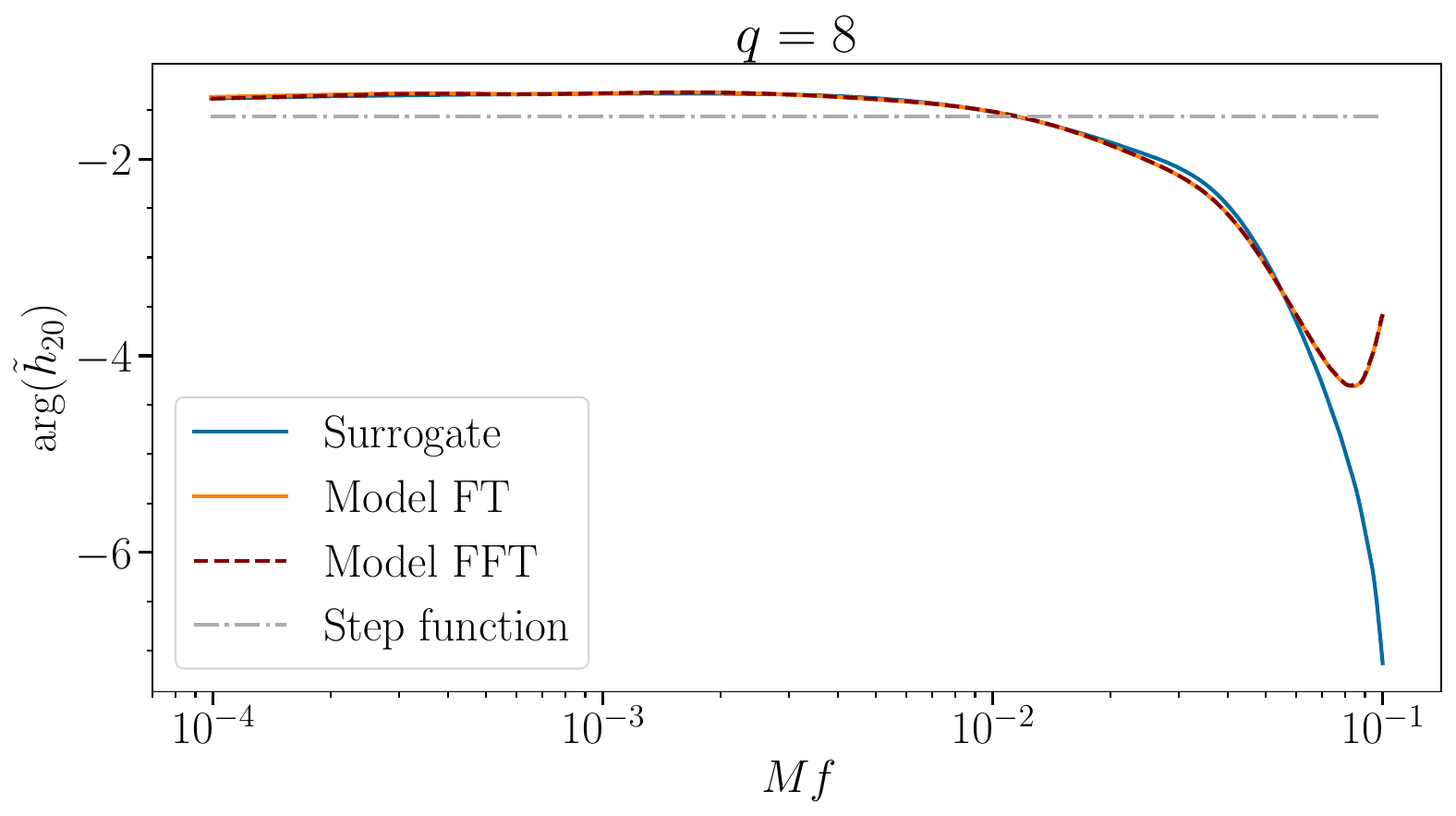}
    \caption{\textbf{Phase of the frequency-domain memory signal versus frequency}:
    The FFT of the surrogate memory signal (solid blue), the analytical FT of the time-domain model (solid orange), the FFT of the time-domain model (dashed maroon), and the step-function model (dashed-dotted gray) are shown for a mass ratio $q=1$ (top panel) and $q=8$ (bottom panel). 
}
    \label{fig:memory_fft_phase}
\end{figure}
In Fig.~\ref{fig:memory_fft_phase}, we show the phase of the FFT of the surrogate memory signal, the FFT and analytical FT of our time-domain model, and the step-function approximation, for mass ratios $q=1$ and $q=8$ in the top and bottom panels, respectively.
The solid, blue curve shows the phase of the FFT of the surrogate memory signal, while the solid orange curve shows the phase of the analytical FT of the memory signal computed from our time-domain memory model in Eq.~\eqref{eq:h20_model_FT}.
The figure also displays the FFT of our time-domain memory model (the dashed, maroon curve), which agrees well with the phase of the analytic FT of the memory model.
Finally, the phase of the step-function approximation in~Eq.~\eqref{eq:step_function_FT} is shown as the dashed-dotted, gray line.
The phases computed through the FFT were computed from signals that were time shifted to have the peak time reside at the end of the time series; otherwise the discrete Fourier time-shift theorem would modify the phase from that of the analytical expression.

As anticipated, at low frequencies, the phase of the FFT of the surrogate memory, the FFT of our model, and the analytic FT all approach the phase of the step-function model.
The convergence of the phase for the mass ratio of $q=8$ is slower than that of the lower mass ratio $q=1$, as it was for the amplitude (as discussed in Sec.~\ref{subsec:FTandFFT}.
The phase evolves by at most roughly one cycle over the three decades in frequency shown in Fig.~\ref{fig:memory_fft_phase}.
At higher frequencies, the phase of the memory signal that was computed directly from the NR surrogate begins to differ from that of our model.
It occurs at a similar frequency to that were the amplitude becomes less reliable, as well.

\section{Ringdown memory model for generic QNMs} \label{app:hlm_ringdown_memory}

In this section, we derive the ringdown memory model for a more generic superposition of QNMs that includes both prograde and retrograde modes.
We denote these QNM frequencies by $\omega^+_{lmn}$ and $\omega^-_{lmn}$, respectively.
The ringdown strain model can be written as a superposition of the two modes
\begin{align}\label{hmodel_rd_+-}
        h&^{\rd}_{\model}(t) =  \nonumber \\
        &\frac{M}{r}\sum_{l,m,n}  \bigg[  C_{lmn}^+ e^{-i\omega_{lmn}^+ (t-t_0)} {}_{-2}{S}_{lm}(\theta,\phi;a\omega^+_{lmn}) \nonumber \\
        & + C_{lmn}^- e^{-i\omega_{lmn}^- (t-t_0)} {}_{-2}{S}_{lm}(\theta,\phi;a\omega^-_{lmn}) \bigg] \, .
\end{align}
We can rewrite this expression in terms of only the prograde frequencies $\omega^+_{lmn}$ by using the following relation between the prograde and retrograde frequencies:
\begin{equation}
    \omega^+_{lmn} = -\bar{\omega}^-_{l-mn} \, .
\end{equation}
When written in terms of the prograde frequencies, Eq.~\eqref{hmodel_rd_+-} becomes
\begin{align}
        h^{\rd}_{\model}{}&(t) =  \nonumber\\
        &\frac{M}{r} \sum_{l,m,n}  \bigg[  C_{lmn}^+ e^{-i\omega_{lmn}^+ (t-t_0)} {}_{-2}{S}_{lm}(\theta,\phi;a\omega^+_{lmn}) \nonumber \\
        & + C_{lmn}^- e^{i\bar\omega_{l-mn}^+ (t-t_0)} {}_{-2}{S}_{lm}(\theta,\phi;-a\bar{\omega}^+_{l-mn}) \bigg]
        \label{eq:h_model_ringdown_allmodes} \, .
\end{align}

We now make use of a few properties of the spin-weighted spheroidal harmonics ${}_{s}S_{lm}(\theta,\phi;c)$ under several discrete transformations that change the sign of the harmonic indices, as well as changes in the sign of the spheroidal parameter $c$, complex conjugation of $c$, and reflections about the equatorial plane.
Specifically, these transformations are
\begin{subequations}
\begin{align}
    {}_{s}\bar{S}_{lm}(\theta,\phi;c) ={}& (-1)^{s+m}{}_{s}S_{l-m}(\theta, \phi; -\bar{c}) , \label{eq:Slmn_transformations_1} \\
    {}_{s}S_{l-m}(\theta,\phi;c) ={}& (-1)^{s+l} {}_{s}\bar S_{lm}(\pi-\theta, \phi; -\bar c) , \label{eq:Slmn_transformations_2}\\
    {}_{-s}S_{lm}(\theta,\phi;c) ={}& (-1)^{l+m} {}_{s}S_{lm}(\pi-\theta, \phi; c) \, .\label{eq:Slmn_transformations_3}
\end{align}
\end{subequations}
%
Relabeling the index $m$ as $-m$ in the sum involving the retrograde amplitudes, and using the property in Eq.~\eqref{eq:Slmn_transformations_2}, we can write the final line in Eq.~\eqref{eq:h_model_ringdown_allmodes} as
\begin{align}
    {}& h^{\rd}_{\model}  (t)=  \nonumber\\
    & \frac{M}{r} \sum_{l,m,n}  \bigg[  C_{lmn}^+ e^{-i\omega_{lmn}^+ (t-t_0)} {}_{-2}{S}_{lm}(\theta,\phi;a\omega^+_{lmn}) \nonumber \\
    & + (-1)^l C_{l-mn}^- e^{i\bar\omega_{lmn}^+ (t-t_0)} {}_{-2}\bar{S}_{lm}(\pi-\theta,\phi;a \omega^+_{lmn}) \bigg] \, .
\end{align}
%
We next use the expansion of the spin-weighted spheroidal harmonics in Eq.~\eqref{eq:Slm_expansion_Ylm} to write ${}_{-2}\bar{S}_{lm}(\pi-\theta,\phi;a\omega^+_{lmn})$ as
\begin{align}\label{eq:Slm_omega-}
    {}_{-2}\bar{S}_{lm}(\pi-\theta,\phi;a\omega^+_{lmn}) ={}&\nonumber\\
    \sum_{\bar{l}} &\bar{A}_{\bar{l}lm}(a\omega^+_{lmn}) {}_{-2}\bar{Y}_{\bar{l}m}(\pi-\theta,\phi)\, .
\end{align}
%
The spin-weighted spherical harmonics satisfy the properties
\begin{subequations}
\begin{align}
{}_{-s}\bar{Y}_{lm}(\theta,\phi) ={}& (-1)^{s+m} {}_{s}Y_{l-m}(\theta,\phi) \, ,\\
{}_{s}\bar{Y}_{l-m}(\pi-\theta,\phi) ={}& (-1)^{l+m} {}_{-s}Y_{l-m}(\theta,\phi) \, .
\end{align}
\end{subequations}
With these results, we can recast Eq.~\eqref{eq:Slm_omega-} as
\begin{align}
    {}_{-2}\bar{S}_{lm}(\pi-\theta,\phi;a\omega^+_{lmn}) =&\nonumber\\
    \sum_{\bar{l}} (-1)^{\bar{l}}& \bar{A}_{\bar{l}lm}(a\omega^+_{lmn}) {}_{-2}\bar Y_{\bar{l}-m}(\theta,\phi) \, .
\end{align}
%
In terms of spin-weighted spherical harmonics, the ringdown memory strain model is
\begin{align}
        h^{\rd}_{\model} {}& (t) = \nonumber\\
     & \frac{M}{r} \sum_{\bar{l},m} \bigg\{\sum_{l,n} \bigg[  C_{lmn}^+ A_{\bar{l}lm}(a\omega^+_{lmn}) e^{-i\omega_{lmn}^+ (t-t_0)} \nonumber \\
    & + (-1)^{l+\bar{l}} C_{lmn}^- \bar{A}_{\bar{l}l-m}(a\omega^+_{l-mn}) e^{i\bar{\omega}_{l-mn}^+ (t-t_0)}\bigg] \bigg\}\nonumber\\
    &\times {}_{-2}Y_{\bar{l}m}(\theta,\phi) \, .
\end{align}
%
We made the substitution of $m \rightarrow -m$ in the second term.
The memory model strain modes $h^{\mem}_{lm}(t)$ can be read-off from this expression as
\begin{align}\label{eq:hlm_model_rd_+-}
    h^{\model}_{lm} {}& (t)  \nonumber\\
    = & \frac{M}{r} \sum_{l',n} \bigg[ C_{l'mn}^+ A_{ll'm}(a\omega^+_{l'mn}) e^{-i\omega_{l'mn}^+ (t-t_0)} \nonumber \\
    & + (-1)^{l+l'} C_{l'mn}^- \bar{A}_{ll'-m}(a\omega^+_{l'-mn}) e^{i\bar{\omega}_{l'-mn}^+ (t-t_0)} \bigg] \, .
\end{align}
%
Because we express the memory only in terms of the prograde frequencies, we drop the $+$ label on the frequency (i.e., $\omega_{lmn}^+\equiv \omega_{lmn}$) and write $A_{\bar l lm}(a\omega_{lmn})\equiv A_{\bar l lm}$.
We compute the memory from Eq.~\eqref{eq:hlmMemory_hlm} by substituting the ringdown model strain modes $h_{lm}^{\model}(t)$ from Eq.~\eqref{eq:hlm_model_rd_+-} into the expression.
We integrate the product of two quasinormal modes, which for a single mode gives
\begin{align}
    \int_t^\infty dt' {}& e^{-i(\omega_{\bar{l}m'\bar{n}}-\bar{\omega}_{\bar{\bar{l}}-m''\bar{\bar{n}}})(t'-t_0)} \nonumber\\
    {}&= \frac{e^{-i(\omega_{\bar{l}m'\bar{n}}-\bar{\omega}_{\bar{\bar{l}}-m''\bar{\bar{n}}})(t-t_0)}}{-i(\omega_{\bar{l}m'\bar{n}}-\bar{\omega}_{\bar{\bar{l}}-m''\bar{\bar{n}}})}.
\end{align}
We used the fact that the imaginary parts of $\omega_{lmn}$ and $\omega_{l-mn}$ are negative, so the integration result vanishes at infinity.
Summing over multiple QNM modes, the ringdown memory $(l,m)$ modes are
\begin{widetext}
\begin{align}
    h_{lm}^{\mem}(t) = {}& i\frac{M^2}{r} \sqrt{\frac{(l-2)!}{(l+2)!}} \sum_{l',l'',m',m''}  (-1)^{m''} C_l(-2,l',m';2,l'',m'')  \nonumber\\
    \times {}& \sum_{\bar{l}, \bar{\bar{l}}, \bar{n}, \bar{\bar{n}}} \bigg[ \bigg(\frac{\omega_{\bar{l}m'\bar{n}} \, \bar{\omega}_{\bar{\bar{l}}-m''\bar{\bar{n}}}}{\omega_{\bar{l}m'\bar{n}}-\bar{\omega}_{\bar{\bar{l}}-m''\bar{\bar{n}}}}\bigg) C^+_{\bar{l}m'\bar{n}} \bar{C}^+_{\bar{\bar{l}}-m''\bar{\bar{n}}} A_{l'\bar{l}m'} \bar{A}_{l''\bar{\bar{l}}-m''} e^{-i(\omega_{\bar{l}m'\bar{n}}-\bar{\omega}_{\bar{\bar{l}}-m''\bar{\bar{n}}})(t-t_0)} \nonumber\\
    +{}& (-1)^{l'+l''+\bar{l}+\bar{\bar{l}}}\bigg(\frac{\bar{\omega}_{\bar{l}-m'\bar{n}} \, \omega_{\bar{\bar{l}}m''\bar{\bar{n}}}}{\bar{\omega}_{\bar{l}-m'\bar{n}} - \omega_{\bar{\bar{l}}m''\bar{\bar{n}}}}\bigg) C^-_{\bar{l}m'\bar{n}} \bar{C}^-_{\bar{\bar{l}}-m''\bar{\bar{n}}} \bar{A}_{l'\bar{l}-m'} A_{l''\bar{\bar{l}}m''} e^{i(\bar{\omega}_{\bar{l}-m'\bar{n}} - \omega_{\bar{\bar{l}}m''\bar{\bar{n}}})(t-t_0)} \nonumber\\
    -{}& (-1)^{l''+\bar{\bar{l}}}\bigg(\frac{\omega_{\bar{l}m'\bar{n}} \, \omega_{\bar{\bar{l}}m''\bar{\bar{n}}}}{\omega_{\bar{l}m'\bar{n}} + \omega_{\bar{\bar{l}}m''\bar{\bar{n}}}}\bigg) C^+_{\bar{l}m'\bar{n}} \bar{C}^-_{\bar{\bar{l}}-m''\bar{\bar{n}}} \bar{A}_{l'\bar{l}-m'} A_{l''\bar{\bar{l}}m''} e^{-i(\omega_{\bar{l}m'\bar{n}} + \omega_{\bar{\bar{l}}m''\bar{\bar{n}}})(t-t_0)} \nonumber\\
    +{}& (-1)^{l'+\bar{l}}\bigg(\frac{\bar{\omega}_{\bar{l}-m'\bar{n}} \, \bar{\omega}_{\bar{\bar{l}}-m''\bar{\bar{n}}}}{\bar{\omega}_{\bar{l}-m'\bar{n}} + \bar{\omega}_{\bar{\bar{l}}-m''\bar{\bar{n}}}}\bigg) C^-_{\bar{l}m'\bar{n}} \bar{C}^+_{\bar{\bar{l}}-m''\bar{\bar{n}}} \bar{A}_{l'\bar{l}-m'} \bar{A}_{l''\bar{\bar{l}}-m''} e^{i(\bar{\omega}_{\bar{l}-m'\bar{n}} - \bar{\omega}_{\bar{\bar{l}}-m''\bar{\bar{n}}})(t-t_0)} \bigg] \, .
\end{align}
\end{widetext}
Setting the amplitudes of the retrograde QNMs equal to zero reproduces the result in the main text.

\section{QNM fit coefficients} \label{app:rd-coeffs}

We give all of the QNM fit coefficients $C_{lmnj}$ in Table~\ref{tab:QNM-coeffs}.
They are given in terms of their modulus and phase.

\begin{table*}
    \centering
    \caption{The coefficients of the QNM fits for the three oscillatory modes with $(l,m)$ equal to $(2,1)$, $(2,2)$ and $(3,2)$.}
    \begin{tabular}{lcr|lcr|lcr}
    \hline
    \hline
    \multicolumn{3}{c|}{$l=2$, $m=1$} & \multicolumn{3}{c|}{$l=2$, $m=2$} & \multicolumn{3}{c}{$l=3$, $m=2$} \\
    \hline
    Coefficient & Amplitude & Phase & Coefficient & Amplitude & Phase & Coefficient & Amplitude & Phase \\
    \hline   
    $C_{2100}$ & $4.82281 \times 10^{-2}$ & $ -1.41047$ & $C_{2200}$ & $3.19296 \times 10^{-1}$ & $-2.77026$ & $C_{3200}$ & $3.74080 \times 10^{-2}$ & $0.289391$ \\
    $C_{2110}$ & $5.67762 \times 10^{-1}$ & $  2.13952 $ & $C_{2210}$ & $1.71980 \times 10^{0}$ & $1.01148$ & $C_{3210}$ & $2.16618 \times 10^{-1}$ & $-2.05806$ \\
    $C_{2120}$ & $3.80294 \times 10^{0}$ & $ -1.10247 $ & $C_{2220}$ & $5.92015 \times 10^{0}$ & $-1.59268$ & $C_{3220}$ & $5.24707 \times 10^{-1}$ & $2.64654$ \\
    $C_{2130}$ & $1.59829 \times 10^{1}$ & $ 1.85066 $ & $C_{2230}$ & $1.58611 \times 10^{1}$ & $1.84418$ & $C_{3230}$ & $3.57704 \times 10^{0}$ & $0.622898$ \\
    $C_{2140}$ & $3.91383 \times 10^{1}$ & $ -1.42840 $ & $C_{2240}$ & $2.88209 \times 10^{1}$ & $-1.19031$ & $C_{3240}$ & $1.06986 \times 10^{1}$ & $-2.26776$ \\
    $C_{2150}$ & $5.31916 \times 10^{1}$ & $ 1.619741 $ & $C_{2250}$ & $3.24851 \times 10^{1}$ & $1.99983$ & $C_{3250}$ & $1.52928 \times 10^{1}$ & $0.964857$ \\
    $C_{2160}$ & $3.71577 \times 10^{1}$ & $-1.58920 $ & $C_{2260}$ & $2.05656 \times 10^{1}$ & $-1.11638$ & $C_{3260}$ & $1.07697 \times 10^{1}$ & $-2.13948$ \\
    $C_{2170}$ & $1.03971 \times 10^{1}$ & $ 1.50051 $ & $C_{2270}$ & $5.55581 \times 10^{0}$ & $2.02049$ & $C_{3270}$ & $3.00686 \times 10^{0}$ & $1.01691$ \\
    $C_{2101}$ & $2.38039 \times 10^{0}$ & $ 2.31646$ & $C_{2201}$ & $9.27953 \times 10^{0}$ & $0.499780$ & $C_{3201}$ & $7.96203 \times 10^{-1}$ & $2.47208$ \\
    $C_{2111}$ & $1.48580 \times 10^{1}$ & $ -0.545106$ & $C_{2211}$ & $4.13466 \times 10^{1}$ & $-2.15539$ & $C_{3211}$ & $4.46748 \times 10^{0}$ & $0.131493$ \\
    $C_{2121}$ & $7.00706 \times 10^{1}$ &  $ 2.24528 $ & $C_{2221}$ & $1.28053 \times 10^{2}$ & $1.31012$ & $C_{3221}$ & $2.12174 \times 10^{0}$ & $2.38641$ \\
    $C_{2131}$ & $2.75410 \times 10^{2}$ & $ -1.27624$ & $C_{2231}$ & $2.91826 \times 10^{2}$ & $-1.67756$ & $C_{3231}$ & $5.97475 \times 10^{1}$ & $-2.14853$ \\
    $C_{2141}$ & $6.79526 \times 10^{2}$ & $ 1.64367 $ & $C_{2241}$ & $4.54998 \times 10^{2}$ & $1.48876$ & $C_{3241}$ & $1.97603 \times 10^{2}$ & $1.04384$ \\
    $C_{2151}$ & $9.34160 \times 10^{2}$ & $-1.62912 $ & $C_{2251}$ & $4.42701 \times 10^{2}$ & $-1.65677$ & $C_{3251}$ & $2.90392 \times 10^{2}$ & $-2.07309$ \\
    $C_{2161}$ & $6.58423 \times 10^{2}$ & $ 1.42574 $ & $C_{2261}$ & $2.44140 \times 10^{2}$ & $1.49805$ & $C_{3261}$ & $2.07484 \times 10^{2}$ & $1.07278$ \\
    $C_{2171}$ & $1.85623 \times 10^{2}$ & $-1.77847 $ & $C_{2271}$ & $5.95055 \times 10^{1}$ & $-1.62020$ & $C_{3271}$ & $5.85327 \times 10^{1}$ & $-2.07471$ \\
    $C_{2102}$ & $6.64711 \times 10^{0}$ & $-1.55498 $ & $C_{2202}$ & $3.31764 \times 10^{1}$ & $-3.10437$ & $C_{3202}$ & $3.27197 \times 10^{0}$ & $-0.783394$ \\
    $C_{2112}$ & $5.68596 \times 10^{1}$ & $ 1.93398 $ & $C_{2212}$ & $1.68339 \times 10^{2}$ & $0.635284$ & $C_{3212}$ & $1.60781 \times 10^{1}$ & $-3.06586$ \\
    $C_{2122}$ & $3.00496 \times 10^{2}$ & $-1.38291 $ & $C_{2222}$ & $5.05646 \times 10^{2}$ & $-2.11001$ & $C_{3222}$ & $1.50722 \times 10^{1}$ & $-2.65323$ \\
    $C_{2132}$ & $1.16339 \times 10^{3}$ & $  1.46882$ & $C_{2232}$ & $1.08443 \times 10^{3}$ & $1.20664$ & $C_{3232}$ & $2.94877 \times 10^{2}$ & $0.736590$ \\
    $C_{2142}$ & $2.80241 \times 10^{3}$ & $-1.86753 $ & $C_{2242}$ & $1.59435 \times 10^{3}$ & $-1.91641$ & $C_{3242}$ & $8.95912 \times 10^{2}$ & $-2.30299$\\
    $C_{2152}$ & $3.78281 \times 10^{3}$ & $  1.15090$ & $C_{2252}$ & $1.45739 \times 10^{3}$ & $1.19785$ & $C_{3252}$ & $1.27318 \times 10^{3}$ & $0.879673$ \\
    $C_{2162}$ & $2.62757 \times 10^{3}$ & $ -2.07400$ & $C_{2262}$ & $7.48094 \times 10^{2}$ & $-1.95375$ & $C_{3262}$ & $8.92803 \times 10^{2}$ & $-2.25037$ \\
    $C_{2172}$ & $7.31855 \times 10^{2}$ & $ 1.00835 $ & $C_{2272}$ & $1.69904 \times 10^{2}$ & $1.205426$ & $C_{3272}$ & $2.48810 \times 10^{2}$ & $0.888774$ \\
    \hline
    \hline
    \label{tab:QNM-coeffs}
    \end{tabular}
\end{table*}

\section{Alternate expression for the frequency-domain memory signal}
\label{app:hdot-calc}

Because the memory signal involves computing an integral of derivatives of $h(t)$, if we compute the Fourier transform of $\dot h(t) = dh/dt$ instead, then the memory can be evaluated from just the Fourier transform of an ``instantaneous'' quantity rather than a ``hereditary'' quantity (i.e., the time integral of the instantaneous integrand).
For a signal that goes to zero as $t$ approaches $\pm \infty$, this can be done with the Fourier integral theorem.
Computing the Fourier transform of quantities with a nonzero late-time value, such as the memory signal, has some subtleties.
We will avoid these by splitting the memory signal into a part that does approach zero at early and late times, and a step-function contribution as follows:
\begin{equation} \label{eq:h_mem_split}
    h_\mathrm{mem}(t) = \Delta h^\mathrm{fit}_\mathrm{mem} \Theta(t) + h_\mathrm{mem}^\mathrm{diff}(t) \, .
\end{equation}
The quantity $h_\mathrm{mem}^\mathrm{diff}(t)$ has the same time dependence as the memory signal for $t<0$; for $t>0$, it has the same time dependence as the memory, but it is offset by $-\Delta h^\mathrm{fit}$ from the memory signal.
Thus, it has a discontinuity at $t=0$, but it smoothly goes to zero as $t\rightarrow\pm\infty$.
Taking the derivative of the expression gives
\begin{equation} \label{eq:dot_h_mem_diff}
    \dot h_\mathrm{mem}(t) = \Delta h^\mathrm{fit}_\mathrm{mem} \delta(t) + \dot h_\mathrm{mem}^\mathrm{diff}(t) \, .
\end{equation}
The fact that the derivative (in a distributional sense) of a step function is a Dirac delta function was used above.
Note that the derivatives of $h_\mathrm{mem}(t)$ and $h_\mathrm{mem}^\mathrm{diff}(t)$ agree everywhere except $t=0$, where $\dot h_\mathrm{mem}(t)$ is well defined, but $\dot h_\mathrm{mem}^\mathrm{diff}(t)$ is singular (with a delta-function singularity that cancels the delta function at $t=0$).

Now we can take the Fourier transform of Eq.~\eqref{eq:h_mem_split} to find that
\begin{equation} \label{eq:h_mem_f_diff}
    \tilde h_\mathrm{mem}(f) = \Delta h^\mathrm{fit}_\mathrm{mem} \left[\frac 12 \delta(f) + \frac{1}{2\pi i f} \right] + \mathcal F[h_\mathrm{mem}^\mathrm{diff}] \, .
\end{equation}
Thus, we would like to evaluate the Fourier transform of $h_\mathrm{mem}^\mathrm{diff}$ in terms of other smooth functions and elementary functions.
Since $h_\mathrm{mem}^\mathrm{diff}(t)$ has a discontinuity at $t=0$, we write the Fourier transform out explicitly as
\begin{align}
    \mathcal F[h_\mathrm{mem}^\mathrm{diff}] = {} & \lim_{\epsilon\rightarrow 0} \int_{-\infty}^{-\epsilon} h_\mathrm{mem}^\mathrm{diff} e^{-i2\pi f t} dt \nonumber \\
    & + \lim_{\epsilon\rightarrow 0} \int_\epsilon^\infty h_\mathrm{mem}^\mathrm{diff} e^{-i2\pi f t} dt \, .
\end{align}
Next we integrate by parts and use the fact that $h_\mathrm{mem}^\mathrm{diff}(t)$ vanishes as $t\rightarrow\pm\infty$ (which removes one set of boundary terms) but has a discontinuity at $t=0$.
This means that there are boundary terms that are nonzero and depend on whether one approaches from $t < 0$ or from $t>0$ .
This gives that
\begin{align}
    \mathcal F[h_\mathrm{mem}^\mathrm{diff}] = {} & \lim_{\epsilon\rightarrow 0}\frac{h_\mathrm{mem}^\mathrm{diff}(\epsilon)-h_\mathrm{mem}^\mathrm{diff}(-\epsilon)}{2\pi i f} \nonumber \\
    & + \frac{1}{2\pi i f} \lim_{\epsilon\rightarrow 0} \int_{-\infty}^{-\epsilon} \dot h_\mathrm{mem}^\mathrm{diff} e^{-i2\pi f t} dt \nonumber \\
    & + \frac{1}{2\pi i f} \lim_{\epsilon\rightarrow 0} \int_\epsilon^\infty \dot h_\mathrm{mem}^\mathrm{diff} e^{-i2\pi f t} dt \, .
\end{align}
The limit of $h_\mathrm{mem}^\mathrm{diff}(\epsilon)-h_\mathrm{mem}^\mathrm{diff}(-\epsilon)$ is just $-\Delta h_\mathrm{mem}^\mathrm{fit}$.
Note that $\dot h_\mathrm{mem}^\mathrm{diff}$ agrees with $\dot h_\mathrm{mem}$ everywhere except $t=0$, but $\dot h_\mathrm{mem}$ is finite at $t=0$.
This allows us to replace the two integrals of $\dot h_\mathrm{mem}^\mathrm{diff}$ over the positive and negative real numbers with a single integral of $\dot h_\mathrm{mem}$ over the entire reals.
This latter integral is just the Fourier transform of $\dot h_\mathrm{mem}$, so we determine that
\begin{equation} 
    \mathcal F[h_\mathrm{mem}^\mathrm{diff}] = \frac{1}{2\pi i f}  \mathcal F[\dot h_\mathrm{mem}] - \frac{1}{2\pi i f} \Delta h^\mathrm{fit}_\mathrm{mem} \, .
\end{equation}
The memory signal's derivative $\dot h_\mathrm{mem}$ has no discontinuities and approaches zero as $t\rightarrow\pm\infty$.
Substituting this expression for $\mathcal F[h_\mathrm{mem}^\mathrm{diff}]$ into Eq.~\eqref{eq:h_mem_f_diff} gives
\begin{equation} \label{eq:hdot-mem}
    \tilde h_\mathrm{mem}(f) = \frac 12 \Delta h^\mathrm{fit}_\mathrm{mem} \delta(f) + \frac{1}{2\pi i f} \mathcal F[\dot h_\mathrm{mem}] \, .
\end{equation}
Thus, aside from a delta function at zero frequency, we can compute the Fourier transform $\tilde h_\mathrm{mem}(f)$ from the Fourier transform for $\dot h_\mathrm{mem}$.
Aside from the delta-function term, the result has the same form as the Fourier integral theorem.

\bibliography{refs}

\end{document}